%% file: paper20200812.tex
\documentclass[iop,apj]{emulateapj}

\usepackage{acronym}
\usepackage{booktabs}
\usepackage{dcolumn}
\usepackage{graphicx}
\usepackage{amsmath,amsfonts,amssymb}
\usepackage{mathrsfs}
\usepackage[utf8]{inputenc}
\usepackage[normalem]{ulem}
\usepackage[breaklinks,colorlinks,citecolor=blue]{hyperref}
\usepackage{graphicx,psfrag}
\usepackage{multirow}
\usepackage{comment}
\usepackage{ulem}
\usepackage{hyperref}
\usepackage{color}

\newcolumntype{d}[1]{D{.}{.}{#1}}
\usepackage{natbib}
\newcommand{\swind}{spiral-wave wind}
\newcommand{\nwind}{$\nu$-component}

\newcommand{\dd}{\mathrm{d}}

\newcommand{\eg}{\textit{e.g.,}}

\newcommand{\be}{\begin{equation}}
\newcommand{\ee}{\end{equation}}
\newcommand{\bea}{\begin{eqnarray}}
\newcommand{\eea}{\end{eqnarray}}
\newcommand{\bel}{\begin{align}}
\newcommand{\eel}{\end{align}}

\def\gccm{{\rm g\,cm^{-3}}}
\def\gcc{{\rm g\,cm^{-3}}}
\def\Msun{{\rm M_{\odot}}}
\def\GMc2{{\rm G M_{\odot} c^{-2}}}

\def\O{\mathcal{O}}
\def\M{\mathcal{M}}

\def\Mo{{\rm M_{\odot}}}

\def\md{M^{\rm d}_{\rm ej}}
\def\vd{v^{\rm d}_\infty}
\def\yd{Y^{\rm d}_{e}}
\def\amd{\md}
\def\avd{\langle\vd\rangle}
\def\ayd{\langle\yd\rangle}
\def\mw{M^{\rm w}_{\rm ej}}
\def\vw{v^{\rm w}_\infty}
\def\yw{Y^{\rm}_{e}}
\def\amw{\mw}
\def\avw{\langle\vw\rangle}
\def\ayw{\langle\yw\rangle}

\usepackage{pifont} 
\newcommand{\cmark}{\ding{51}}%
\newcommand{\xmark}{\ding{55}}%

\DeclareOldFontCommand{\rm}{\normalfont\rmfamily}{\mathrm}
\DeclareOldFontCommand{\sf}{\normalfont\sffamily}{\mathsf}
\DeclareOldFontCommand{\tt}{\normalfont\ttfamily}{\mathtt}
\DeclareOldFontCommand{\bf}{\normalfont\bfseries}{\mathbf}
\DeclareOldFontCommand{\it}{\normalfont\itshape}{\mathit}
\DeclareOldFontCommand{\sl}{\normalfont\slshape}{\@nomath\sl}
\DeclareOldFontCommand{\sc}{\normalfont\scshape}{\@nomath\sc}
\DeclareRobustCommand*\cal{\@fontswitch\relax\mathcal}
\DeclareRobustCommand*\mit{\@fontswitch\relax\mathnormal}

\def\arraybackslash{\let\\\tabularnewline}

\definecolor{cyan}{rgb}{0,0.9,0.9}
\definecolor{orange}{rgb}{0.9,0.5,0}
\definecolor{magenta}{rgb}{1,0,1}
\definecolor{purple}{rgb}{0.8,0.4,0.8}
\definecolor{darkgreen}{rgb}{0.0,0.5,0.0}
\definecolor{gray}{rgb}{0.8242,0.8242,0.8242}
\definecolor{cadmiumgreen}{rgb}{0.0, 0.42, 0.24}

\newcommand{\red}[1]{\textcolor{red}{#1}}

\begin{document}

\title{Numerical Relativity Simulations of the Neutron Star Merger GW170817: \\
Long-Term Remnant Evolutions, Winds, Remnant Disks, and Nucleosynthesis}

\author{Vsevolod \surname{Nedora}$^{1}$,
  Sebastiano \surname{Bernuzzi}$^{1}$,
  David \surname{Radice}$^{2,3,4}$,
 Boris \surname{Daszuta}$^{1}$,
 Andrea \surname{Endrizzi}$^{1}$,
 Albino \surname{Perego}$^{5,6}$,
 Aviral \surname{Prakash}$^{2,3}$,
 Mohammadtaher \surname{Safarzadeh}$^{7}$,
 Federico \surname{Schianchi}$^{1}$,
 Domenico \surname{Logoteta}$^{8,9}$\\
$\quad$ \\
\textit{${}^1$Theoretisch-Physikalisches Institut, Friedrich-SchillerUniversit\"{a}t Jena, 07743, Jena, Germany\\
  ${}^2$Institute for Gravitation \& the Cosmos, The Pennsylvania State University, University Park, PA 16802, USA\\
  ${}^3$Department of Physics, The Pennsylvania State University, University Park, PA 16802, USA\\
  ${}^4$Department of Astronomy \& Astrophysics, The Pennsylvania State University, University Park, PA 16802, USA\\
  ${}^5$Dipartimento di Fisica, Universit\`{a} di Trento, Via Sommarive 14, 38123 Trento, Italy\\
  ${}^6$INFN-TIFPA, Trento Institute for Fundamental Physics and Applications, via Sommarive 14, I-38123 Trento, Italy\\
  ${}^7$Center for Astrophysics, Harvard \& Smithsonian, 60 Garden Street, Cambridge, MA, USA\\
  ${}^8$Dipartimento di Fisica, Universit\`{a} di Pisa, Largo Pontecorvo 3, 56127 Pisa, Italy\\
  ${}^9$Istituto Nazionale di Fisica Nucleare (INFN), Largo Pontecorvo 3,56127 Pisa, Italy}
}

\begin{abstract} 
  We present a systematic numerical-relativity study of the dynamical
  ejecta, winds and nucleosynthesis in neutron
  star merger remnants. Binaries with the chirp mass compatible with GW170817,
  different mass ratios, and five microphysical
  equations of state (EOS) are simulated with an approximate neutrino transport 
  and a subgrid model for magnetohydrodynamics turbulence up to $100$ milliseconds
  postmerger.
  Spiral density waves propagating from the neutron star remnant to the disk
  trigger a wind with mass flux ${\sim}0.1{-}0.5\,\Msun/s$ persisting for
  the entire simulation as long as the remnant does not collapse to
  black hole. This wind has average electron fraction $\gtrsim 0.3$
  and average velocity ${\sim}0.1-0.17\,$c 
  and thus is a site for the production of weak $r$-process
  elements (mass number $A<195$).
  Disks around long-lived remnants have masses ${\sim}0.1-0.2\,\Msun$,
  temperatures peaking at $\lesssim10\,{\rm MeV}$ near the inner edge,
  and a characteristic double-peak distribution in entropy resulting from 
  shocks propagating through the disk.
  The dynamical and spiral-wave ejecta computed in our targeted
  simulations are not compatible 
  with those inferred from AT2017gfo using two-components kilonova models.
  Rather, they indicate that multi-component kilonova models including
  disk winds are necessary to interpret AT2017gfo.   
  The nucleosynthesis in the combined dynamical ejecta and \swind{} in the
  comparable-mass long-lived mergers robustly accounts for all the 
  $r$-process peaks, from mass number ${\sim}75$ to actinides
  in terms of solar abundances. Total abundandes are weakly dependent
  on the EOS, while the mass ratio affect the production of first peak elements. 
\end{abstract}

\section{Introduction}

The mass ejection of neutron-rich matter from binary neutron star (BNS)
mergers has been theoretically studied since the `70s as a possible site for
$r$-process nucleosynthesis 
\citep{Lattimer:1974a,Symbalisty:1982a,Rosswog:1998hy,Freiburghaus:1999,Rosswog:2005su}. 
The radioactive decay of $r$-process elements produces a
characteristic electromagnetic (EM) transient in the UV/optical/NIR
bands, called kilonova (kN) \citep{Li:1998bw,Kulkarni:2005jw,Metzger:2010sy,Roberts:2011xz,Kasen:2013xka},
that was observed as counterpart of the gravitational-wave event GW170817 \citep{Abbott:2017wuw,Abbott:2017oio,Abbott:2018wiz,Abbott:2018hgk}
and it was named AT2017gfo
\citep{Arcavi:2017xiz,Coulter:2017wya,Drout:2017ijr,Evans:2017mmy,Hallinan:2017woc,Kasliwal:2017ngb,Nicholl:2017ahq,Smartt:2017fuw,Soares-santos:2017lru,Tanvir:2017pws,Troja:2017nqp,Mooley:2018dlz,Ruan:2017bha,Lyman:2018qjg}.
The NIR luminosity of AT2017gfo peaked at several days after the
merger \citep{Chornock:2017sdf}, and it is consistent with the expectation
that the opacities of expanding $r$-process material are dominated by 
lanthanides and possibly actinides \citep{Kasen:2013xka}.  
The UV/optical luminosity peaked instead less than one day after the
merger \citep{Nicholl:2017ahq}, and it originates from ejected material
that experienced only a partial $r$-process nucleosynthesis
\citep{Martin:2015hxa}.

The ejecta masses inferred from observations \citep{Cowperthwaite:2017dyu,Villar:2017wcc,Tanvir:2017pws,Tanaka:2017qxj,Perego:2017wtu,Kawaguchi:2018ptg} are not compatible with those predicted by numerical simulations with targeted neutron star (NS) masses, and several questions remain open. 
In particular, understanding the early blue kN remains a challenging
aspect to most models.
Both semi-analytical and radiation transport
models require large ejecta velocities and electron fractions ($Y_e$'s),
different from those found in simulations,
\citep[\eg][]{Fahlman:2018llv,Nedora:2019jhl}.
The late red kN component requires ejecta masses generally
not observed for the dynamical ejecta computed in numerical relativity (NR) simulations \citep{Radice:2018pdn}. 
In addition, the number of components and the geometry of the emission can have a significant 
effect on the ejecta parameters 
\citep{Perego:2017wtu,Kawaguchi:2018ptg}. Also, it is important to note that the photon diffusion and
emission is often simplified in semi-analytical kN models \citep[\eg][]{Villar:2017wcc,Perego:2017wtu,Siegel:2019mlp}, and 
the more accurate radiation transfer computations may alter the inferred ejecta parameters 
\citep{Kawaguchi:2018ptg,Korobkin:2020spe}. However, photon radiation transfer simulations often employ ad-hoc, simplified ejecta that are not computed from ab-initio simulations.

Key for interpreting BNS electromagnetic emissions is the detailed modeling of
the mass ejection from BNS mergers, which must include general
relativity, a microphysical equation of states (EOS) of strongly interacting
matter, relativistic (magneto-)hydrodynamics, and neutrino
transport. NR simulations performed so far
mostly focused on the dynamical ejecta that are launched during merger
by tidal torques (tidal component) and by the shocks generated by the
bounce of the NS cores (shocked component),
\citep[\eg][]{Hotokezaka:2013iia,Bauswein:2013yna,Wanajo:2014wha,Sekiguchi:2015dma,Radice:2016dwd,Sekiguchi:2016bjd,Radice:2018pdn,Vincent:2019kor}. In
equal-mass mergers, the shocked component is found to be a factor
${\sim}10$ more massive than the tidal component. This is in
contrast to early works that employed Newtonian gravity and in which
the tidal component dominated the ejecta due to the weaker gravity and
stiffer EOS employed in those simulations
\citep{Ruffert:1996by,Rosswog:1998hy,Rosswog:2001fh,Rosswog:2003rv,Rosswog:2003tn,Rosswog:2003ts,Oechslin:2006uk,Rosswog:2013kqa,Korobkin:2012uy}.
However, even the dynamical ejecta found in NR simulations cannot account alone for the
bright blue and late red components of the observed kN in AT2017gfo \citep{Siegel:2019mlp}.

Winds originating from the merger remnant on timescales of
$\O(0.1-1\,{\rm s})$ can unbind $O(0.1\, \Msun)$ from the remnant and
represent (if present) the largest contribution to the kilonova signal
\citep{Dessart:2008zd,Fernandez:2014bra,Just:2014fka,Lippuner:2017bfm,Siegel:2017nub,Fujibayashi:2017puw,Radice:2018xqa,Fernandez:2018kax,Janiuk:2019rrt,Miller:2019dpt,Fujibayashi:2020qda,Mosta:2020hlh}.
Thus far, these winds have been mostly studied by means of long-term
Newtonian simulations of neutrino-cooled disks, assuming simplified
initial conditions, \citep[\eg]{Metzger:2008av,Beloborodov:2008nx,Lee:2009uc,Fernandez:2012kh}.
Ab-initio (3+1)D NR simulations of the merger with weak-interactions and
magnetohydrodynamics are not yet fully developed at sufficiently long timescales \citep{Sekiguchi:2011zd,Wanajo:2014wha,Sekiguchi:2015dma,Palenzuela:2015dqa,Radice:2016dwd,Lehner:2016lxy,Sekiguchi:2016bjd,Foucart:2016vxd,Bovard:2017mvn,Fujibayashi:2017puw,Fujibayashi:2017xsz,Radice:2018xqa,Nedora:2019jhl,Vincent:2019kor,Bernuzzi:2020txg}.
These simulations are essential to interpret AT2017gfo and future events.
For example, long-term (up to $100$~ms postmerger) NR simulations 
pointed out the existence of \swind{} in which there are
favourable conditions (large ejecta mass, high-velocity and not extremely neutron rich conditions) 
for the early emission from lanthanide poor material \citep{Nedora:2019jhl}.
Such mass ejection can also be boosted by global large-scale magnetic
stresses \citep{Metzger:2018uni,Siegel:2017jug,Siegel:2017nub},
although significant mass fluxes can only be achieved by fine-tuning
initial configuration or setting unrealistic strength of the magnetic
field \citep[\eg][]{Ciolfi:2020hgg,Mosta:2020hlh}.
A third contribution can come from neutrino-driven winds of mass ${\sim} 10^{-4}-10^{-3}M_{\odot}$ originating above the remnant, 
but their mass cannot account for bright signals \citep{Dessart:2008zd,Perego:2014fma,Just:2014fka}.

The nucleosynthesis from BNS mergers is believed to provide a major
contribution to the $r$-process material in the Universe.
However, whether or not BNS mergers are the only source is still debated and 
possible additional $r$-process sites, such as collapsars, 
jet-driven supernovae, and neutron star implosions, have been proposed
\citep{Argast:2003he,Duan:2010af,Winteler:2012hu,Nishimura:2015nca,Hirai:2015npa,Bramante:2016mzo,Nishimura:2016hak,Fuller:2017uyd,Mosta:2017geb,Siegel:2018zxq,Ji:2019ssk,Bartos:2019twj,vandeVoort:2019bxg,Wehmeyer:2019ovu,Vassh:2019cey}.
In particular, it is not clear if BNS mergers can explain r-process enriched ultra-faint dwarf galaxies,
classical dwarf galaxies \citep{Ji:2015wzg,Bramante:2016mzo,Safarzadeh:2018ent,Safarzadeh:2018fdy,Skuladottir:2019bjz,Bonetti:2019fxj}, 
and the evolution of $r$-process abundances both at early and late times \citep{Safarzadeh:2017riw,Safarzadeh:2018ent,Bonetti:2018nwo,Cote:2018qku,Hotokezaka:2018aui,Banerjee:2020eak}

In this work we address the problem of the remnant evolution on the viscous timescale by means of ab-initio (3+1)D NR simulations.   
We present new simulations performed with five microphysical EOS, a M0 neutrino
transport scheme and a subgrid model for the magnetohydrodynamics turbulence.
We compute dynamical ejecta and \swind{}, and we calculate the nucleosynthesis of the resulting unbound mass.
The simulations and analysis methods are detailed in Sec.~\ref{sec:methods}.
Section~\ref{sec:overview} gives an overview of the remnants dynamics,
describing the main features in terms of the binary parameters.
The properties of the dynamical ejecta are summarized in
Sec.~\ref{sec:dynej}, where we compare with simple models for AT2017gfo.
Sections~\ref{sec:spiralw} and \ref{sec:wind:nu} describe the mechanism powering the
\swind{} and \nwind{} in long-lived remnants. This mechanism is a combination of
$m=2$ and $m=1$ modes in the remnant powering spiral-density waves in
the disk. 
A polar component of the \swind{} is powered by neutrino heating above
the remnant.
The properties of the remnant disk, including thermodynamical
quantities, are discussed in Sec.~\ref{sec:remdisk}. The
composition of the disk at the end of the simulations is charaterized by 
double peaks in the entropy and electron fraction profiles. 
Section~\ref{sec:nucleo} presents nucleosynthesis calculations on the
combined dynamical and wind ejecta. The combined yields in the ejecta 
of long-lived remnants show a good fit to the solar abundance patterns for all
$r-$process peaks.
Throughout the text we discuss the implications of our results for 
AT2017gfo. 

\section{Methods}
\label{sec:methods}

Within (3+1)D NR we solve the equations of
general-relativistic hydrodynamics for a perfect fluid coupled to
the Z4c free evolution scheme for Einstein's
equations \citep{Bernuzzi:2009ex, Hilditch:2012fp}. The interactions
between the neutrinos radiation and the fluid 
are treated with a leakage scheme in the optically thick
regions \citep{Ruffert:1995fs,Galeazzi:2013mia,Neilsen:2014hha} while free-streaming
neutrinos are evolved according to the M0 scheme \citep{Radice:2018pdn}.
The effect of large-scale magnetic fields are simulated with 
the general-relativistic large eddy simulations method (GRLES) for
turbulent viscosity \citep{Radice:2017zta}. 

\subsection{Matter and radiation treatement}
\label{sec:method:formalism}

We write the fluid's stress-energy tensor as 
\begin{equation}
    T_{\mu\nu} = \rho h u_{\mu} u_{\nu} + Pg_{\mu\nu}
\end{equation}
where $\rho=m_{\rm b} n$ is the baryon rest-mass density, $n$ the baryon number density, $m_{\rm b}$
the baryonic mass, $h=1+\epsilon + P/\rho$ the specific enthalpy, $\epsilon$ the specific internal energy,
$u^{\mu}$ the fluid 4-velocity and $P$ the pressure. The fluid
satisfies the Euler's equations: 
\be
\nabla_\nu T^{\mu\nu} = Q u^\mu \ ,
\ee
where $Q$ is the net energy exchange rate due to the
absorption and emission of neutrinos, given by Eq.~(11) of
\citet{Radice:2018pdn}. The above system of equations is 
closed by a finite-temperature ($T$), composition-dependent EOS in the form
$P=P(\rho,Y_e,T)$, and by the evolution equations for the proton
and neutron number densities:
\be
\nabla_\nu (n_p u^\mu) = R_p^\mu \ \ , \ \ 
\nabla_\nu (n_n u^\mu) = R_n^\mu \ .
\ee
where the total proton fraction is computed as $n_p=Y_e n$, 
$n_p + n_n = n$, and $R_p=-R_n$ is the
net lepton number exchange rate due to the absorption and emission
of neutrinos and anti-neutrinos. 

We treat compositional and energy changes in the material due to weak
reactions using the leakage scheme presented in \citet{Galeazzi:2013mia,Radice:2016dwd}; 
see also \citet{Vanriper:1981mko,Ruffert:1995fs,Rosswog:2003rv,OConnor:2009iuz,Sekiguchi:2010ep,Neilsen:2014hha,Perego:2015agy,Ardevol-pulpillo:2018btx,Gizzi:2019awu} for other implementations.
We track reactions involving electron neutrinos ($\nu_e$) and
antineutrinos ($\bar{\nu}_e$) separately, and treat heavy-lepton
neutrinos in a single effective species ($\nu_x$). 
The production rates $R_{\nu}$, $\nu \in \{ \nu_e, \bar{\nu}_e, \nu_x \}$,
the associated production energies $Q_{\nu}$, and neutrino absorption
$\kappa_{\nu,a}$ and scattering $\kappa_{\nu,s}$ opacities are
computed from the reactions listed in table 3 of \citet{Perego:2019adq}. 
Neutrinos are then split into a trapped component with number density $n_\nu^{\rm trap}$ 
and a free-streaming component $n_\nu^{\rm fs}$. 
The latter are emitted according to the effective rate $R_\nu^{\rm eff}$
\citep{Ruffert:1995fs} \citep[see][Eq.~(4)]{Radice:2018pdn} and with
average energy $Q_{\nu}^{\rm eff}/R_{\nu}^{\rm eff}$ and then evolved
according to the M0 scheme of \cite{Radice:2018pdn}.
The M0 scheme evolves the number density of the
free-streaming neutrinos assuming that they move along radial null
rays, and estimates the free-streaming neutrino energy, $E_\nu$, under the
additional assumption of stationary metric.
Note that the pressure due to the trapped neutrino component is
neglected, since it is found to be important at a level $\lesssim 5\%$
in the conditions relevant for BNS mergers \citep{Galeazzi:2013mia,Perego:2019adq}.

Our simulations do not include magnetic fields but we simulate the
angular momentum transport due to magnetohydrodynamics turbulence by
using an effective viscosity and the GRLES scheme
\citep{Radice:2017zta,Radice:2020ids}.
The subgrid model employed in this work is
described in \cite{Radice:2020ids}, and it is 
designed based on the results of the high-resolution general relativistic magnetohydrodynamics simulations results of a BNS merger of \cite{Kiuchi:2017zzg}.
This GRLES subgrid model has been already used in \cite{Perego:2019adq,Endrizzi:2019trv,Nedora:2019jhl,Bernuzzi:2020txg}.

\subsection{EOS models}
\label{sec:method:eos}

We consider five different nuclear EOS models: 
BLh, DD2, LS220, SFHo and SLy4 \citep[see][table~1]{Perego:2019adq}
where DD2, LS220 and SFHo are summarized).
All these EOS include neutrons ($n$),
protons ($p$), nuclei, electrons, positrons, and photons as relevant
degrees of freedom.  Cold, neutrino-less
$\beta$-equilibrated matter described by these microphysical EOSs
predicts NS maximum masses and radii within the range allowed by
current astrophysical constraints, including the recent GW constraint
on tidal deformability
\citep{TheLIGOScientific:2017qsa,Abbott:2018wiz,De:2018uhw,Abbott:2018exr}.
All EOS models have symmetry energies at saturation density within
experimental bounds. However, LS220 has a significantly steeper
density dependence of its symmetry energy than the other models
\citep{Lattimer:2012xj,Danielewicz:2013upa}, 
and it could possibly underestimate the symmetry energy below saturation density.
In the considered models thermal effects enter in a quite different way. 
In particular particle correlations beyond the mean field approximation 
are included only in the BLh EOS. Such effects play an important role 
in the thermal evolution of neutron star matter. In the other models these 
effects are mainly encoded in the nucleon effective mass which is a density and 
temperature dependent quantity. At fixed entropy, the smaller is the effective mass 
the higher is the temperature.

The BLh EOS is a new finite temperature EOS derived in the framework
of non-relativistic many-body Brueckner-Hartree-Fock (BHF) approach (Logoteta et
al, in preparation). The zero temperature, $\beta$-equilibrated version of this EOS
was first presented in \cite{Bombaci:2018ksa} and applied to BNS
mergers in \cite{Endrizzi:2018uwl}; the finite temperature extension
was employed in \cite{Bernuzzi:2020txg} where a more detailed description can be found.
The interactions between nucleons is described through a potential derived
perturbatively in chiral-effective-field
theory \citep{Machleidt:2011zz}. It consists in a two-body part 
\citep{Piarulli:2016vel} calculated up to next to-next to-next
to-leading (N3LO) order and three-nucleon interation calculated up to
N2LO \citep{Logoteta:2016nzc}. At low densities ($n \leq
0.05~{\rm fm^{-3}} $) it is smoothly connected to the SFHo EOS \citep{Bernuzzi:2020txg}.

The DD2 and the SFHo EOSs are based on relativistic mean field (RMF) theory 
of high density nuclear matter \citep{Typel:2009sy,Hempel:2009mc}. 
Both the EOSs contain neutrons, protons, light nuclei such as deuterons, helions, tritons, alpha
particles and heavy nuclei in nuclear statistical equilibrium \citep{Steiner:2012xt}. 
DD2 and SFHo use different parameterizations of the covariant Lagrangian which models the mean-field
nuclear interactions. The resulting RMF equations are solved in Hartree's approximation.  
In particular, DD2 uses linear, but density
dependent coupling constants \citep{Typel:2009sy}, while the RMF parametrization
of SFHo employs constant couplings adjusted to reproduce neutron star radius measurements from low-mass X-ray binaries 
(see \cite{Steiner:2012rk} and references therein).
The DD2 is the stiffest EOS model considered in the present work and it is not in very good agreement 
with the so-called flow-constraint \citep{Danielewicz:2002pu}.

The LS220 \citep{Lattimer:1991nc} and the SLy4 EOSs are based on a liquid droplet model of Skyrme 
interaction.
The LS220 EOS includes surface effects and models $\alpha$-particles as
an ideal, classical, non-relativistic gas. 
Heavy nuclei are treated using the single nucleus approximation (SNA).
LS220 does not satisfy the constraints from Chiral effective field theory \citep{Hempel:2017ikt}. 
The SLy4 Skyrme parametrization was originally introduced
in \cite{Douchin:2001sv} for cold nuclear and NS matter. In this work
we employ the finite temperature extension presented
in \cite{daSilvaSchneider:2017jpg} using an improved version of the
LS220 model that includes non-local isospin asymmetric terms. In this EOS version it is also introduced a 
better and more consistent treatment of both nuclear surface properties and the size of heavy nuclei. 

\subsection{Computational setup}
\label{sec:method:comp_setup}

We prepare irrotational BNS initial data in quasi-circular orbit with
NS at an initial separation of $45\, {\rm km}$, corresponding to
${\sim}3-4$ orbits before merger. 
Initial data are computed using the
\texttt{Lorene} multidomain pseudospectral library \citep{Gourgoulhon:2000nn}. The EOS
used for the initial data are constructed from the minimum temperature
slice of the EOS table used for the evolution assuming neutrino-less
beta-equilibrium.

Initial data are evolved with the \texttt{WhiskyTHC} code
\citep{Radice:2012cu,Radice:2013hxh,Radice:2013xpa} for general
relativistic hydrodynamics that implements the approximate neutrino
transport scheme developed in \cite{Radice:2016dwd,Radice:2018pdn} and
the GRLES for turbulent viscosity \citep{Radice:2017zta} described above.
The M0 scheme is switched on
shortly before the two NS collide, when neutrino matter
interactions become dynamically important. The equations for the M0
scheme are solved on a uniform spherical grid 
extending to ${\simeq} 756\, {\rm km}$ and having
$n_r \times n_\theta \times n_\phi = 3096 \times 32 \times 64$ grid
points.

\texttt{WhiskyTHC} is implemented within the \texttt{Cactus}
\citep{Goodale:2002a,Schnetter:2007rb} framework and coupled to an
adaptive mesh refinement driver and a metric solver. 
The Z4c spacetime solver is implemented in the \texttt{CTGamma} code \citep{Pollney:2009yz, Reisswig:2013sqa}, which
is a part of the \texttt{Einstein Toolkit} \citep{Loffler:2011ay}. We use
fourth-order finite-differencing for the metric's spatial derivatives and
the method of lines for the time evolution of both metric and fluid variables.
We adopt the optimal strongly-stability preserving third-order
Runge-Kutta scheme \citep{Gottlieb:2009a} as time integrator. The
timestep is set according to the speed-of-light Courant-Friedrich-Lewy
(CFL) condition with CFL factor $0.15$. While numerical stability
requires the CFL to be less than $0.25$, the smaller value of $0.15$
is necessary to guarantee the positivity of the density when using the
positivity-preserving limiter implemented in \texttt{WhiskyTHC}. 

The computational domain is a cube of 3,024~km in diameter whose
center is at the center of mass of the binary. Our code uses
Berger-Oliger conservative AMR \citep{Berger:1984zza} with
sub-cycling in time and refluxing \citep{Berger:1989a, Reisswig:2012nc}
as provided by the \texttt{Carpet} module of the \texttt{Einstein
  Toolkit} \citep{Schnetter:2003rb}. We setup an AMR grid structure
with $7$ refinement levels. The finest refinement level covers both NS
during the inspiral and the remnant after the merger and has a typical
resolution of $h\simeq 246$~m (grid setup named {\tt LR}), $h \simeq 185$~m
({\tt SR}) or $\simeq 123$~m ({\tt HR}). 

\subsection{Postprocess analysis}
\label{sec:method:analysis}

To study the dynamical modes in the remnant we follow previous
work \citep{Paschalidis:2015mla,Radice:2016gym,East:2016zvv} and define
a complex azimuthal mode decomposition of 
the rest-mass density as 
\begin{equation}\label{eq:modes}
    C_m = \int\rho W e^{-i m \phi} \sqrt{\gamma} \text{d}x \text{d} y \, ,
\end{equation}
where 
$\gamma$ is the determinant of the three-metric and $W$ is the
Lorentz factor between the fluid and the Eulerian observers. 
Note that the above quantities are gauge dependent.

Following a common convention, we define remnant disk the baryon
material either outside the black hole (BH) apparent horizon or with rest mass
density $\rho\lesssim10^{13}$~$\gccm$ around a NS remnant. The baryonic mass of
the disks is computed as the volume integral of the conserved rest-mass
density $D=\sqrt{\gamma}~W\rho$ from 3D snapshots of the simulations in
postprocessing.
The threshold $\rho\sim10^{13}~\gccm$ corresponds to the point in the remnant where
the angular velocity profiles becomes approximately Keplerian, \citep[\eg][]{Shibata:2005ss,Shibata:2006nm,Hanauske:2016gia,Kastaun:2016elu}.

We make use of mass-averaged quantities and for a quantity $f$ they are computed as 
\be
\langle f \rangle = \frac{\sum_i f(m_i)m_i}{\sum_i m_i}
\ee
where $m_i$ is the mass contained in the $i$-th bin.

The fluid's angular momentum analysis in the remnant and disk is performed
assuming axisymmetry. That is, we assume $\phi^{\mu} = (\partial_{\phi})^{\mu}$ to be a Killing
vector. Accordingly, the conservation law
\begin{equation}
  \partial_t(T^{\mu\nu}\phi_{\nu}n_{\nu}\sqrt{\gamma}) -
  \partial_i(\alpha T^{i \nu}\phi_{\nu}\sqrt{\gamma}) = 0 \ ,
\end{equation}
where $n^\mu$ is the normal vector to the spacelike hypersurfaces of
the spacetime's $3+1$ decomposition, 
implies the conservation of the angular momentum
\begin{equation}
    J = 
    -\int \,
    T_{\mu\nu}n^{\mu}\phi^{\nu}\,\sqrt{\gamma}\, \dd^3 x\ .
\end{equation}
In the cylindrical coordinates $x^i=(r,\phi,z)$ adapted to the symmetry
the angular momentum density is  
\begin{equation}
    j = 
        \rho h W^2 v_{\phi} \ ,
    \label{eq:intro:ang_mom}
\end{equation}
and the angular momentum flux is 
\begin{equation}
    \alpha\sqrt{\gamma}T^r _{\nu}\phi^{\nu} =
    \alpha\sqrt{\gamma}\rho h W^2 (v^{r}v_{\phi}) .
\end{equation}

All considered mass ejecta are calculated on a coordinate sphere at $R \simeq
294$km. The dynamical ejecta is computed assuming the fluid elements 
to follow unbound geodesics,  
$-u_t > 1$ 
and to reach an asymptotic velocity 
$\upsilon_{\infty} \simeq \sqrt{2E_{\infty}} = \sqrt{(1-u_t ^2)}$.
Wind ejecta are instead computed according to the Bernoulli criterion 
$-hu_t > 1$ 
and the associated asymptotic velocity is calculated as 
$\upsilon_{\infty} \simeq \sqrt{2 (h (E_{\infty}+1)-1)}$. 
Note that the geodesic criterion above neglects the fluid's pressure and might
underestimate the ejecta mass. The Bernoulli criterion assumes that the (test
fluid) flow is stationary, so that there is a pressure gradient that can
further push the ejecta.  We find that both criteria predict very similar dynamical
ejecta mass if applied to extraction spheres at large coordinate sphere;
differences between the two criteria are instead present if they are applied to
matter volumes \citep[cf.][]{Kastaun:2014fna}.

\subsection{Simulations}
\label{sec:method:sim}

We discuss simulations of $37$ binaries with chirp mass
$\mathcal{M}_c=1.188\,\Msun$ compatible to the source of GW170817,
total gravitational mass spanning the range $M\in[2.73, 2.88]\Msun$ and mass
ratio values $q=M_A/M_B\in[1,1.8]$.
Summary data for the simulations is collected in Tab.~\ref{tab:sim}. 
Most of the binaries are simulated at both grid resolutions {\tt LR}
and {\tt SR} and 
$16$ binaries are simulated also at {\tt HR} for a total of
$76$ simulations. We follow the evolution of long-lived remnants up
to $\sim100$~ms postmerger.
Note a subset of simulations are performed without GRLES scheme in
order to asses the the effect of turbulent viscosity; they are
  indicated with ``*'' in the following.
The short-term evolution of the largest
mass ratio binaries has been already presented in \cite{Bernuzzi:2020txg}.
Together with our previous data these simulations form the largest
sample of merger simulations with microphysics available to date
\citep{Bernuzzi:2015opx,Radice:2016dwd,Radice:2016rys,Radice:2017lry,Radice:2018xqa,Radice:2018pdn,Perego:2019adq,Endrizzi:2020lwl,Bernuzzi:2020txg}.



\section{Overview of the remnant dynamics}
\label{sec:overview}

The early (dynamical) postmerger phase is driven by the GW emission,
which removes about twice as much energy as the whole 
inspiral-to-merger phase in ${\sim}10-20$~ms \citep{Bernuzzi:2015opx}.
After this GW postmerger transient at kiloHertz frequencies, 
the GW emission drops significantly and removes angular 
momentum only on timescales of a few seconds \citep{Radice:2018xqa}.
The remnant evolution on timescales $\O(100)$~ms is then driven by
viscous and weak-interactions.
Merger remnants after the GW-driven phase have a significant excess of
angular momentum and gravitational mass if compared to
zero-temperature rigidly rotating equilibrium with the same baryonic
mass \citep{Radice:2018xqa}.
Temperature and composition effects are key to determine
if the remnant evolves towards an axisymmetric stationary NS close to
the mass shedding or collapses to BH.
The new simulations presented here allows us to investigate these
timescales with the relevant physics effects.

\input{tabSim.tex}

\begin{figure}[t]
  \centering 
  \includegraphics[width=0.49\textwidth]{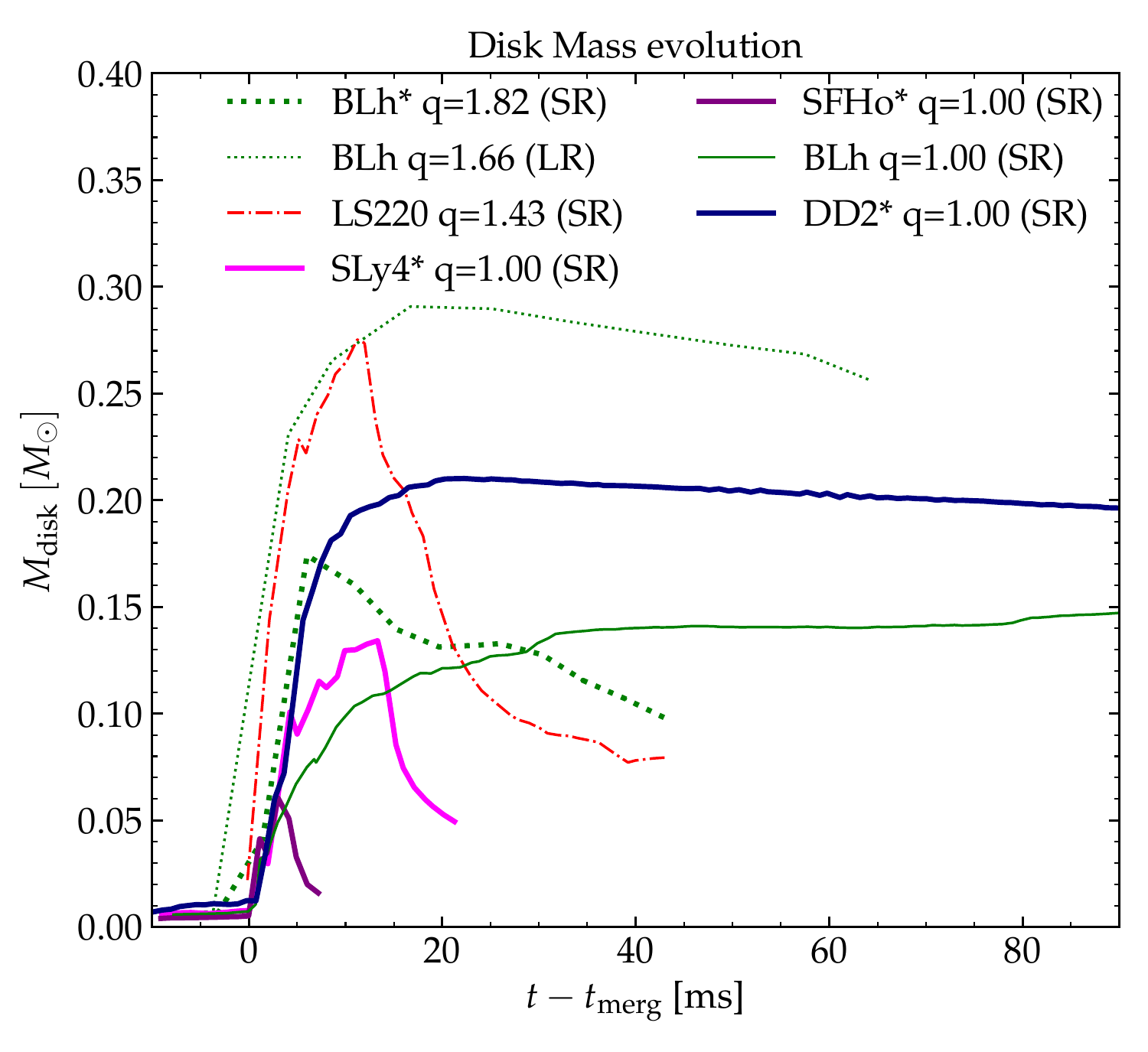}
  \caption{Time evolution of the total disk mass for a few selected
    short-lived and long-lived cases. The former show a rapid 
    accretion right after disk formation. The plots show
    distinct difference in dynamical evolution after disk formation: accretion onto
    the newly formed BH (short-lived remnants) or accretion onto the NS
    remnant (DD2 $q=1$) with possible continuous mass shedding from the remnant
    into the disk (BLh* $q=1$).} 
  \label{fig:disk_mass_evo}
\end{figure}

The short-term dynamics of ten of these BNS has been previously discussed
in \cite{Bernuzzi:2020txg}, in the context of prompt collapse of large
mass ratio binaries~\footnote{We call here prompt
  collapse mergers in which the central density increases monotonically
  and there is no core bounce \citep{Radice:2020ddv,Bernuzzi:2020tgt,Bernuzzi:2020txg}.}. 
Indeed, the only merger remnants that promptly collapse in
the simulated sample are those with $q\gtrsim1.67$. The collapse in the 
mergers of BLh, LS220, SFHo and SLy with $q=1.67,1.8$
is induced by the accretion of the
companion (less massive) onto the primary star NS. In these cases, the BH
remnant is surrounded by an accretion disk formed by the tidal tail of
the companion. The disk is thus composed of
very neutron-rich material $Y_e\sim 0.1$ and with baryon masses at
formation ${\sim}0.15\,\Msun$, significantly heavier than the
remnant disks in equal-masses prompt collapse mergers.
Examples of disk mass evolution are shown in 
Fig.~\ref{fig:disk_mass_evo} for representative BNS. 
These high-$q$ mergers
launch dynamical ejecta of mass ${\sim}0.01\,\Msun$ that also originate
from the tidal disruption of the companion. The dynamical ejecta are
neutron rich and expand from the orbital plane with a crescent-like
geometry different from the more isotropic dynamical ejecta of the
equal-mass mergers \citep{Bernuzzi:2020txg}.

Among the comparable-mass ($q\lesssim1.4$) mergers, the merger outcome is either a
short-lived or a long-lived NS remnant. The former collapses to BH
within few dynamical periods set by the NS remnant's rotation, the
latter does not collapse within the simulated time. In practice, the 
short-lived remnants of LS220 $q=1,1.1,1.2$, SFHo $q=1,1.1,1.4$ and SLy
$q=1,1.1,1.4$ collapse within $20$~ms postmerger. 
The exact time of the collapse is strongly dependent on the
simulated physics and also on numerical errors. For example,
the inclusion turbulent viscosity \citep{Radice:2017zta} or changes in the
resolution can accelerate or delay the collapse.

The remnant disk originates from the matter expelled by tidal torques
and shocks produced at the collisional interface of the NS cores during merger. 
Starting at merger, the NS remnant sheds mass and angular momentum
outwards through spiral density waves streaming from the shock interface \citep{Bernuzzi:2015opx,Radice:2018xqa}.
The maximum temperatures are experienced in these streams that however
rapidly decrease because of the expansion and neutrino emission. 
The electron fraction is reset by an initial excess of electron antineutrino emission
and electron neutrino absorption, while the entropy per baryon varies between 
$3$ and $\lesssim 10\,$$k_{B}$/baryon \citep{Perego:2019adq}.
In the short-lived cases, the process quickly shuts down at BH
formation: the disk rapidly accretes at early times around the
newly formed BH and then reaches a steady state, Fig.~\ref{fig:disk_mass_evo}.
The resulting configuration is approximately axisymmetric and Keplerian; it is 
characterized by neutron-rich $Y_e\sim0.1$ and hot $T\sim10\,$MeV
material in the inner part ($\rho\sim10^{13}$~$\gcc$) and colder and
reprocessed material near the edge $Y_e\sim0.4$.
The maximum disk masses (at formation) are generically larger for
stiffer EOS and higher mass ratio.
The disk mass can be described within the numerical uncertainties by a
quadratic function of the mass ratio and the reduced tidal
parameters (see Sec.~\ref{sec:remdisk}). 
In particular, the most massive disks are formed in case of 
highly asymmetric BLh $q=1.82$ binary and of the softer EOS LS220 but less asymmetric $q=1.43$ binary.
In the latter case the quick collapse of the remnant removes more then half of the 
disk mass within $40$~ms postmerger.

In the long-lived cases, the disk (now defined by the material with
$\rho\lesssim10^{13}$~$\gcc$) is more massive and extended than the disk
around BH remnants \citep{Perego:2019adq}. 
In general, the maximum disk mass is larger for stiffer EOS and
higher mass ratio. For example, the DD2 $q=1$ remnant has disk mass
${\sim}0.2\,\Msun$ while the BLh $q=1$ has ${\lesssim}0.15\,\Msun$. The disk of the BLh
$q\sim1.4-1.5$ remnant is up to a factor two more massive than the latter.
The long-term disk evolution is determined by its interaction
with the central object. On the one hand the gravitational pull and
the neutrino cooling causes the material to accrete. On the other
hand the spiral density waves continuously feed the disk with
centrifugally supported material, and the angular momentum transport
caused by the turbulence favors its expansion.
Thus, the disk looses its mass by accretion if the central object is a BH,
but can either acquire or loose mass if the central object is a NS.
The latter cases are visibile in Fig.~\ref{fig:disk_mass_evo} for the
the BLh EOS and the DD2 EOS.
In particular, the BLh* $q=1$ postmerger configuration is such 
that the mass-shedding by the remnant exceeds the mass accretion. 
This behaviour is believed to be set by a combination of the EOS softness  
and the treatment of the thermal effects within the BLh EOS.
The former implies stronger postmerger remnant oscillations than the DD2 EOS, the latter higher remnant average temperature.  

In terms of disk structure, the inclusion of turbulence appears to smoothen the 
mass distribution of disk properties, such as $Y_e$, $s$, $T$,
making them slightly broader. However, detailed quantitative study requires 
more runs at several resolutions to separate finite-grid from 
subgrid turbulence effects \citep{Bernuzzi:2020txg,Radice:2020ids}.

\begin{figure}[t]
  \centering 
  \includegraphics[width=0.49\textwidth]{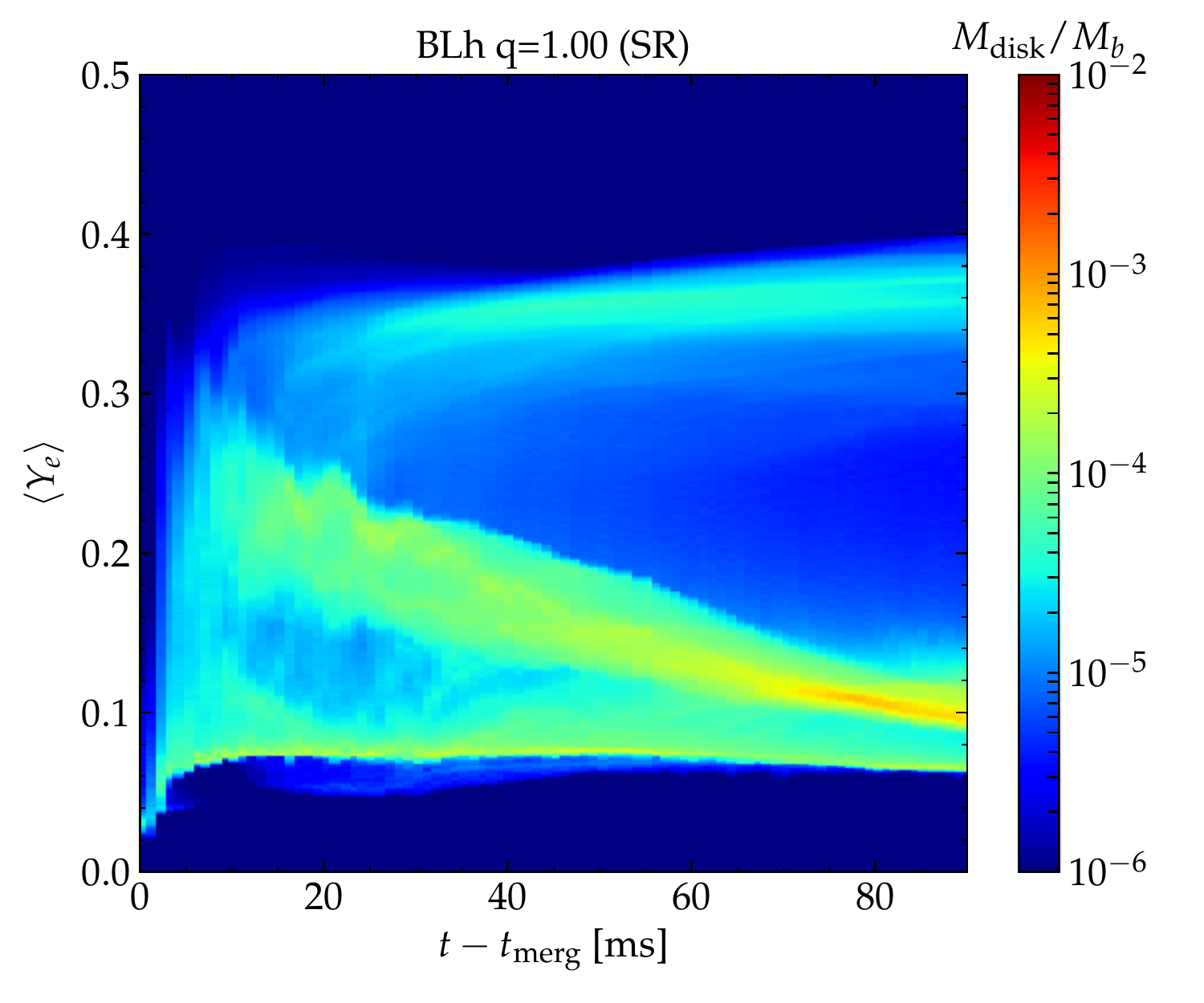}
  \includegraphics[width=0.49\textwidth]{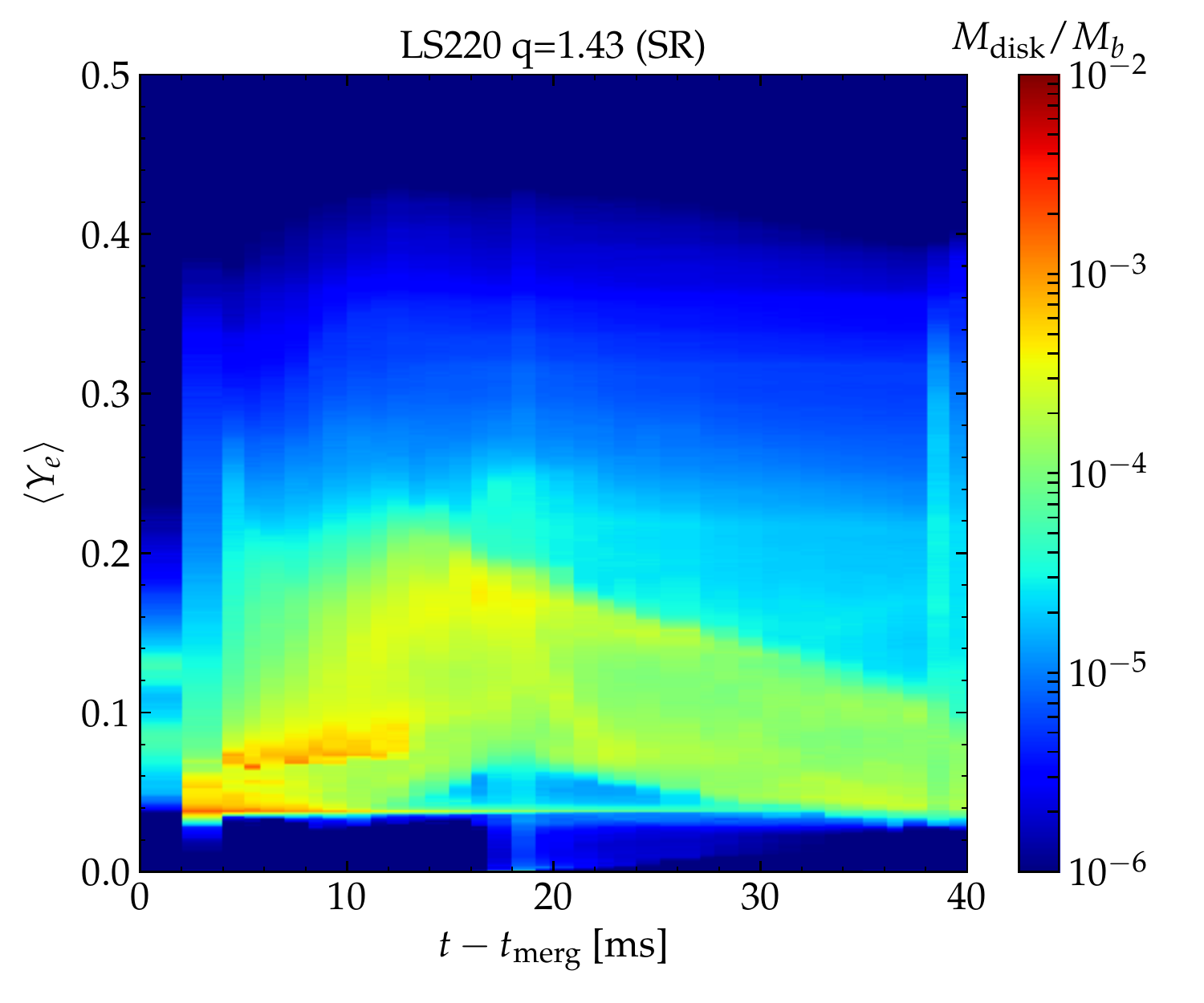}
  \caption{Evolution of the disk mass-averaged electron fraction with
    time for a long-lived (top) and a short-lived (bottom)
    remnant. The plot shows that with time the bulk of the disk lowers
    its $Y_e$ via cooling, while a small fraction in terms of mass
    gains a high $Y_e$, which relates to the highly 
    irradiated surface of the disk.}
  \label{fig:total_disk_time_corr_Ye_Blh_q1}
\end{figure}

Disks around long-lived remnant are also more optically thick than
disks around BH. The top panel of Figure~\ref{fig:total_disk_time_corr_Ye_Blh_q1}
shows the evolution of the mass-weighted electron fraction for
the case of BLh $q=1$ up to $90$~ms.
At early times a fraction of fluid elements have $Y_e\sim0.25$ as a
results of the shock and spiral waves during formation. After about
${\sim}40$~ms from merger, most of the matter comprises a neutron-rich bulk at $Y_e\lesssim0.1$. Neutrinos irradiate the
disk edge (Fig. \ref{fig:slice:heating_hu}, density contours) that at
${\sim}40$~ms reaches $Y_e\sim0.4$. Note that neutrinos in merger
remnants decouple at $\rho\sim10^{11}$~$\gccm$ \citep{Endrizzi:2020lwl}. 
While we expect this picture to be qualitatively correct, the gap at
intermediate $\langle Y_e \rangle \simeq 0.15$ might be a artifact of the M0
which assumes radial propagation of neutrinos and cannot correctly capture the
reabsorption of neutrinos emitted from the midplane of the disk.
In the case of BH remnant (bottom panel of Figure~\ref{fig:total_disk_time_corr_Ye_Blh_q1}), 
the more compact disk still emits efficiently neutrinos, but
due to the lack of emission from the massive NS neutrino absorption at the disk edge is not relevant
and the average electron fraction is systematically lower.

If the disk expands outwards sufficiently far, recombination of nucleons
into alpha particles provide enough energy to unbind the outermost material and
generate mass outflows \citep{Beloborodov:2008nx,Lee:2009uc,Fernandez:2013tya}.  
On the simulated timescales, mass is ejected from the remnant due to 
the \swind{} \citep{Nedora:2019jhl} and the neutrino-driven wind (\nwind;
\citealt{Dessart:2008zd,Perego:2014fma,Just:2014fka}). The former is powered by a
hydrodynamical mechanism that preferentially ejects material at low latitudes.
The \swind{} can have a mass up to
a few $10^{-2}\,\Mo$ and velocities ${\sim}0.2$~c. The ejecta
have electron fraction typically larger than ${\sim} 0.25$ since they are 
partially reprocessed by hydrodynamic shocks in the expanding arms. 
The \nwind{} is driven by neutrino heating above the remnant. It generates outflows
with smaller masses ${\sim} 10^{-4}M_{\odot}$ and larger $Y_e$
than the \swind{}. 
Differently from \swind{} the mass flux of the \nwind{} in our simulations subsides 
before the end of out simulations, due to rapid baryon loading of the 
polar region.
The \swind{} will be discussed in detail in Sec.~\ref{sec:spiralw}.

\begin{figure}[t]
  \centering 
  \includegraphics[width=0.49\textwidth]{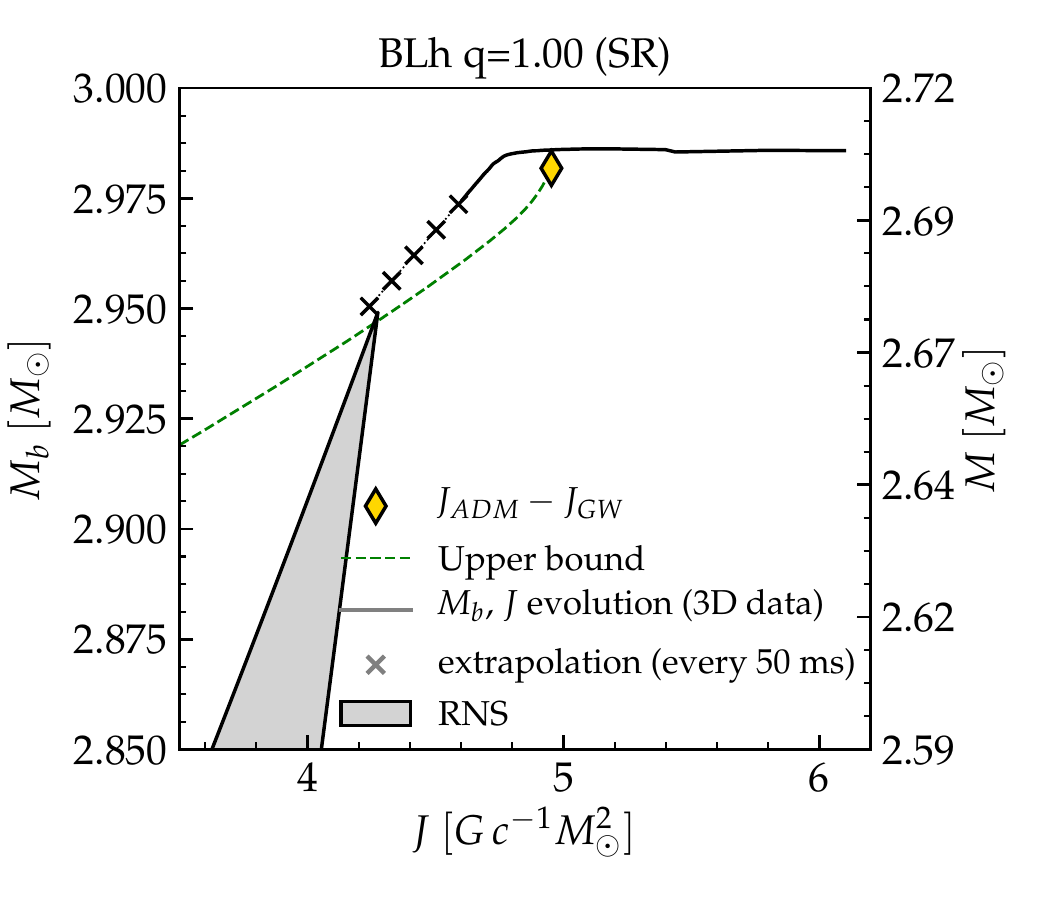}
  \caption{Baryon mass vs angular momentum diagram for the BLh $q=1$ remnant.
    The colored diamond marks the baryonic mass and angular momentum at the end
    of the dynamical gravitational-wave dominated phase.
    After the GW phase, the evolution is driven by the massive outflows.
    The solid black line is the $M_b$ and $J$ estimated from the 3D data
    integrals under the assumption of axisymmetry.
    The green dashed line is a conservative estimate
    of the mass ejection and a possible trajectory for the viscous
    evolution as estimated in \citet{Radice:2018xqa}. The crosses are
    a linear extrapolation in time of the solid black line. The gray
    shaded region is the region of stability of rigidly rotating NS equilibria.}
  \label{fig:total_j_mb_rns_blh}
\end{figure}

The fate of the long-lived remnant beyond the simulated
timescale is difficult to predict without longer, ab-initio
simulations in $(3+1)D$ with complete physics.
To illustrate this aspect we discuss the representative case of BLh
$q=1$ that is one of our 
longest runs of binaries with baryon mass larger than the one supported
by the zero-temperature beta-equilibrated rigidly rotating equilibrium
single NS configurations.
Figure~\ref{fig:total_j_mb_rns_blh} shows the evolution
of the remnant in the baryon mass vs.~angular momentum diagram. 
The total baryon mass of the system is conserved, and in absence of
ejecta (\eg~ during the inspiral) the binary evolves along curves of
constant baryonic mass but loses angular momentum due to emission of
GWs. The latter is computed from the multipolar GW following
\citet{Damour:2011fu,Bernuzzi:2012ci,Bernuzzi:2015rla}, in particular
taking the difference between the Arnowitt-Deser-Misner intial angular
momentum of the initial data and the angular momentum carried away by
the gravitational waves by the end of the simulations.
After the GW losses becomes inefficient, the remnant remains to 
the right with respect to the rigidly rotating equilibria region, marked as the
gray shaded area in Fig.~\ref{fig:total_j_mb_rns_blh}.
This indicate that the remnant has angular momentum in excess with respect to the relative (same baryon mass) NS equilibrium, and it is a generic features of all
the simulated binaries \citep{Zappa:2017xba,Radice:2018xqa}. Additionally,
the baryon mass of the remnant after the GW-driven phase is larger
than the maximum baryon mass for rigidly rotating equilibria. This is
usually called a hypermassive NS remnant,  according to a
classification based on zero-temperature EOS equilibria
\citep{Baumgarte:1999cq}, and it is thus expected to collapse to BH at
finite time. After the dynamical GW-dominated phase (yellow diamond)
we compute the angular momentum and mass evolution under the assumption
of axisymmetry (black solid curve)~\footnote{Note that the angular momentum estimated
  from the GW and from the integral of Eq.~\eqref{eq:intro:ang_mom} assuming
  axisymmetry are compatible within the errors made in the
  latter estimate.}.
Massive ejecta beyond the simulated time can drive the remnant
evolution to the stability limit, in contrast with the naive expectation of BH
collapse.
Indeed, both the extrapolation of the data at longer
timescales (black crosses) and a conservative upper bound estimate 
\citep{Radice:2018xqa} (green dashed line) are compatible with a
possible massive NS remnant close to the Keplerian limit.
A linear extrapolation of the final trend indicates that if about
$\approx0.05\,\Msun$ ($\approx40$\% of the disk mass at the final evolution
time) of the disk evaporates at the same rate, then the remnant 
would be close to mass-shedding limit of rigidly rotating equilibria
at about ${\sim 300}$~ms postmerger. Note this simulations
is with viscosity, but magnetic stresses could further boost ejecta 
\citep{Metzger:2006mw,Bucciantini:2011kx,Siegel:2017nub,Fernandez:2018kax,Ciolfi:2020hgg}.

A similar outcome is obtained for other binaries. 
In case of DD2, however, remnants lay below the cusp of the equilibria region, having
an excess of angular momentum but not of baryonic mass.
The evolution towards the stability is slower in these cases. 
More asymmetric models are formed with larger excess in the total 
angular momentum and must shed a larger amount of mass to reach the 
equilibrium.
We estimate that the amount of ejected mass required to reach stability
lies between ${\sim}0.05M_{\odot}$ and $0.2M_{\odot}$ for the $q=1$
and $q=1.4$ binaries respectively, again corresponding to $\lesssim40\%$ of the disk mass.

\section{Dynamical Ejecta}
\label{sec:dynej}

The mechanisms behind dynamical ejecta and results for our simulations
have been extensively discussed in recent papers
\citep{Radice:2018pdn,Bernuzzi:2020txg}. Here, we focus on the overall
properties of the mass ejecta of our set of targeted simulations
 and provide approximate fitting formulae for the average
mass, velocity and electron fraction. Then, we discuss on the applicability of 
these results for the kN AT2017gfo,
associated with the gravitational wave event GW170817.

\begin{figure*}[t]
    \centering 
    \includegraphics[width=0.32\textwidth]{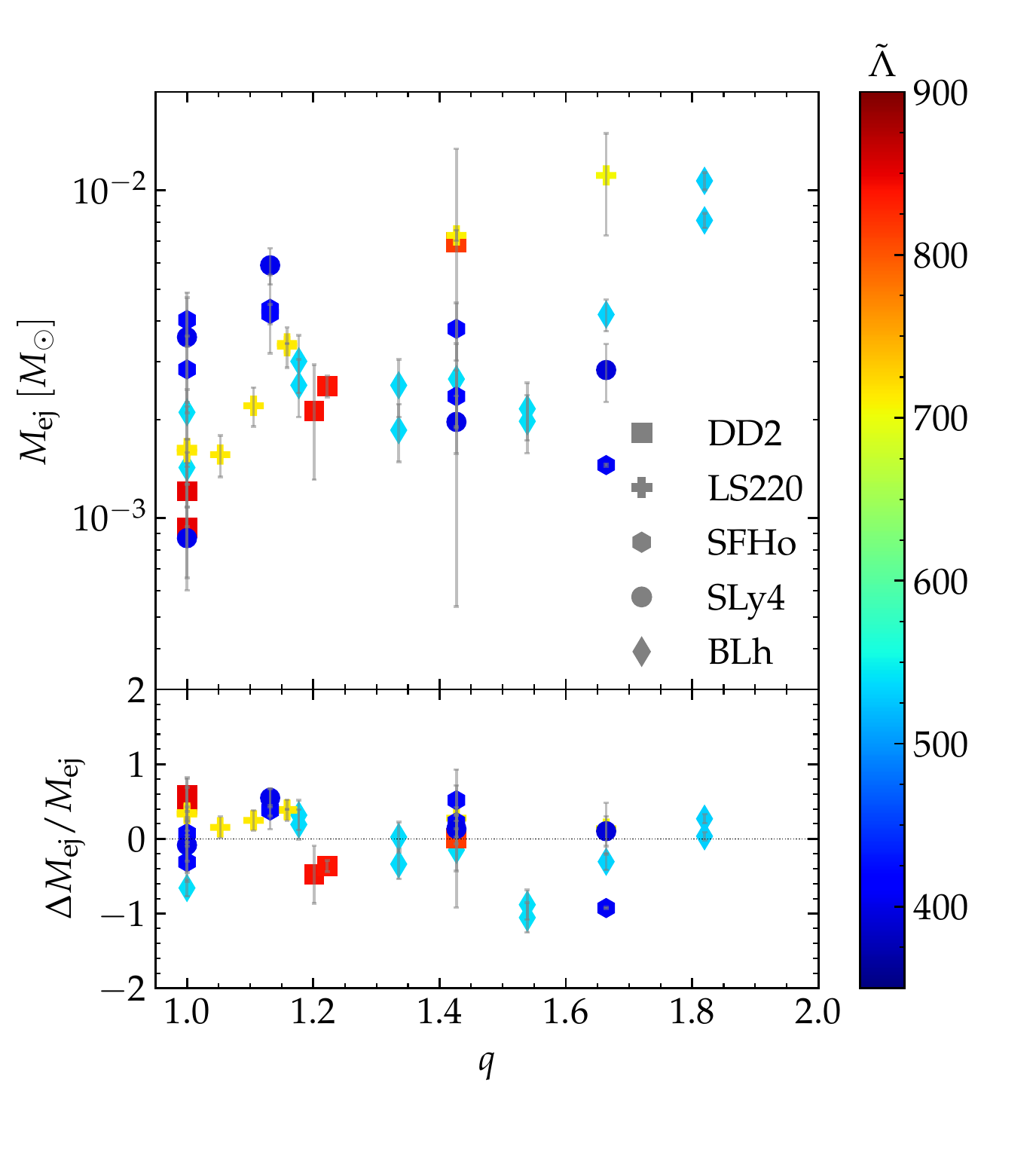}
    \includegraphics[width=0.32\textwidth]{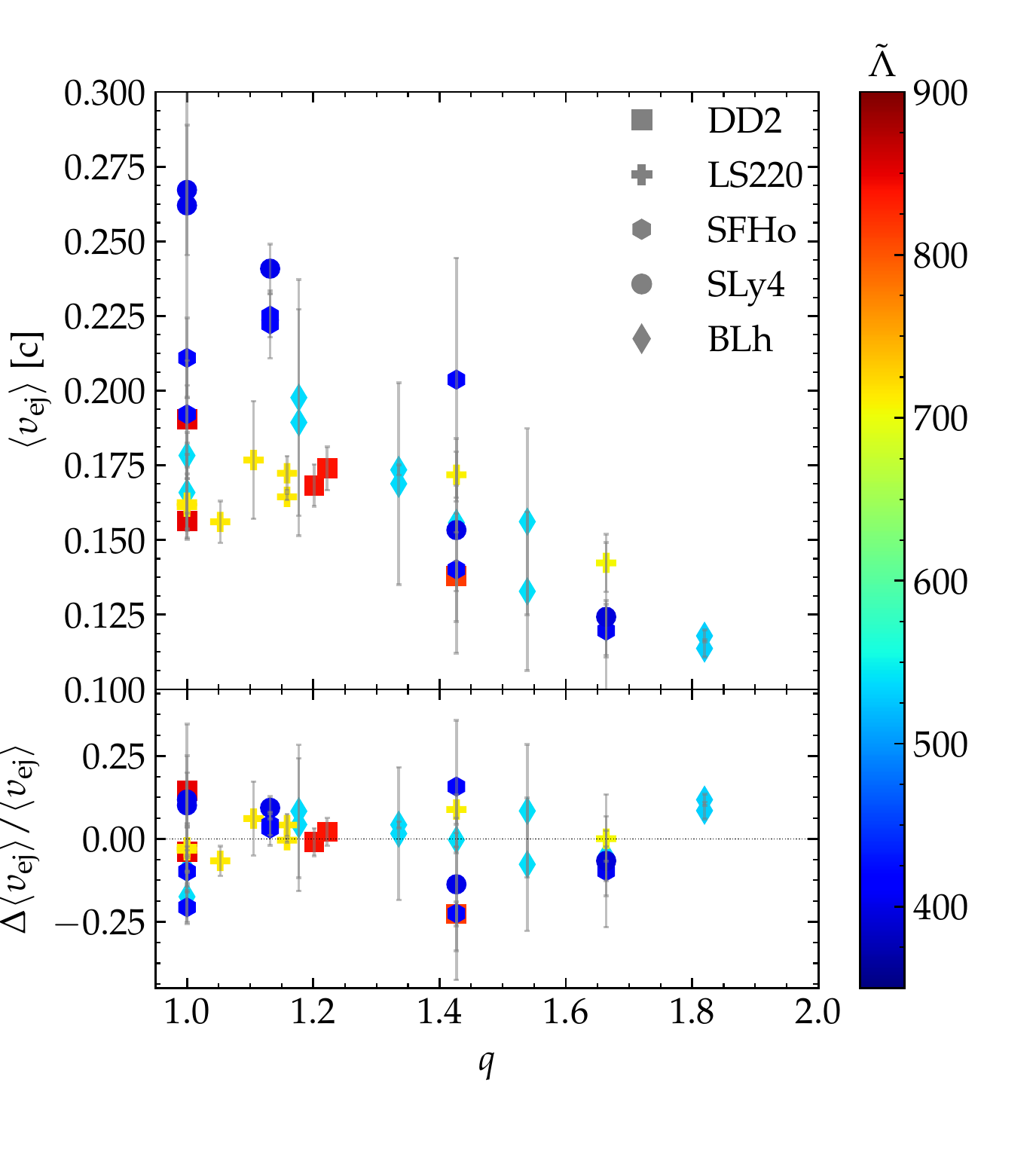}
    \includegraphics[width=0.32\textwidth]{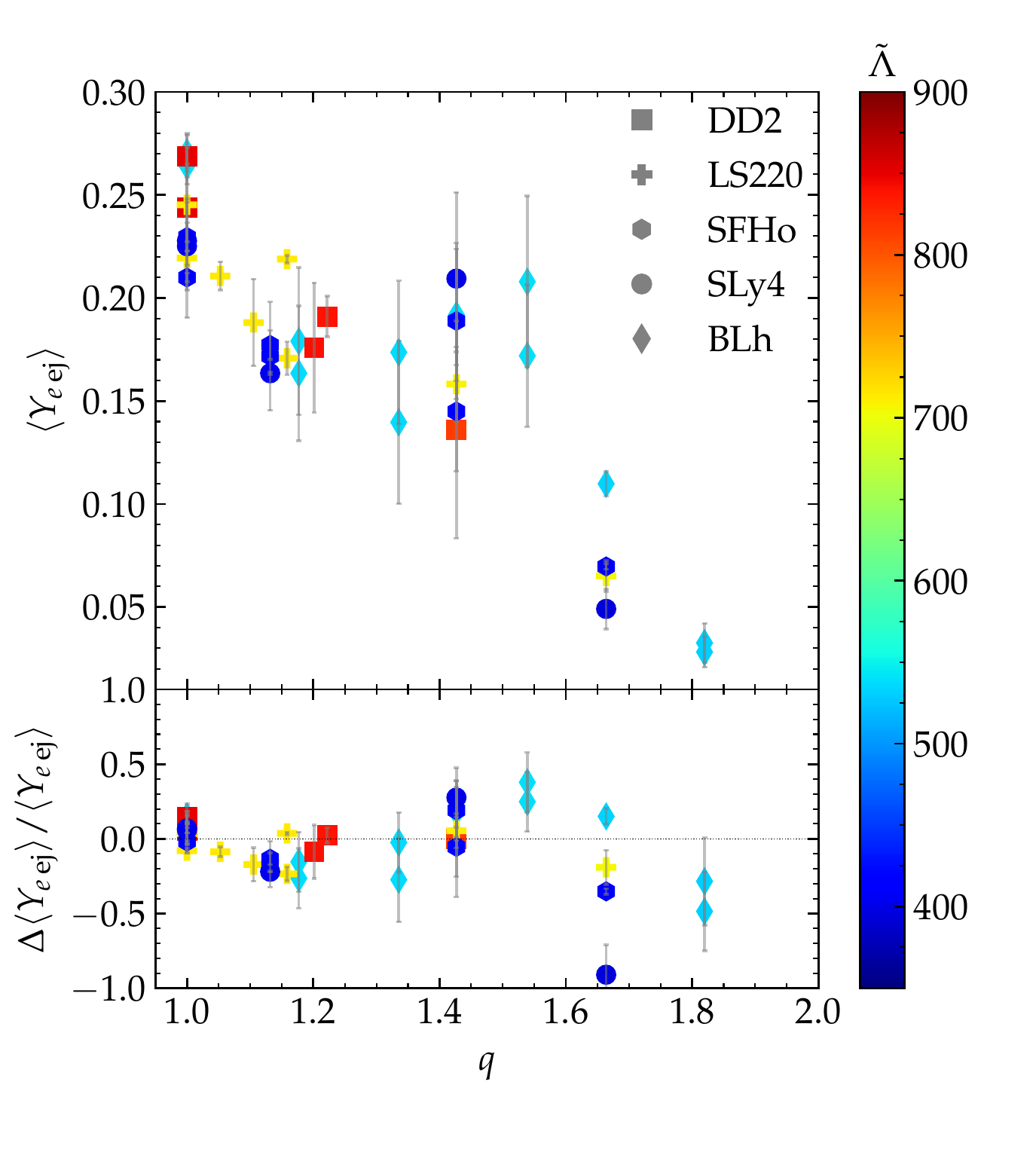}
    \caption{Dynamical ejecta properties as a function of mass ratio
      and reduced tidal parameter. The dependency on the latter is
      color coded. From left to right the main panels show the total
      mass, the mass-averaged velocity and the electron fraction.
      The bottom panels show the relative difference between the data
      and the fit polynomial fit discussed in the text.}
    \label{fig:ejecta:dyn:ds}
\end{figure*}

The data presented in this work is obtained with the M0 and GRLES
schemes and span a
significant range in mass ratio but a smaller range in the reduced tidal
parameter $\tilde\Lambda$ than our previous dataset of
\cite{Radice:2018pdn}, where most of the simulations were
performed with the leakage scheme only. 
Comparing the data obtained with leakage and those with the M0, we observe 
that neutrino absorption leads to not only an increased average
electron fraction but also to larger total ejected
mass and velocity. 
For example, the mass averaged over the simulations from Tab.~\ref{tab:sim} is 
$\overline{\amd} = (3.442 \pm 2.495)\times 10^{-3}\,M_{\odot}$ (where
hereafter we report also the standard deviation), while the same
quantity calculated for data of \cite{Radice:2018pdn} 
is $\overline{\md} = (1.352\pm 1.250)\times 10^{-3}\,\Msun$.
The mass-averaged terminal velocity of the dynamical ejecta 
ranges between $0.1$c and $0.3$c, in a good agreement with 
\cite{Radice:2018pdn}.
The mass-averaged velocity, averaged over all the simulations, is 
$\overline{\avd} = (0.172\pm0.038)\,{\rm c}$.
The new data at fixed chirp mass shows
a correlation of $\avd$ with the tidal parameter $\tilde{\Lambda}$: 
the lower the $\tilde{\Lambda}$ the higher the velocity. This is a consequence 
of the fact that dynamical ejecta in comparable-mass mergers is dominated 
by the shocked component and that the shock velocity is larger the more compact the binary
is~\footnote{Note that in the definition of prompt collapse we
    adopted, there is no shocked ejecta.}.
On the contrary, for high mass ratios $q\gtrsim1.5$, the ejecta is
dominated by the tidal component and a larger $q$ leads to a smaller
$\avd$. 
The mass-averaged electron fraction in our simulations 
varies between $0.1$ and $0.3$ and averaged among the simulations is $\overline{\ayd} = 0.175 \pm 0.063$.
The range is broader than what previously reported in
\cite{Radice:2018pdn}, where the upper limit was $\approx0.2$ and the
lower was $0.1$. The main difference for this result is the use of the
M0 scheme, as noted above.
The average electron fraction of our models with
M0 neutrino transport is very similar to the ones obtained with M1
scheme of \citet{Sekiguchi:2016bjd} and \citet{Vincent:2019kor}.
Moreover, the high-$q$ simulations where the dynamical ejecta is dominated by the tidal
component contribute to the lower boundary of $\ayd$.
The comparison between simulations with and without the GRLES scheme
does not indicate a strong effect on the dynamical ejecta; the effect is
comparable to the effect of finite grid resolution, \citep{Bernuzzi:2020txg,Radice:2020ids}.

 \input{tabFitsPoly22.tex}

Overall, we find that the properties of the ejecta depend strongly on
mass-ratio and the EOS softness, that can be parameterized by the
reduced tidal parameter.
Figure~\ref{fig:ejecta:dyn:ds} shows the dynamical ejecta
properties as a function of the mass ratio and (color coded) $\tilde\Lambda$.
We can fit of our data at fixed chirp mass using a second
order polynomial in these two 
parameters, 
\begin{align}
  &P_2(q, \tilde\Lambda) = b_0 + b_1q + b_2 \tilde\Lambda + b_3q^2 + b_4q\tilde\Lambda + b_5\tilde\Lambda^2 \ .
  \label{eq:fit:poly22}
\end{align}
Fitting coefficients are reported in Tab.~\ref{tab:fitpoly22coefs} for
all the quantities; fit residuals are displayed in the bottom panel of
Fig.~\ref{fig:ejecta:dyn:ds}. 
We have explored several fitting functions, including
several proposals in the literature, and find that overall the choice in 
Eq.~\eqref{eq:fit:poly22} is simple and robust; these results will be 
reported elsewhere.

\begin{figure*}[t]
  \centering 
  \includegraphics[width=0.48\textwidth]{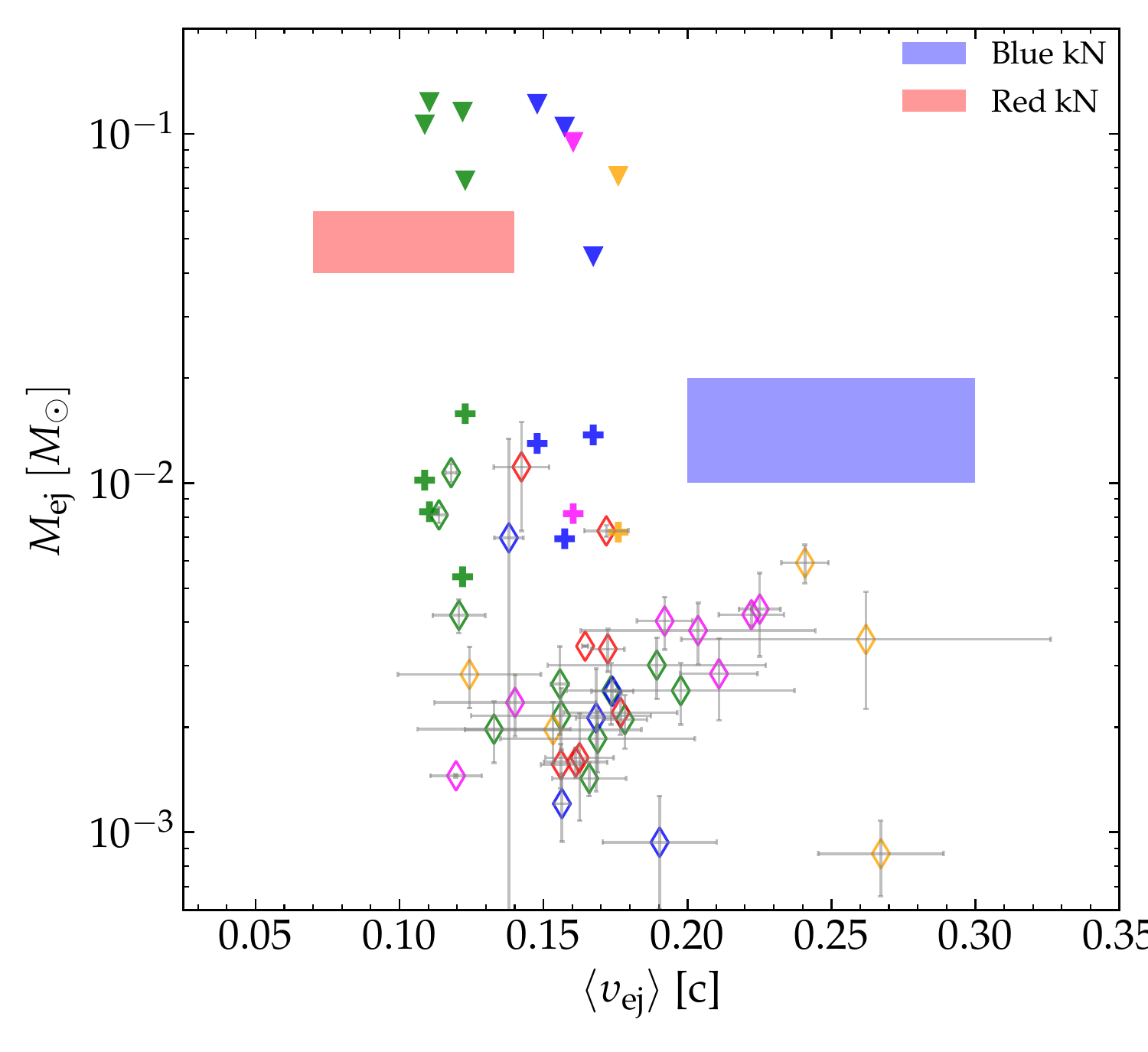}
  \includegraphics[width=0.48\textwidth]{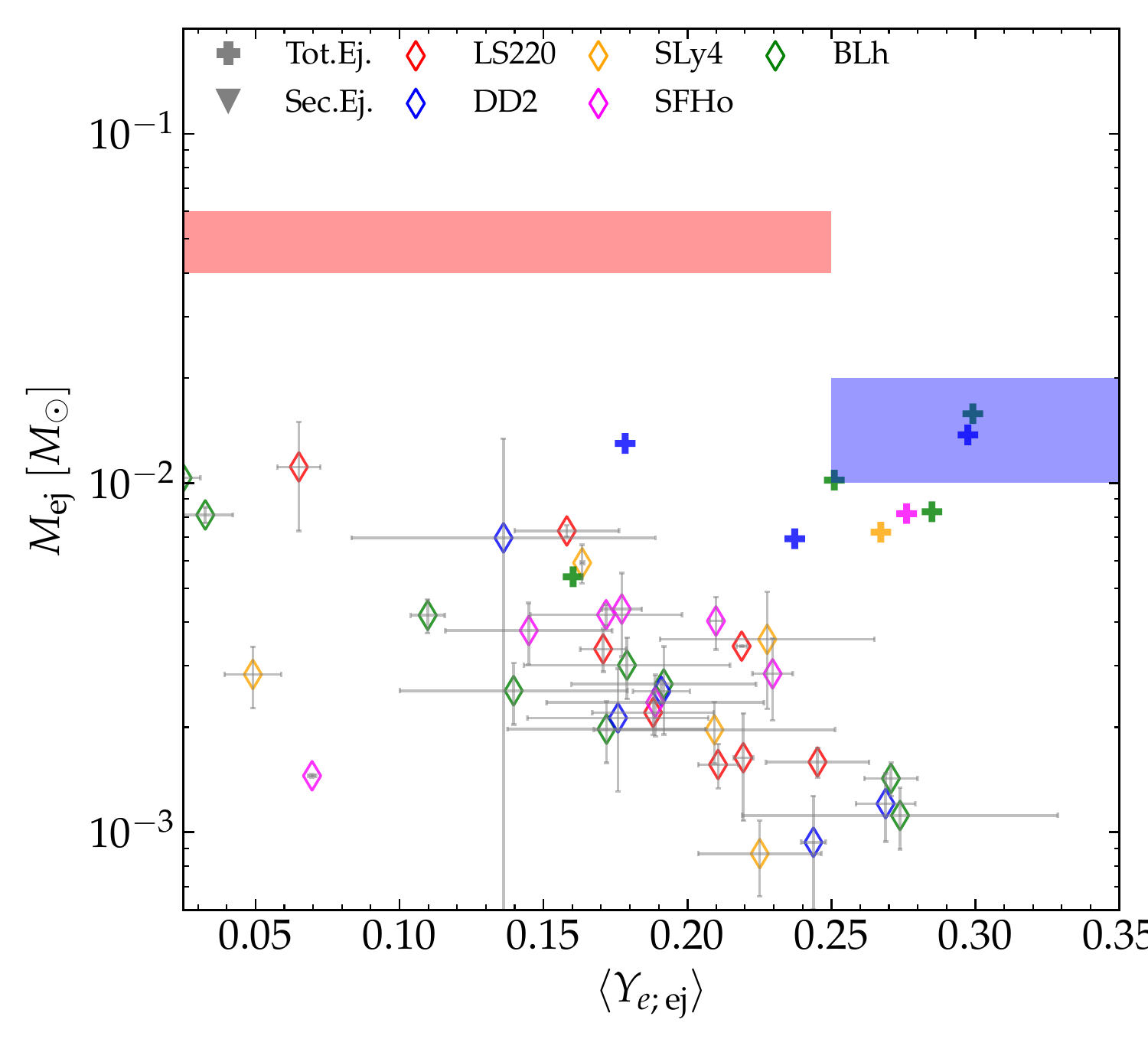}
  \caption{
    Summary of the ejecta properties of our models.
    Diamonds mark the dynamical ejecta, crosses include the
    contribution of the \swind{} for the long-lived models, 
    triangles are an estimate of the total ejecta mass on a secular
    timescale, assuming $40\%$ of the disk mass is unbounded on
    secular timescales.         
    The ejecta mass is shown is terms of the mass-averaged velocity
    (left) and of the averaged electron fraction (right).
    The filled blue and red patches are the expected values of
    ejecta mass and velocity for blue and red components of
    AT2017gfo compiled by \cite{Siegel:2019mlp}, based on
    \cite{Villar:2017wcc}. }
  \label{fig:ejecta:dyn:ds_sww}
\end{figure*}

Let us discuss an application of our results to GW170817.
We apply the best-fits using the 90\% credible intervals estimated of
$q$ and $\tilde\Lambda$ from the LIGO-Virgo GW analysis \citep{TheLIGOScientific:2017qsa,Abbott:2018wiz,De:2018uhw,Abbott:2018exr},
i.e.~$\tilde{\Lambda}=300_{-190}^{+500}$ and $q\in[1., 1.37]$. 
We find $\amd = 3.44^{+2.50}_{-2.50} \times 10^{-3}\: M_{\odot}$,
$\avd = 0.15^{+0.01}_{-0.15}$c and 
$\ayd \lesssim 0.12$. 
These values are not compatible with the ejecta
properties inferred from AT2017gfo using spherical two-components kN
models \citep{Villar:2017wcc}. \cite{Siegel:2019mlp} estimates that
the various fitting models predict 
$M_{\rm ej}^{\rm red}\in(4, 6)\times10^{-2}M_{\odot}$ and
$\upsilon_{\rm ej}^{\rm red}\in(0.07, 0.14)$ for the red component, while
$M_{\rm ej}^{\rm blue}\in(1, 2)\times10^{-2}M_{\odot}$ and $\upsilon_{\rm ej}^{\rm blue}\in(0.2, 0.3)$
for the blue component. Thus, neither component can be explained
with the dynamical ejecta from our simulations.
In Fig.~\ref{fig:ejecta:dyn:ds_sww} we show the ejecta properties from
all our models (diamonds) and the parameters inferred from the
observations as red and blue boxes. 
Despite the fact that $\ayd\sim0.15-0.25$ for comparable masses BNS, 
none of our models has dynamical ejecta massive enough to account for
the red component fit. The NR data also have significantly higher
velocities then the one inferred by the two-component kN model.
This indicates that additional ejecta components should be considered
in order to robustly associate the kN to the ejecta mechanisms 
\citep{Perego:2017wtu,Kawaguchi:2018ptg,Nedora:2019jhl}. The analysis of AT2017gfo with
realistic ejecta models and possibly more realistic radiation
transfer simulations is beyond the scope of this work, and will
be performed in future work.
We will refer to  Fig.~\ref{fig:ejecta:dyn:ds_sww} throughout the text
when discussing the \swind{} and possible winds from the remnant disks.

\section{Spiral-waves wind}
\label{sec:spiralw}

\begin{figure*}[t]
  \centering 
  \includegraphics[width=0.49\textwidth]{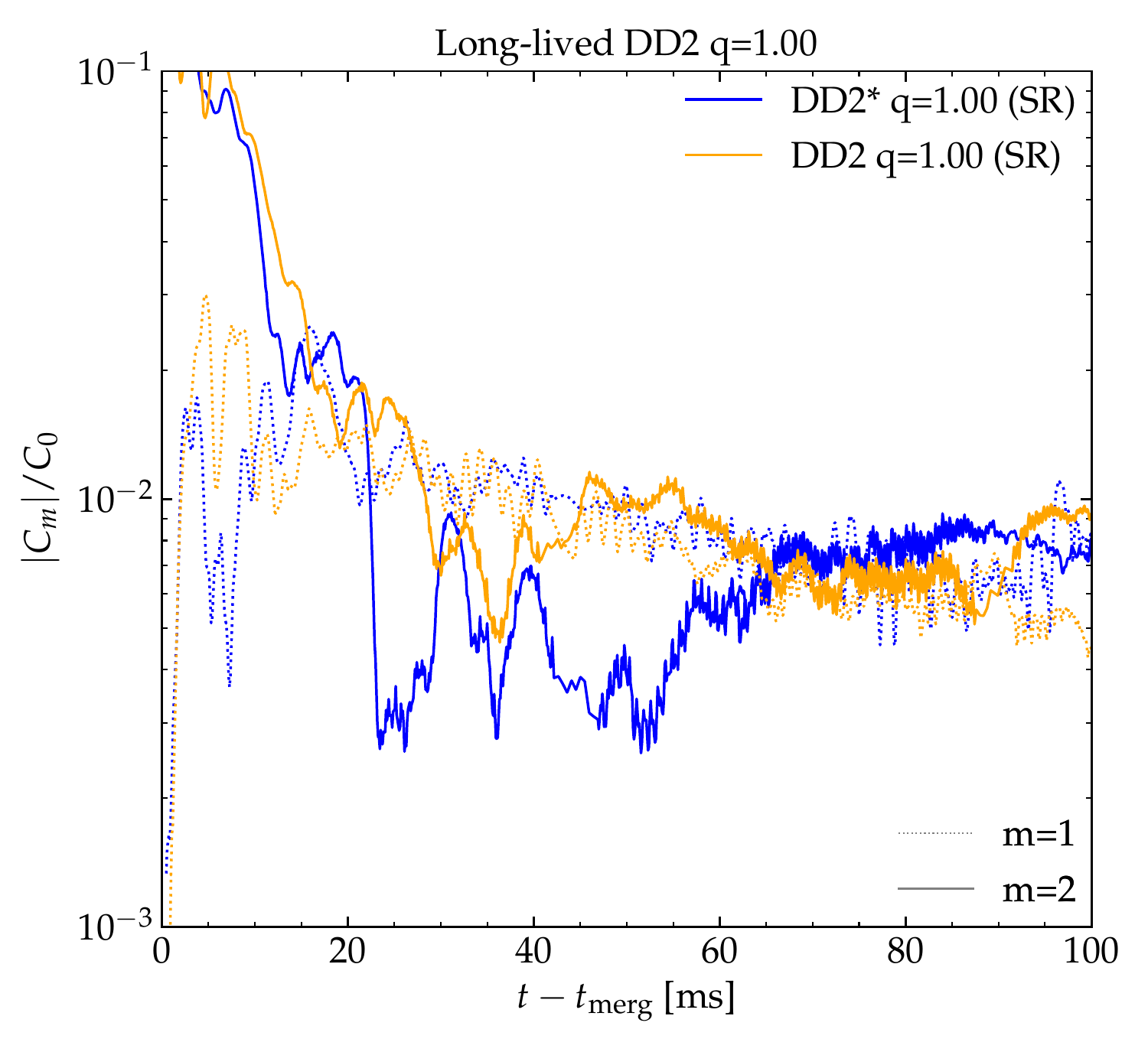}
  \includegraphics[width=0.49\textwidth]{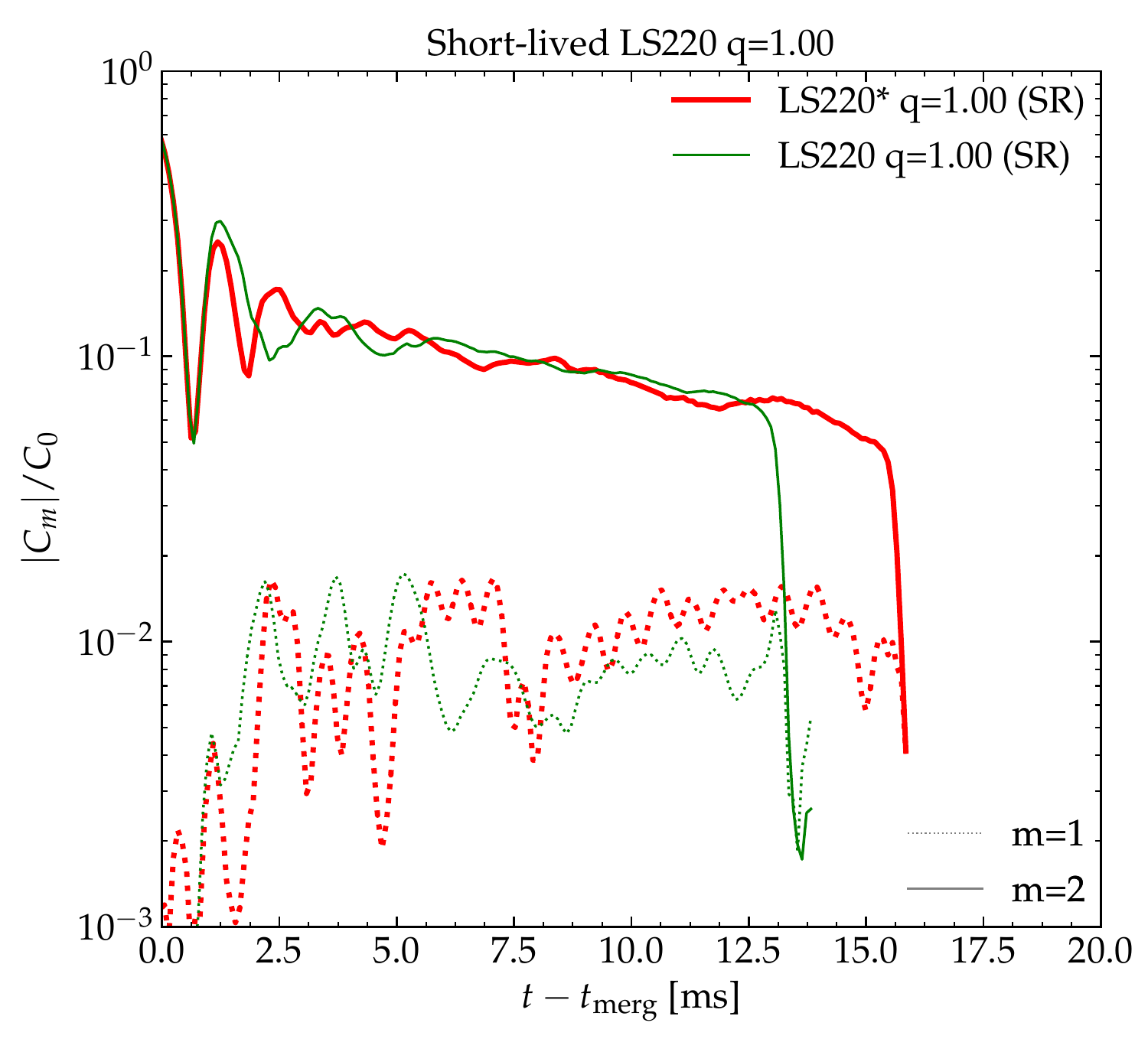}
  \caption{Modes analysis for exemplary equal-mass long-live and short-lived
  remnants. The evolution of the $m=2$ and the $m=1$ monitored by
  Eq.~\eqref{eq:modes} is shown for the DD2 and LS220 remnant with and
  without turbulent viscosity. The $m=2$ mode in the long-lived
  remnant is strongly damped by the emission of gravitational
  radiation and becomes comparable to the $m=1$ mode on a timescale of
  ${\gtrsim}20\,$ms. Turbulent viscosity sustain the $m=2$ mode for
  a longer period. The $m=2$ mode is instead dominant to collapse in
  the short-lived remnant.}
\label{fig:dens_modes}
\end{figure*}

In this section we discuss in detail the spiral waves dynamics and the associated \swind{}.
We postprocess the simulations to compute the hydrodynamical 
modes of the NS remnants using the method discussed in Sec.~\ref{sec:method:analysis}. 
The mode analysis for few representative cases is shown in
Fig.~\ref{fig:dens_modes}. The remnant NS is strongly deformed with
the characteristic spiral arms developing from the cores' shock
interface and expanding outwards
\citep{Shibata:1999wm,Shibata:2006nm,Bernuzzi:2013rza,Kastaun:2014fna,East:2015vix,Paschalidis:2015mla,Radice:2016gym,Lehner:2016wjg}. At
early times the main deformation is a $m=2$ bar-shaped mode, while at
later times a $m=1$ mode become the dominant deformation \citep{East:2015vix,Paschalidis:2015mla,Radice:2016gym,Lehner:2016wjg,Bernuzzi:2013rza,Kastaun:2014fna}.
In the short-lived LS220 $q=1$ binary, the $m=1$ mode is subdominant
with respect to the $m=2$, and it reaches a maximum close to the
collapse \citep[cf.][]{Bernuzzi:2013rza}.
Instead, in the long-lived remnant DD2 $q=1$ the $m=1$ mode 
becomes at least comparable to the $m=2$ mode at ${\sim}20$~ms and
persists throughout the remnant's lifetime, while the $m=2$ mode 
efficiently dissipates via GW emission \citep{Bernuzzi:2015opx,Radice:2016gym}.
With respect to the mass ratio we observe that the magnitude of the $m=1$ mode 
increases with $q$. In particular,
BLh $q=1.43$ and LS220 $q=1.22$ show the largest $C_{m=1}$. 
Thus remnants from asymmetric binary mergers exhibit stronger $m=1$
modes, which in turn leads to a larger \swind{} mass flux.
Regarding $C_{m=2}$, we observer no clear trend in $q$.
This is in agreement with what reported by \citet{Lehner:2016wjg}.

The spiral arms in a remnant are a hydrodynamics effect that is present also in
simulations with polytropic EOS and without weak
interactions \citep{Bernuzzi:2013rza,Radice:2016gym}. 
However, the quantitative development of these
modes in a remnant is affected by the physics input.
For example, Fig.~\ref{fig:dens_modes} highlights that turbulent
viscosity in the DD2 remnant helps sustaining the $m=2$ mode in time,
thus boosting angular momentum transport into the disk.
By contrast, the $m=1$ modes are not significantly affected by
viscosity.
On the other hand, viscosity effects are not significant on short
timescales after merger, and do not affect the dynamics of the
LS220 remnant that collapses to BH at ${\sim}15$~ms. 

We compute the angular momentum of the NS remnant and the disk under
the assumption of axisymmetry and integrating
Eq.~\eqref{eq:intro:ang_mom} using $\rho=10^{13}$~$\gccm$ as a cutting
density. We observe that, for all long-lived remnants, ${\sim}50\%$ of the angular
momentum available at formation is transported into the disk during the
first ${\sim}20$~ms. Henceforth, the disk contains about half of the
total angular momentum budget,
and the remnant settles on a quasi-stationary evolution track (see Sec.~\ref{sec:overview}).
Similarly, we estimate that 
spiral density modes inject ${\sim}0.1-0.4\,\Msun$ of baryon mass into the
disk during the first ${\sim}20$ms. 
For the same mass and mass ratio $q=1$, the DD2 remnant sheds a larger
mass into the disk than the BLh remnant, suggesting that the
process might be more efficient for stiffer EOS. 
Unequal mass binaries form a larger disks than equal mass binaries;
compare, for instance, BLh* $q=1.82$ and LS220* $q=1.43$ on the Fig.~\ref{fig:disk_mass_evo}. 

\begin{figure}[t]
  \centering 
  \includegraphics[width=0.50\textwidth]{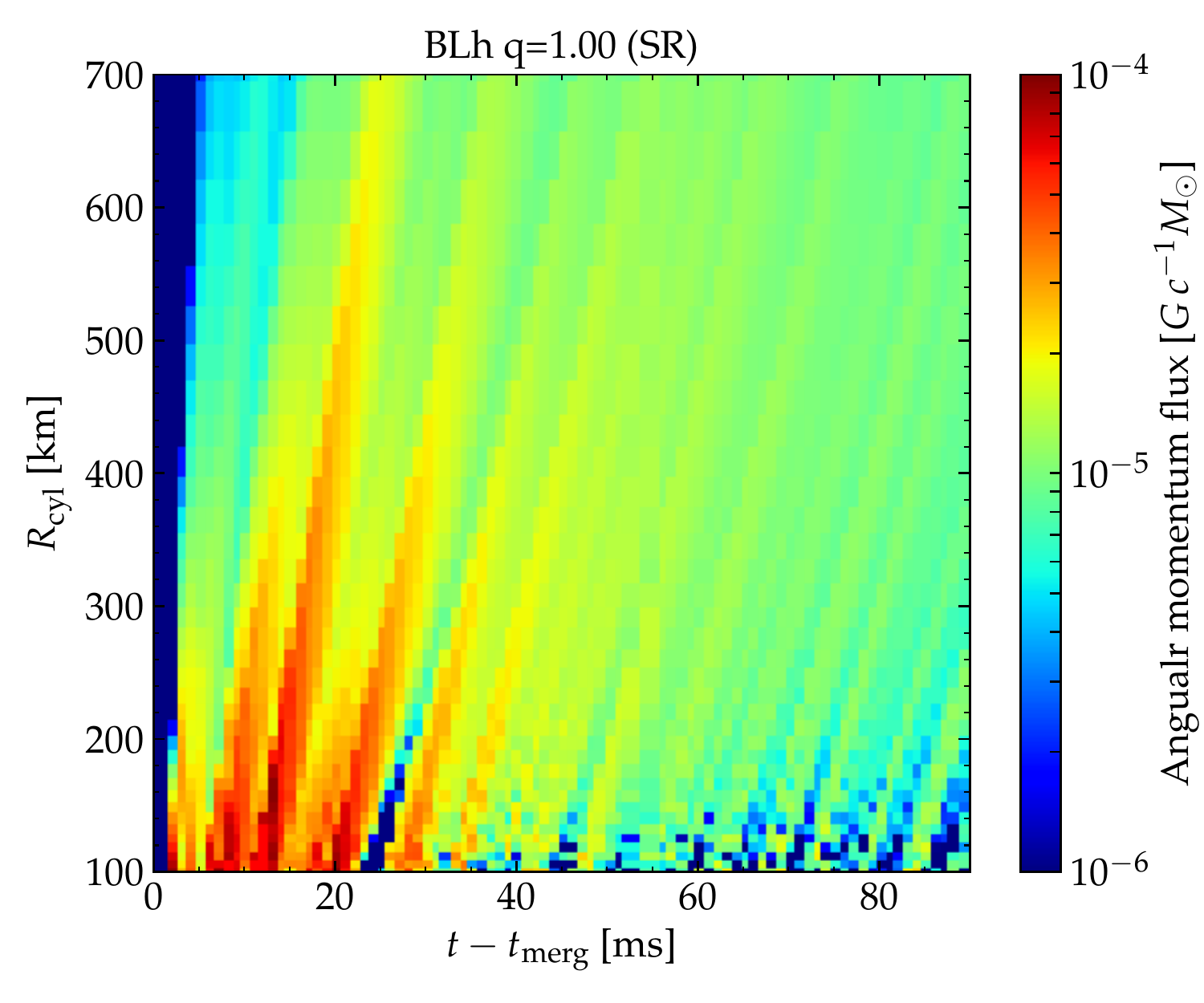}
  \includegraphics[width=0.50\textwidth]{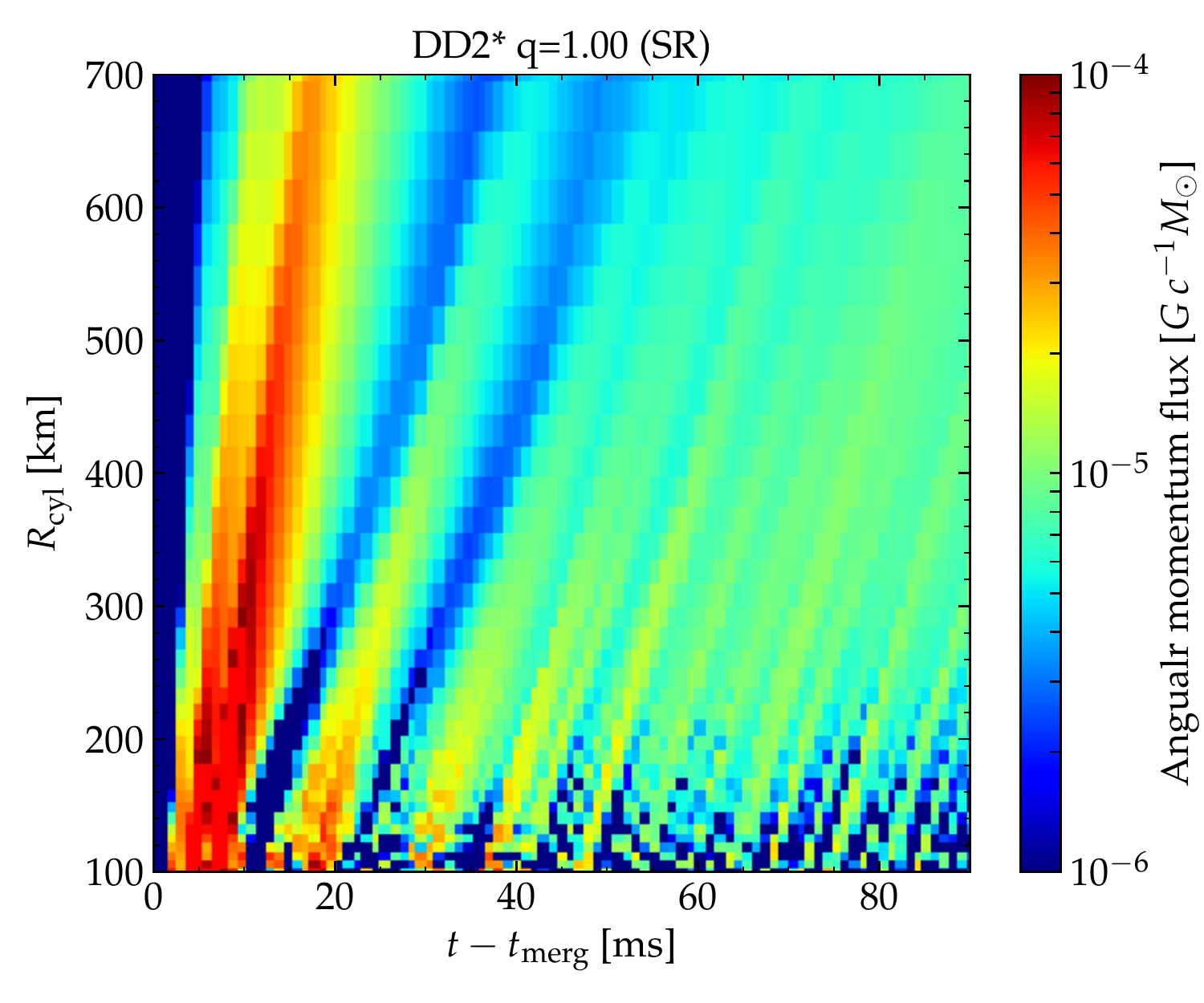}
  \caption{Angular momentum flux through consecutive cylindrical
    surfaces identified by cylindrical radii from $R_{\rm cyl}=100$ to $R_{\rm cyl}=500$. The
    plot shows the angular momentum transport into the disk.}
  \label{fig:disk_ang_mom_flux_map_blh_q1}
\end{figure}

The angular momentum transported into the disk is shown in
Fig.~\ref{fig:disk_ang_mom_flux_map_blh_q1} for the DD2 and BLh $q=1$
remnants. Angular momentum is transported by waves propagating in the disk.
These correspond to the spiral
density waves in the remnant with $m=1,2$ geometry described above.
The angular momentum transported during the first waves is larger for
the more massive DD2 disk than for the BLh. DD2* and BLh* show qualitative
differences in their evolution starting at ${\sim}20$~ms postmerger. While the DD2*
remnant continues to accrete and its disk decreases in mass, the BLh* remnant keeps on shedding
more material into the disk than what it accretes. 
See Fig.~\ref{fig:disk_mass_evo} and discussion in
Sec.~\ref{sec:overview}. The reason is the strong angular
momentum flux from the central region in the BLh* case as well as 
the larger temperature reached in this model, which lowers the rotational frequency at
which mass shedding takes place \citep{Kaplan:2013wra}. 
More simulations of the long-lived remnant evolution are required to 
investigate the effects of mass-ratio and subgrid turbulence.

\begin{figure}[t]
  \centering 
  \includegraphics[width=0.50\textwidth]{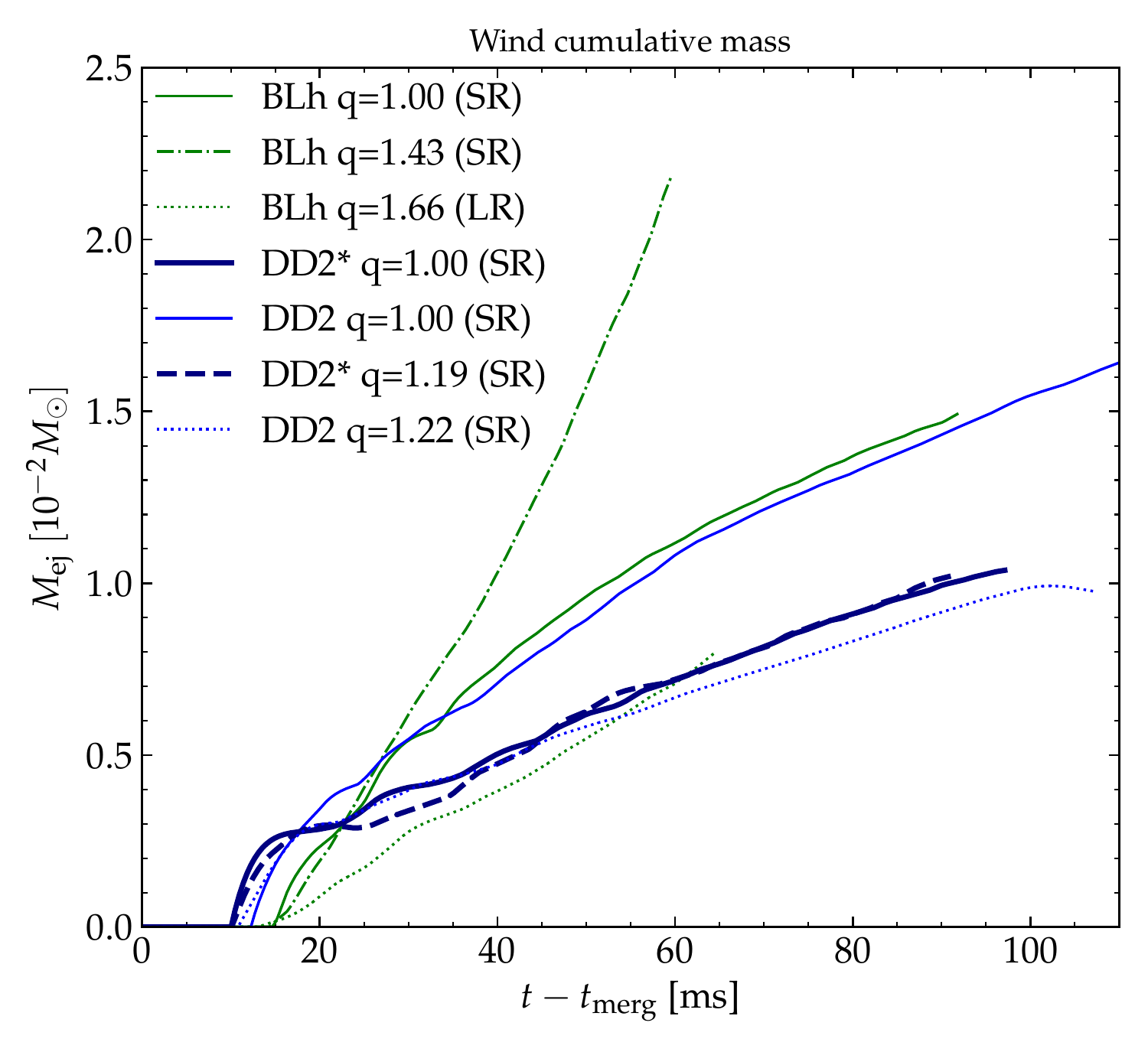}
  \caption{Cumulative mass of the \swind{} from long-lived
    remnants. The wind persists on timescales of $\O(100)\,$~ms with
    mass fluxes ${\sim}0.33-1.23\,\Msun/s$.}
  \label{fig:mej:bern}
\end{figure}

\begin{figure*}[t]
  \centering 
  \includegraphics[width=0.99\textwidth]{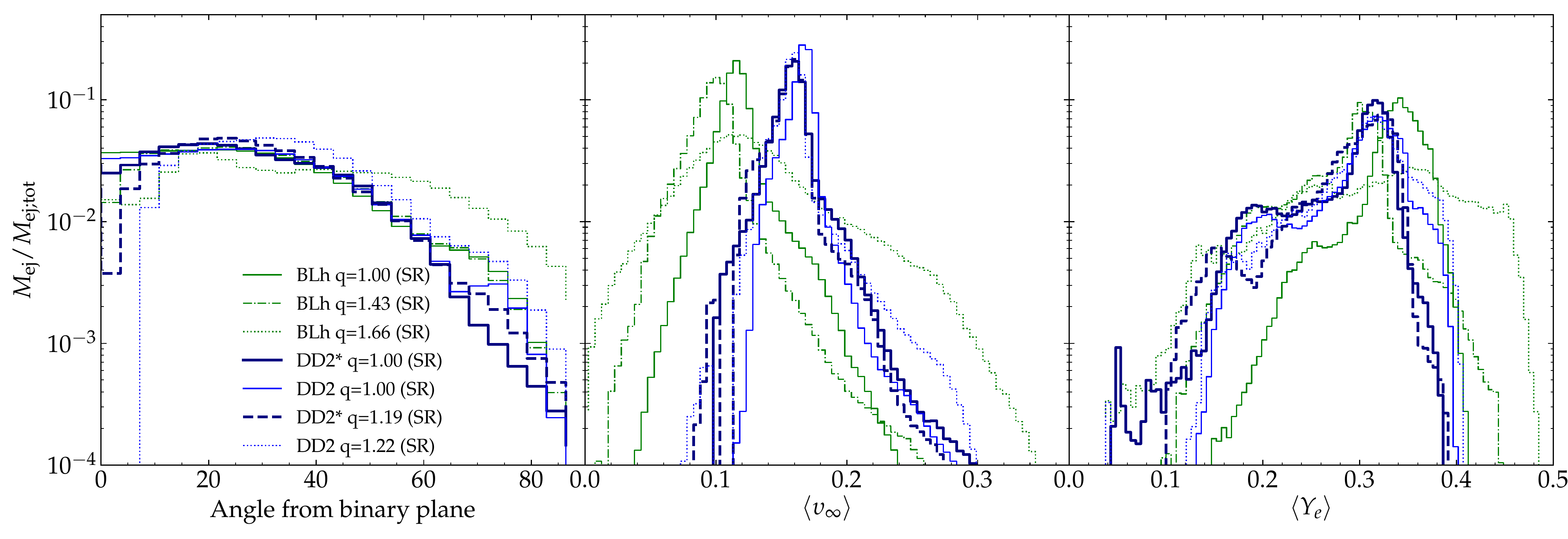}
  \caption{Mass-averaged histograms of the \swind{} for a selected
    subset of long-lived remnant. From left to right: ejecta angular
    distribution, ejecta terminal velocity and electron
    fraction. Remnants from more asymmetric binaries produce winds
    with broader angular distribution.
    The \swind{} from the DD2 EOS remnants has larger velocities
    then the winds from the softer BLh EOS. The electron fraction
    peaks at ${\sim}0.3$ and it is distributed from $0.1$ to $0.4$.}
  \label{fig:ejecta:bern:hist}
\end{figure*}

\input{tabWind.tex}

Spiral-density waves in long-lived remnants trigger a massive \swind{}
\citep{Nedora:2019jhl}. The \swind{} is computed
with the Bernoulli criterion described in
Sec.~\ref{sec:method:analysis}. Summary data are reported in Tab.~\ref{tab:spiralwavewind}.
Figure~\ref{fig:mej:bern} shows the total wind 
unbound mass as a function of time. The wind is monitored after the
mass flux of the dynamical ejecta (computed according to the geodesic
criterion) has saturated. Mass outflows due to the \swind{} continue for all
the duration of the simulations with no indication of saturation.
Indeed, while mass and angular momentum injection from
the high-density core of the remnant into the disk decreases with time
as the system becomes more stationary, the mass ejection is expected
to continue for as long as the spiral waves persist.
Because the $m=1$ modes are not efficiently damped \citep{Paschalidis:2015mla,Radice:2016gym,Lehner:2016wjg,East:2016zvv},
the ejection can in principle continue for the timescales that the
system needs to reach equilibrium or to collapse to BH (Sec.~\ref{sec:overview}).
 
The largest wind masses are obtained for asymmetric binaries like BLh
$q=1.67$ and LS220 $q=\red{1.4}$ that in about ${\sim}50$~ms unbind
${\sim}0.02\,\Msun$ with the rate of ${\sim}0.5\, \Msun/{\rm s}$. 
We find that models with softer EOS, achieve higher mass flux at lower mass-ratios,
\textit{i.e.,} the mass flux of BLh* $q=1.66$ is achieved by LS220* with $q=1.22$. 
This might be attributed to softer EOS models having a stronger $m=1$ modes in the remnant
(see Sec. \ref{sec:remdisk}). However, if these remnants collapse,
the spiral-wave mechanism shuts down and the outflow terminates. 
Thus the total ejected mass via \swind{}
depends directly on the lifetime of the remnant in
addition to the binary parameters, EOS and mass ratio.

Thermal effects play an important role in determining the outflow properties,
because high thermal pressures result in more extended disks with 
material that is more easy to unbind. The highest
temperatures in our simulations are found for the BLh EOS.
On longer timescales than those simulated, the \swind{} 
from the remnants with stiffer EOS might be larger,
also in relation to the larger disk masses (Sec.~\ref{sec:overview}).
Overall, the \swind{} from the long-lived remnant has a mass flux $\geq 0.4\, \Msun/{\rm s}$.

The property of the \swind{} are found to be remarkably uniform across our
simulated sample of remnants. In Fig.~\ref{fig:ejecta:bern:hist}, we show
mass-histograms of the wind angular distribution, velocity and 
electron fraction. The ejecta mass is distributed around the orbital plane in
a large solid angle, similarly to the dynamical ejecta. 
The electron fraction is broadly distributed in $0.1\lesssim
\ayw\lesssim0.4$ and peaks around ${\sim}0.35$. Notably, the neutron rich tail
of the distribution is determined by the early time \swind{}, before the 
quasi-steady state outflow sets in. 
The velocity peaks above ${\sim}0.1\,$c for a softer EOS and around
${\sim}0.2\,$c for a stiffer EOS.
If this picture is confirmed by future simulations, this would imply 
an EOS dependent distinct feature in the electromagnetic counterpart. 
In particular, the observation of a fast blue kN given by the \swind{}
should be associated to a stiff EOS.

Assuming that the source of AT2017gfo was a long-lived remnant surviving
for at least $\O(100)$~ms, the \swind{} would significantly
contribute the kN.
In Fig.~\ref{fig:ejecta:dyn:ds_sww} we report the total
(dynamical+\swind{}) ejecta mass and mass-averaged velocity for the
simulated long-lived BNS (crosses).
The ejecta mass and electron fraction in BLh $q=1.18, 1.42$ and DD2 is $q=1$ are compatible with the blue component inferred using the two-component kN
fit \citep{Villar:2017wcc}. However, the velocity is significantly lower than that estimated using \cite{Villar:2017wcc} models.  
Note that a multi-component fitting model that explicitly accounts
for the \swind{} can fit the early blue
emission from AT2017gfo \citep{Nedora:2019jhl}.
The emission from lanthanide-rich ejecta, however, cannot be explained
by the ejecta launched within the first ${\sim}100\,$ms of the remnant
evolution. It is thus necessary to consider mass outflows on a longer
timescales, as we shall discuss below \citep{Lee:2009uc,Fernandez:2015use,Siegel:2017nub,Fujibayashi:2017puw,Fernandez:2018kax,Radice:2018xqa}.

\section{Neutrino-driven wind}
\label{sec:wind:nu}

\begin{figure*}[t]
  \centering 
  \includegraphics[width=0.49\textwidth]{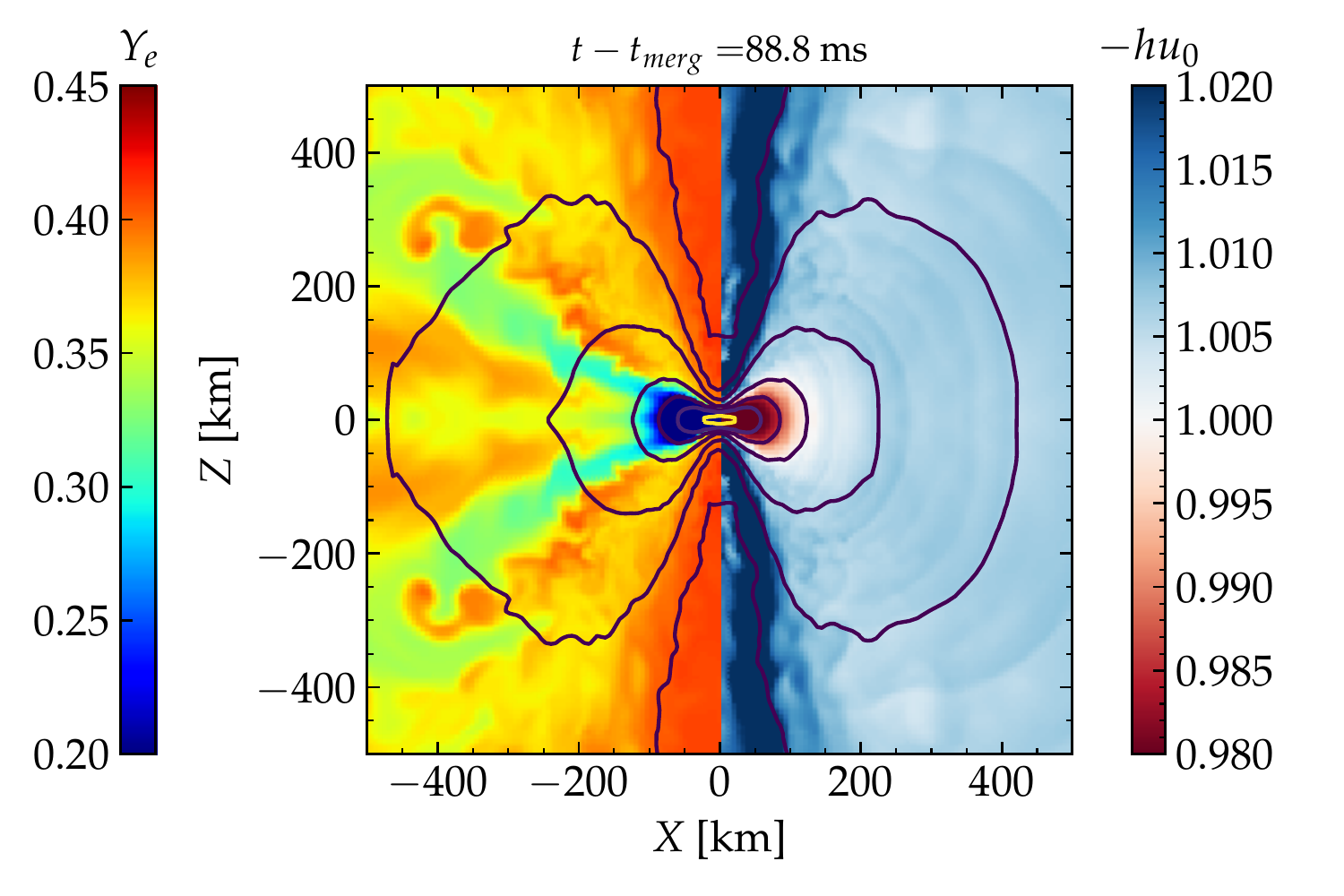}
  \includegraphics[width=0.49\textwidth]{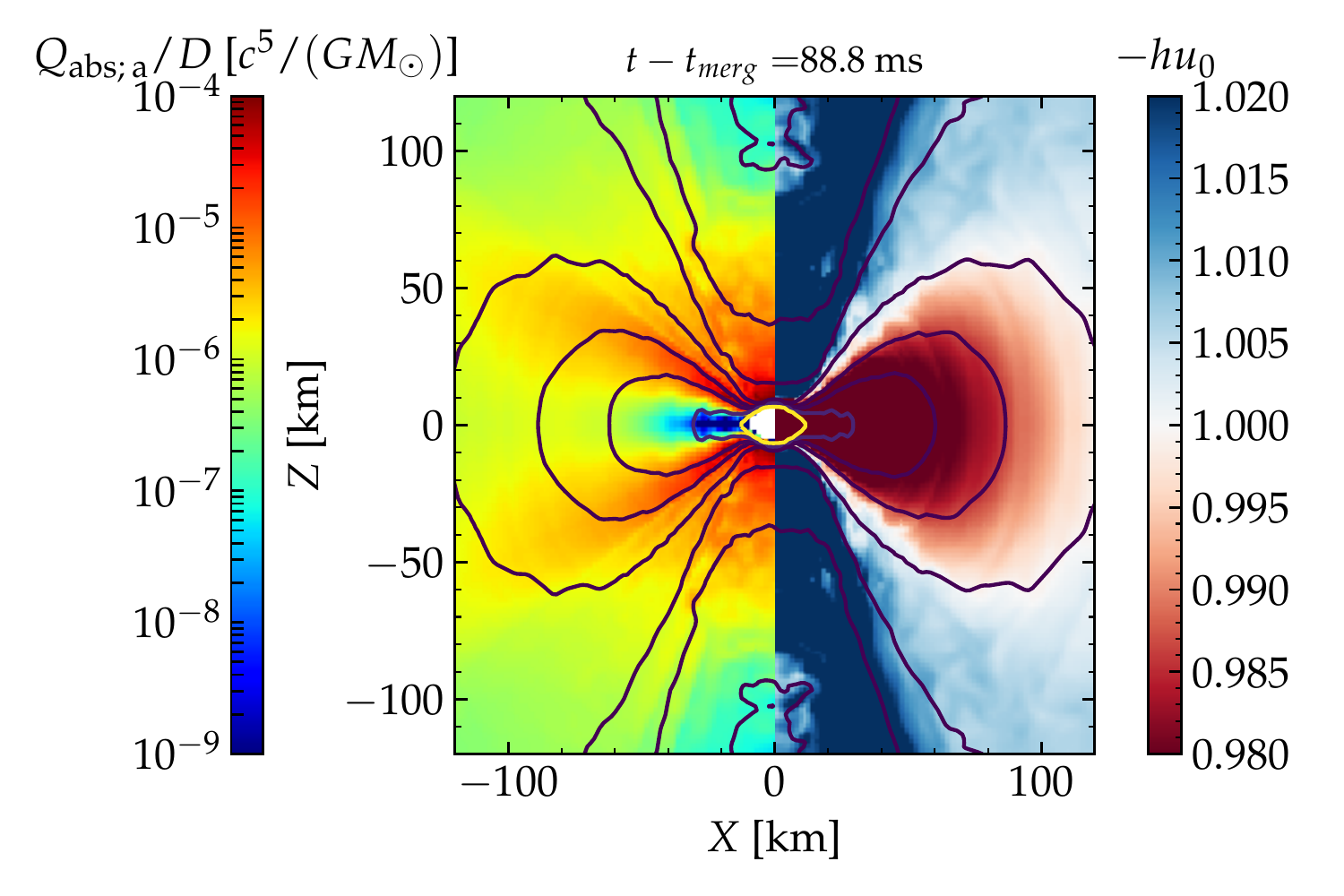}
  \includegraphics[width=0.49\textwidth]{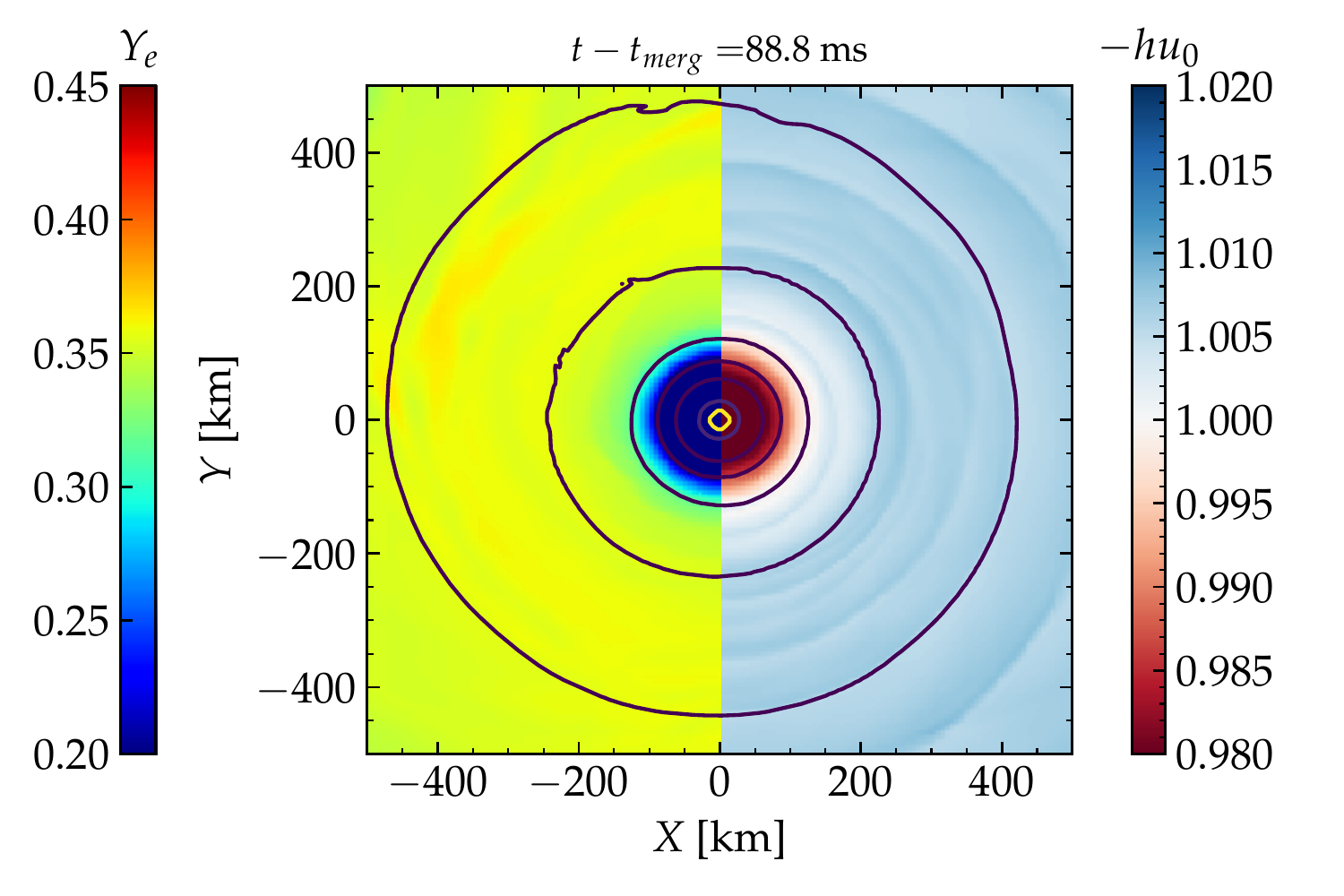}
  \includegraphics[width=0.49\textwidth]{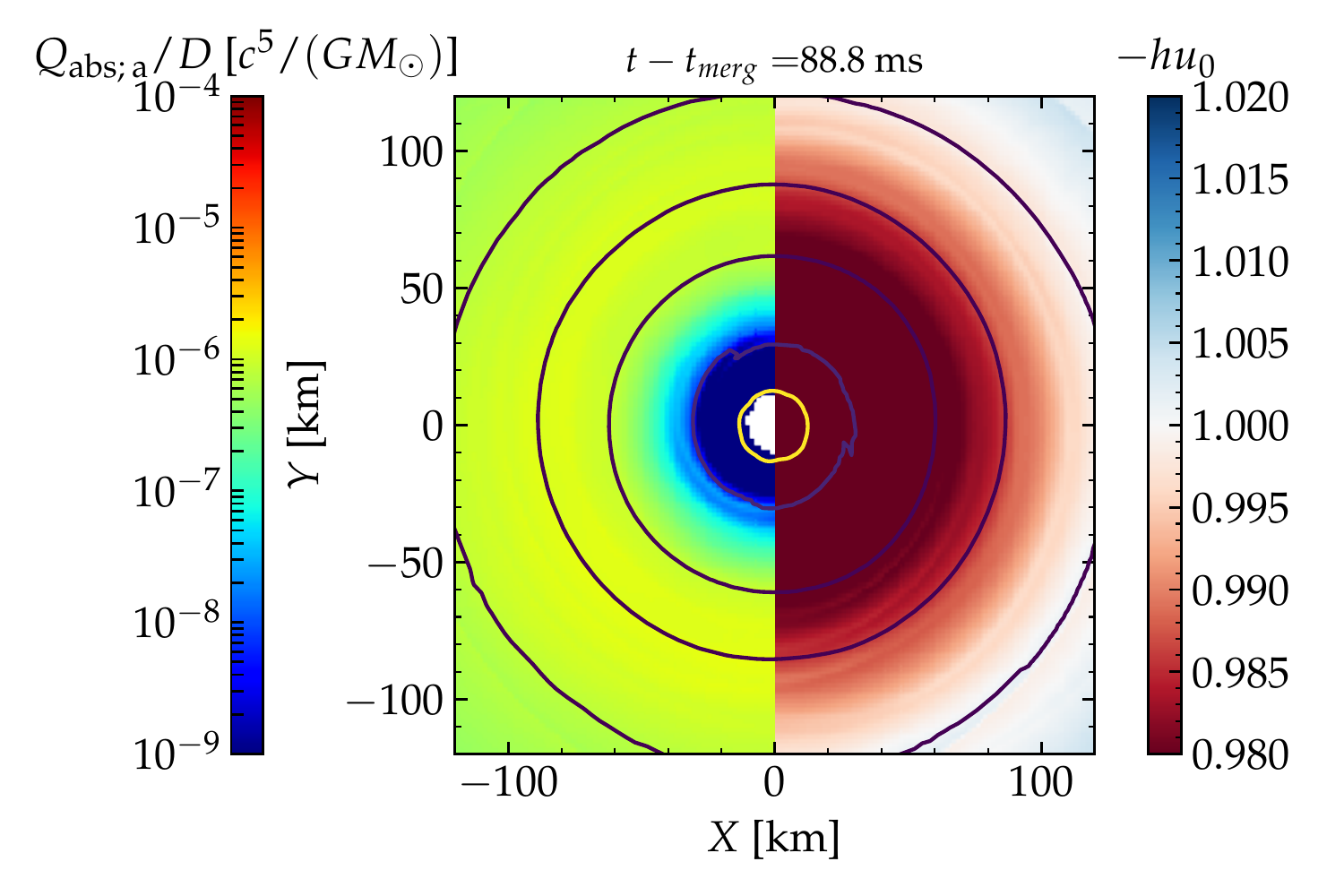}
  \caption{Snapshot of the $(x,z)$ and $(x,y)$ slices of the BLh $q=1$ model at 
    ${\sim}89\,$ms after merger. Left panels: electron fraction and
    $-hu_0$. High $Y_e$ values indicate neutrino
    postprocessing and irradiation. The $-hu_0>1$ indicates the
    material that gains enough energy to become unbound at
    infinity. 
    Right: $-hu_0$ and the absorption energy rate $Q_{\rm abs;\:\bar{\nu}_e}$ 
    of electron antineutrinos normalized to the fluid density $D$.
  }
  \label{fig:slice:heating_hu}
\end{figure*}

We study in more detail the polar component of the Bernoulli ejecta and suggest
that the outflow above the remnant is mostly driven by neutrino
absorption rather than by the spiral wave mechanisms.
Neutrino interactions above the remnant produce a baryonic outflow that
develops parallel to the rotational axis on timescales of ${\sim}\O(10)$~ms 
postmerger \citep{Perego:2014fma}. Inside this wind, rotational support
creates a funnel around the rotational axis as shown in Fig.~\ref{fig:slice:heating_hu}.
In the figure we present the electron fraction, the Bernoulli
parameter $-h u_t$ 
and the heating energy rate due to electron anti-neutrino absorption 
$Q_{\rm abs;\:\bar{\nu}_e}$ divided by with $D=W\rho\sqrt{\gamma}$ (
fluid's conserved rest-mass density) 
for the BLh $q=1$ remnant. 
We consider both the $(x,z)$ and the $(x,y)$ planes, while in the right panels
we focus on the innermost part of the remnant.
The electron fraction in the polar region with angle from binary plane
$\theta>60^{\circ}$ reaches $Y_e\sim0.35$  
due to the absorption of electron-type neutrinos.
Neutrino heating is maximal close to the bottom of the funnel where the 
\nwind{} originates.
This corresponds to densities $\rho\sim10^{11}$~$\gccm$, in the vicinity of
neutrino decoupling region \citep{Endrizzi:2020lwl}.
Large magnetic fields can further boost and stabilize the collimated
outflow in the polar region \citep{Bucciantini:2011kx,Ciolfi:2020hgg,Mosta:2020hlh}. 

We confirm that the high latitude outflows constitute a \nwind{} by studying the
correlation between the Bernoulli parameter $-h u_t$ and $E_\nu/D$.
Moreover, we verified that
simulations without neutrino heating (i.e. employed only a leakage
scheme) do not have this mass ejecta in the polar region.
A robust distinction between the \nwind{} and the \swind{} is impossible
to draw at intermediate latitudes ($\theta \sim 45^{\circ}$), where both mechanisms are at work.
The mass of the \nwind{} can be estimated either taking the ejected
material with $\theta>60^{\circ}$ or selecting $Y_e > 0.35$. 
Contrary to the main component of the \swind{} we find that, for both criteria, 
the mass flux of \nwind{} is time-dependent, exhibiting strong growth
after merger with a rapid decay in time. For most models, 
by the end of the run, the mass flux saturates, resulting in a total
of ${\sim}10^{-3}-10^{-4}M_{\odot}$ being ejected. We backtrace the cause of this
flow interruption to the presence of high density material that is lifted 
by thermal pressure from the disk and pollutes the polar regions.
The properties of this outflow are qualitatively similar to the ones discussed 
in e.g. \citet{Dessart:2008zd,Perego:2014fma,Fujibayashi:2020dvr}. In some of these
models the \nwind{} develops over longer timescales than those considered here, it 
achieves a quasi-steady state and it possibly unbinds larger masses. 
These differences could 
result from the conservative choices we have done in isolating the \nwind{} contribution
and in the lack of \swind{} in the other models. Moreover, it could be that the right conditions 
for the formation of a steady \nwind{} might not have been reached in our simulations yet.

\section{Remnant disk structure}
\label{sec:remdisk}

\begin{figure}[t]
  \centering
  \includegraphics[width=0.49\textwidth]{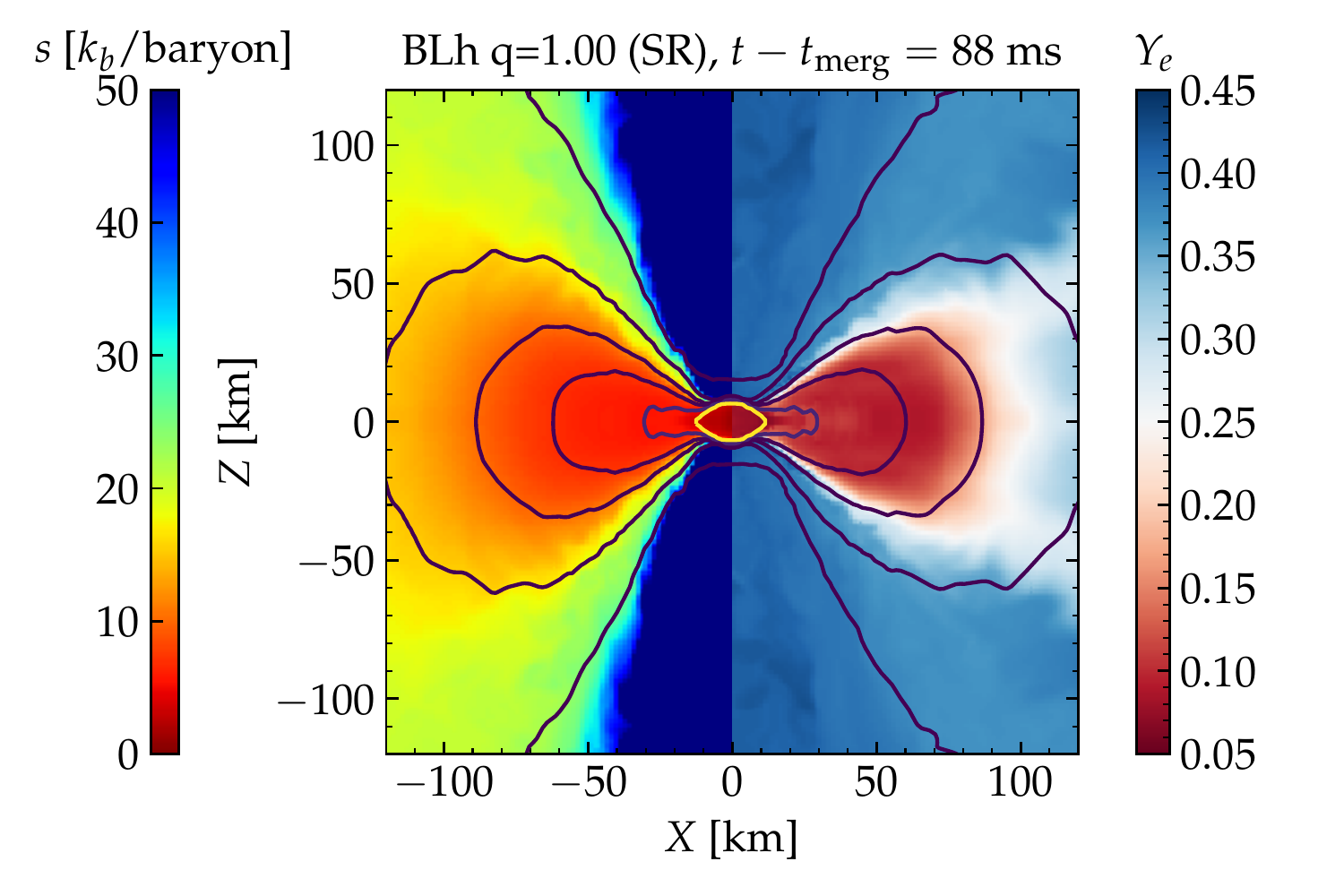}
  \includegraphics[width=0.49\textwidth]{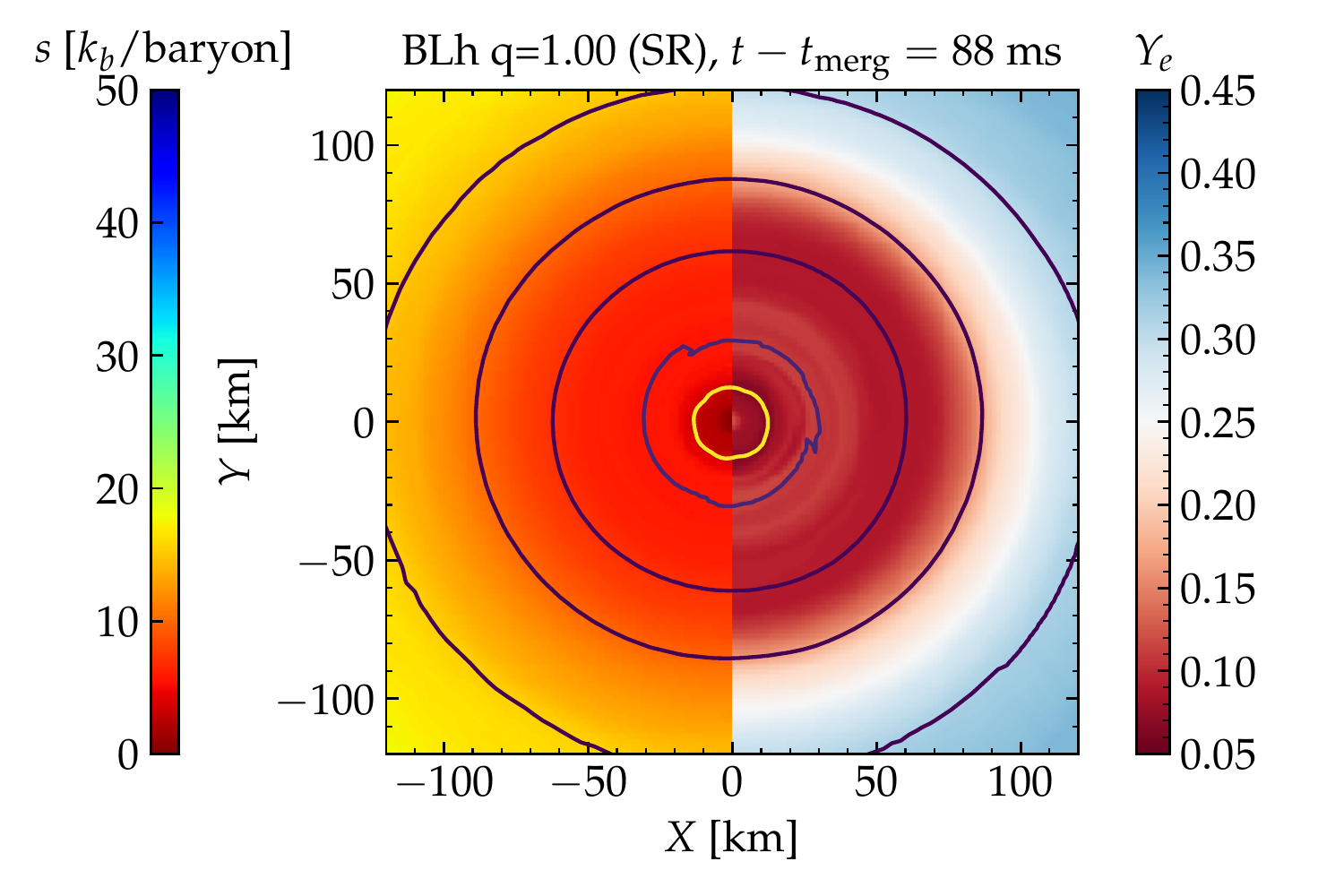}
  \caption{Entropy and electron fraction on the $(x,z)$ (top) and
    $(x,y)$ planes (bottom) for the remnant of BL $q=1$ at the end
    of the simulation. Each plot is divided vertically, with entropy
    being color-coded on the right and electron fraction on the
    left. Solid contours stand for rest muss density. Counting from
    the center, the values are $[10^{13}, 10^{12}, 10^{11}, 10^{10},
      10^{9}]$ g cm$^{-3}$, with the inner most contour encompassing
    the remnant.}  
  \label{fig:snapshots_xy_ye_entr}
\end{figure}

\begin{figure*}[t]
  \centering 
  \includegraphics[width=0.95\textwidth]{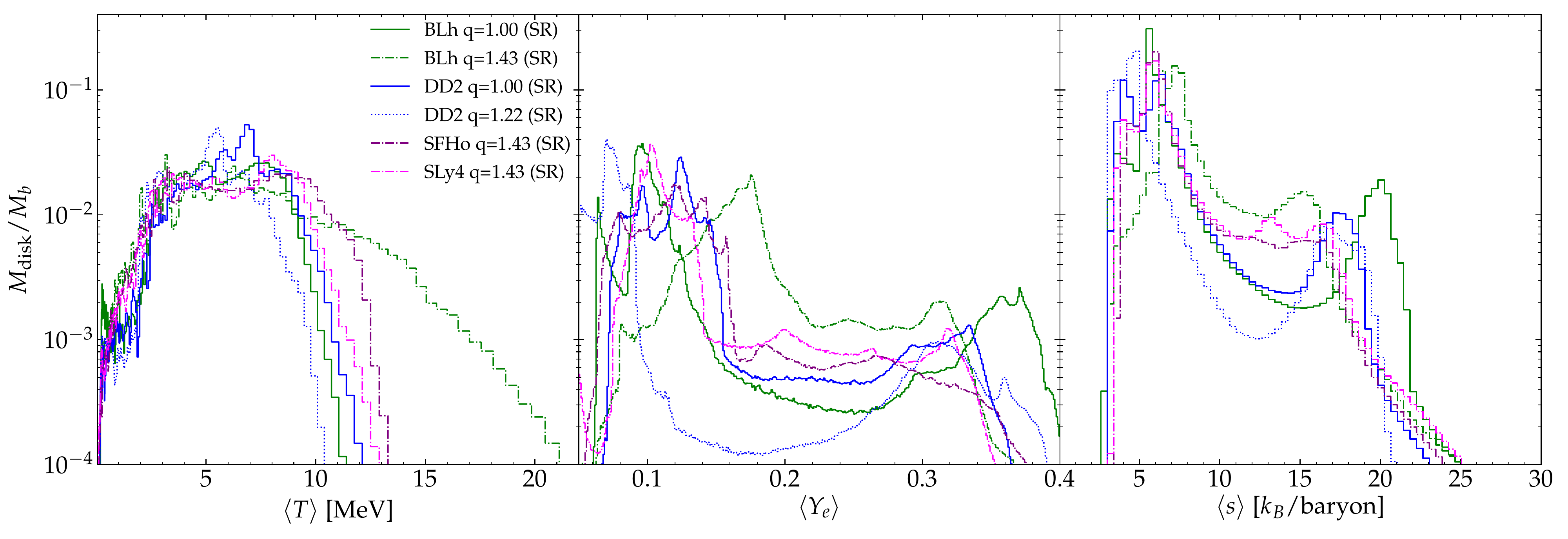}
  \caption{Composition of the disks at the end of the long-lived
    remnants simulations. The histograms refer to the temperature $T$ (left), entropy $s$
    (middle) and electron fraction $Y_e$ (right).}
  \label{fig:final_disk_struct_hist_long}
\end{figure*}

We now discuss the disk structure in long-lived remnants at the
end of our simulations, namely at ${\sim}60{-}100$~ms postmerger, 
and the final disk masses of all our models.

We find that disks around remnant are geometrically thick, 
with a RMS opening angle of $\langle\theta\rangle_{\rm rms}\sim60^{\circ}$, 
rather independent of the EOS and $q$.
Meanwhile, the radial extend is larger for softer EOS and for larger $q$. 
The final disk masses range between ${\sim}0.1M_{\odot}$ and ${\sim}0.4M_{\odot}$ 
(see Tab.~\ref{tab:sim}); smaller masses are obtained for short-lived
remnants and for equal-masses binaries.
The mean value and standard deviation are $\overline{M}_{\rm
  disk}=(0.161 \pm 0.083)M_{\odot}$.
Similarly to what we did for the dynamical ejecta, we
fit the disk masses with a second order polynomial in
$(q,\tilde{\Lambda})$. The coefficients of Eq.~\eqref{eq:fit:poly22}
for this fit are given in Tab.~\ref{tab:fitpoly22coefs}.
A more detailed study with various fitting formulas and extended
datasets from the literature is reported in a companion paper.

The disk composition at ${\sim}60{-}100$ms postmerger is not uniform,
and we study it using the mass-weighted histogram reported in
Fig.~\ref{fig:final_disk_struct_hist_long}. 
The entropy and the electron fraction show a bimodal distribution
which is more prominent for equal-mass binaries and less prominent for large
$q$ ones. The mass-weighted distribution of the entropy shows a dominant 
peak at low entropy $s\sim5-10k_B/$baryon. This peak is rather EOS and $q$
independent and it correspond to the inner, mildly shocked material. 
The second, subdominant peak is located at larger
entropies, $s\sim15-22k_B/$baryon,and it is more depended on the EOS
model: for softer EOSs a larger amount of mass reaches a larger entropy, while for
more asymmetric binaries the second peak is centered around lower values of the entropy. 
Similarly, we observe a first peak in the $Y_e$ distribution, around $Y_e\sim0.1$,
that corresponds to the neutrino-shielded bulk of the disk. The
second (subdominant in mass) peak is at $Y_e\sim0.3-0.4$ and it corresponds to the irradiated
disk surface. We stress that, both for the entropy and the electron fraction, the two peaks
refer to different regions inside the disk, as visible in Fig.~\ref{fig:snapshots_xy_ye_entr}.
Most of the matter in the disk has temperatures in the range $T\sim 1-10\,$MeV.
The inner part of the disk is hotter than the edge. The temperature distribution
is also weakly independent of the EOS and mass ratio. \\

Nuclear recombination is expected to unbind a fraction of the disk mass
on secular timescales of a few seconds, longer than those simulated here.
Simulations and analytical estimates indicate that up to ${\sim}40\%$
of the disk would become unbound due to viscous processes, with typical
velocities of the order of ${\lesssim}0.1\,$c
\citep{Lee:2009uc,Fernandez:2015use,Wu:2016pnw,Siegel:2017nub,Fujibayashi:2017puw,Fernandez:2018kax,Radice:2018xqa,Fujibayashi:2020dvr}. 
Assuming these values, the mass of the secular wind from our simulated
remnant disks would amount to ${\sim}0.05\, M_{\odot}$. 
We include this secular wind estimate in
Fig.~\ref{fig:ejecta:dyn:ds_sww} for the long-lived remnants (lower
triangles). The estimated mass is sufficient to explain
the red component of AT2017gfo, as inferred from the two-components kN
models of \cite{Villar:2017wcc}.

\section{Nucleosynthesis}
\label{sec:nucleo}

\begin{figure*}[t]
    \centering 
    \includegraphics[width=0.45\textwidth]{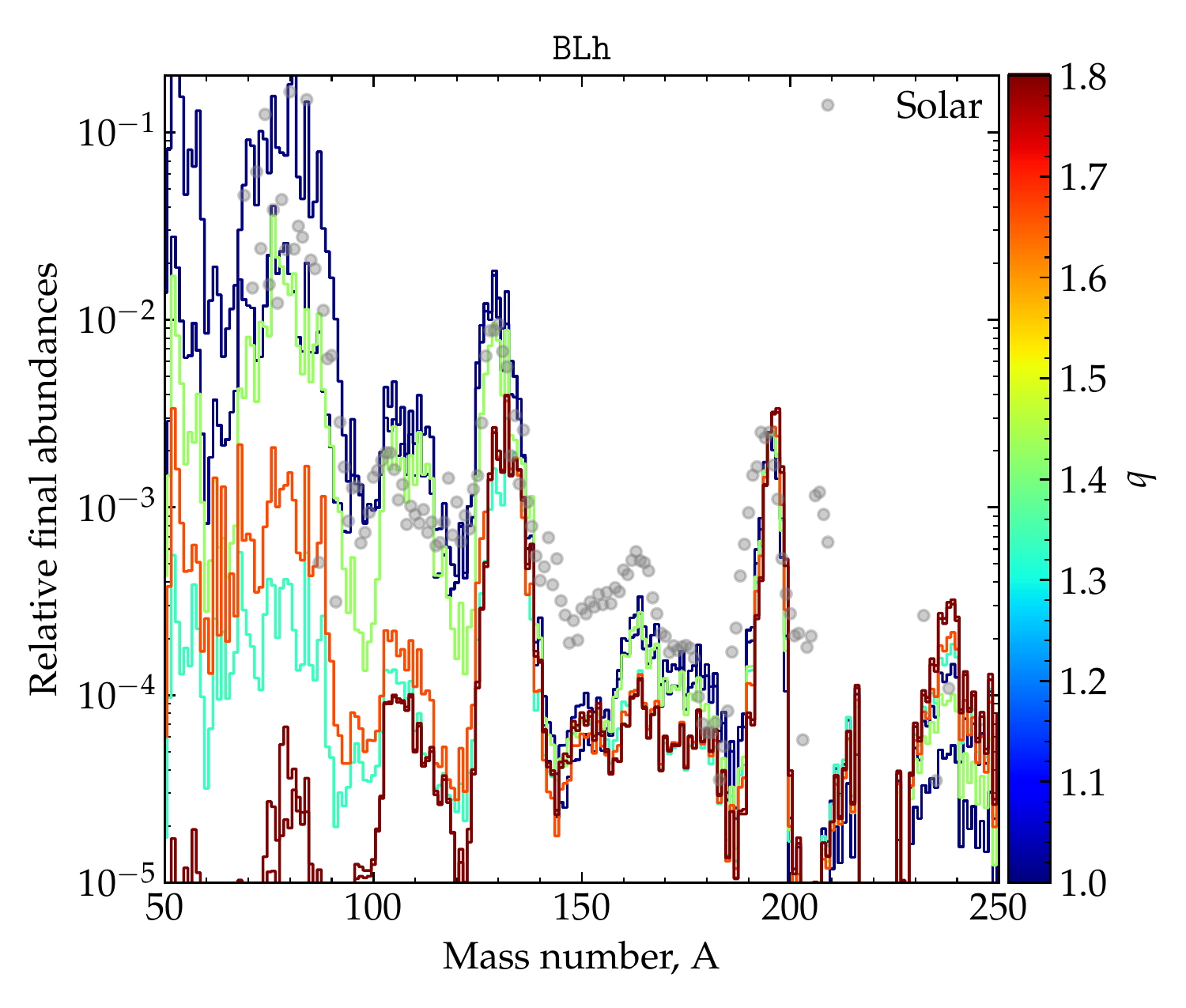}
    \includegraphics[width=0.45\textwidth]{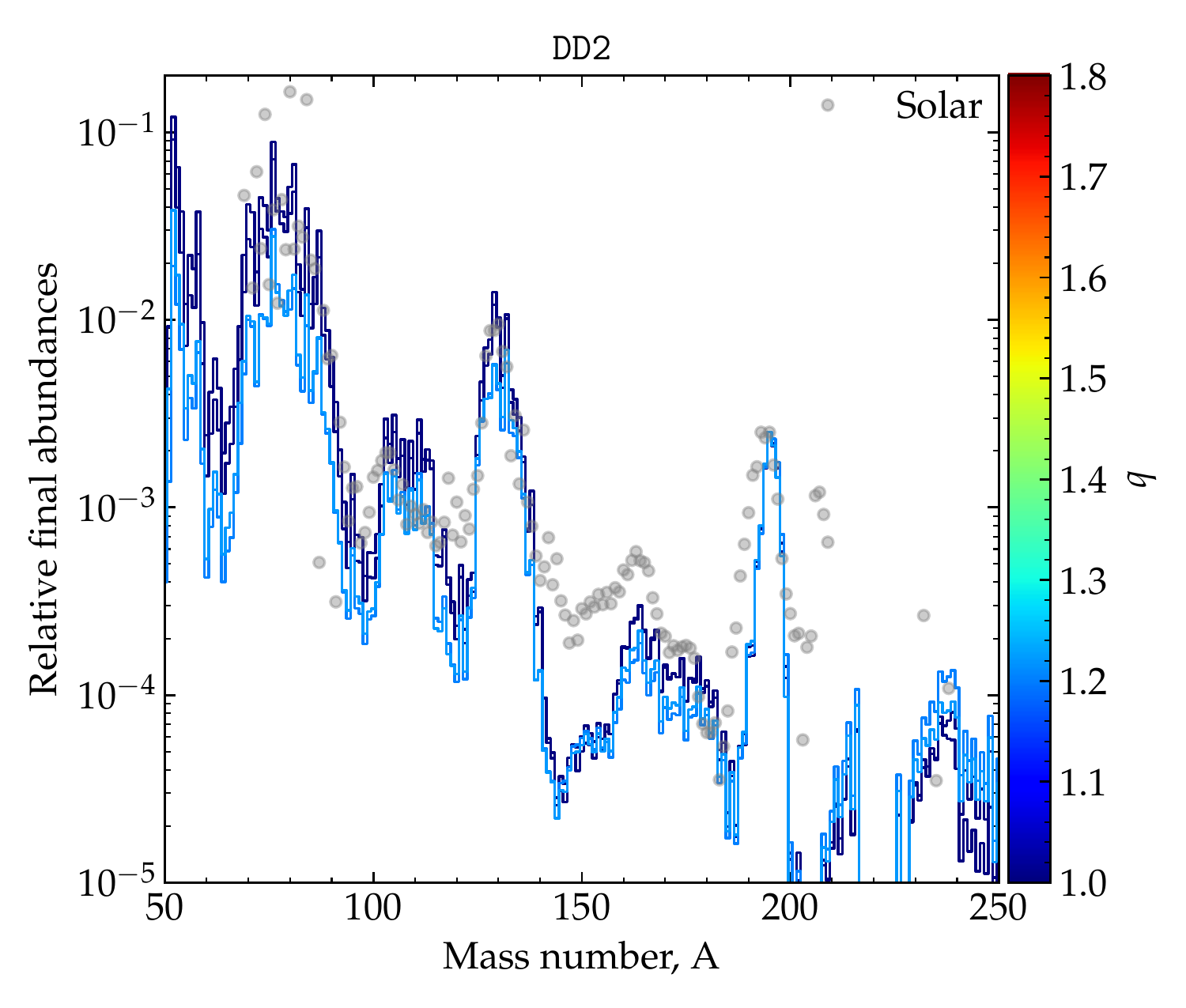}
    \includegraphics[width=0.45\textwidth]{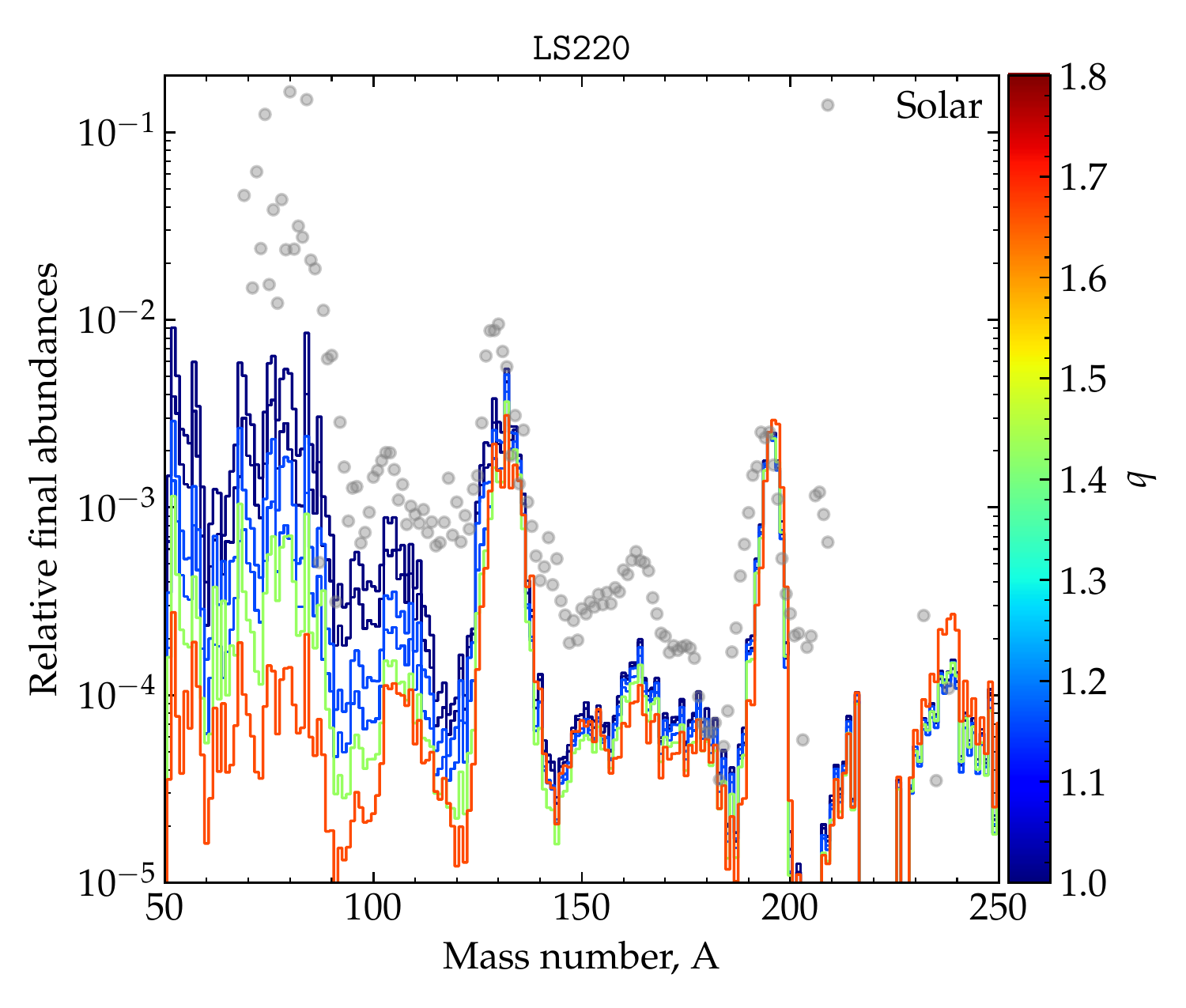}
    \includegraphics[width=0.45\textwidth]{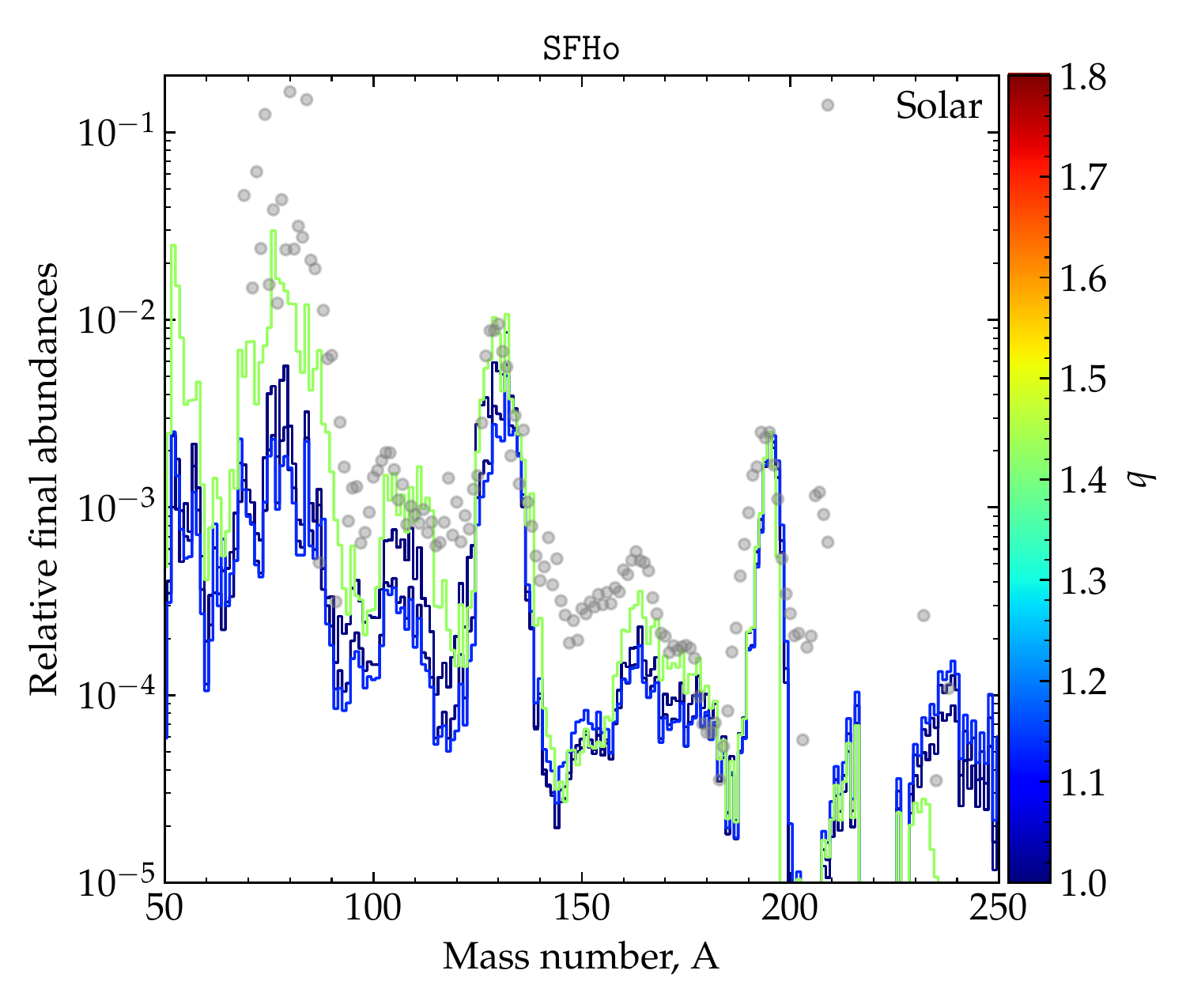}
    \includegraphics[width=0.45\textwidth]{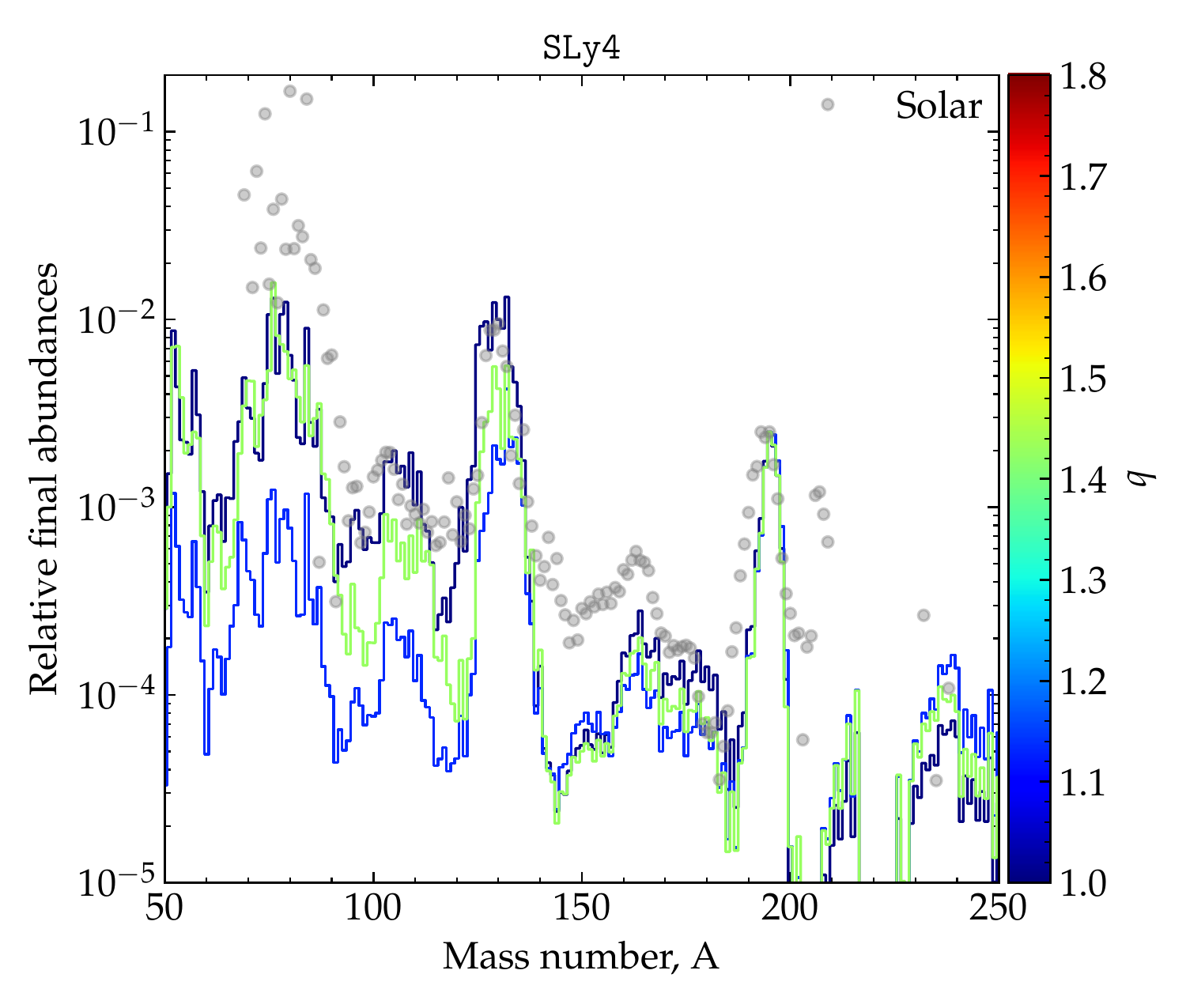}
    \includegraphics[width=0.42\textwidth]{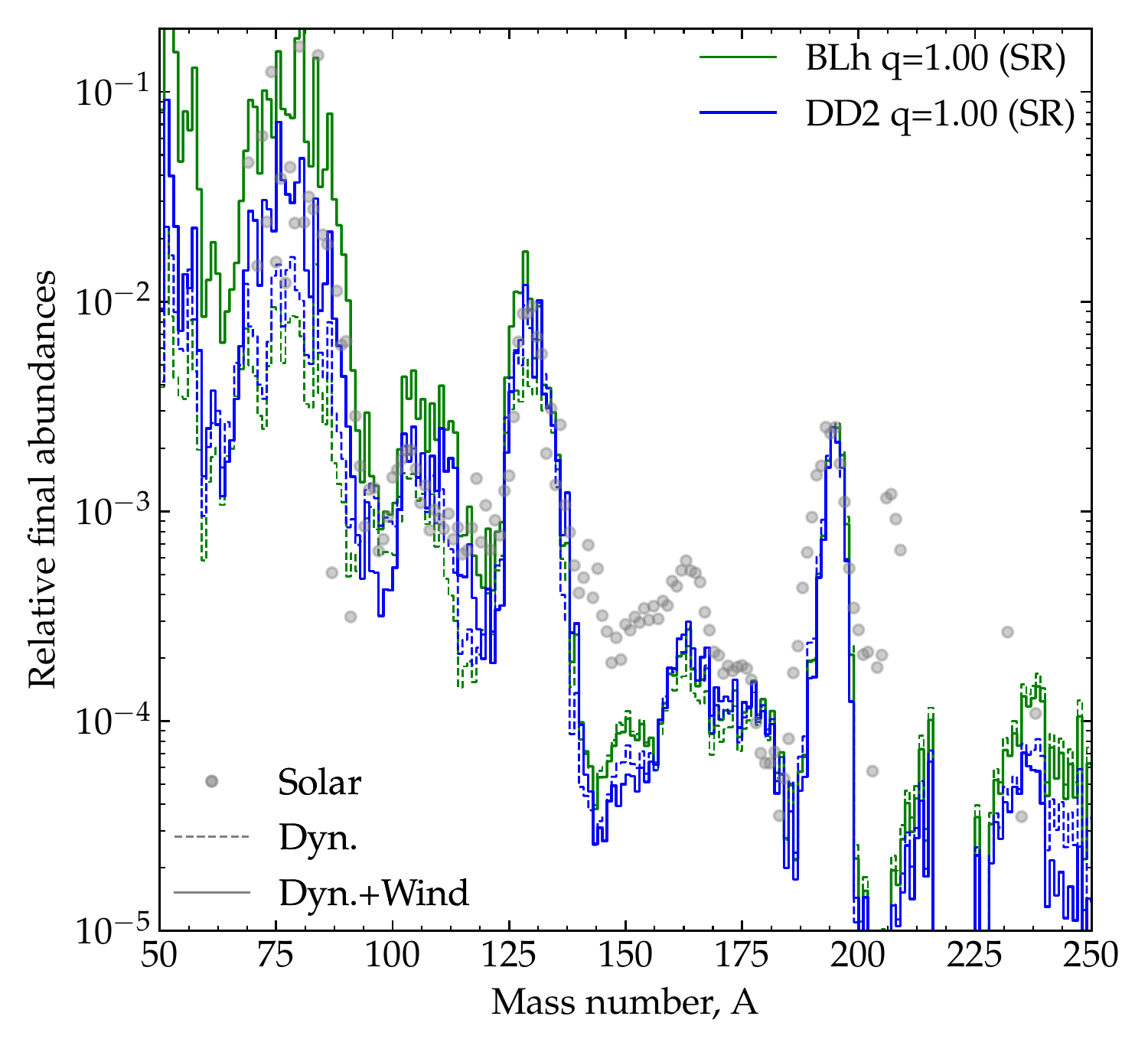}
    \caption{Nucleosynthesis yields for all simulations. Each 
      of the first five panels 
      shows a different EOS and the scale color the dependency on the
      mass ratio. The nucleosynthesis is computed on the total ejecta
      computed during the simulations and 
      composed of the dynamical (all models) plus the \swind{} 
      (for the long-lived remnants listed in
      Tab.~\ref{tab:spiralwavewind}.).
      The last (bottom-right) panel compares the nucleosynthesis in
      the dynamical ejecta and \swind{} for the long-lived
      remnants. The inclusion of the \swind{} contributes to 
      improve the agreement with solar data for elements around the first peak.}
    \label{fig:nucle:totalyields}
\end{figure*}

The nucleosynthesis calculations are performed in postprocessing
following the same approach as in \cite{Radice:2016dwd,Radice:2018pdn}.
We report the abundances as a function of the mass number $A$ of the different 
isotopes synthesized by the $r$-process 32 years  
after the merger in the material ejected from the system.
Comparing to our previous study \citep{Radice:2018pdn}, the new
simulations allow us to investigate more in detail the nucleosynthesis
in presence of neutrino absorption, the contribution of the \swind{}
in long-lived remnant, and the effect of mass ratio up to $q\sim1.8$.

Figure~\ref{fig:nucle:totalyields} shows the nucleosynthesis yields
from the dynamical ejecta (short-lived remnants) and from the
dynamical ejecta + wind (long-lived remnants).
We compare the abundances inferred from the simulations with up-to-date solar residual 
$r$-process abundances from
\cite{Prantzos2020} (for a review of the solar system abundances see
\eg~\citealt{Pritychenko:2019xvf}). 
To compare the different distributions, we shift the 
abundances from our models such that they are always the same as the solar one 
for $A=195$.
Notably, all the $r$-process peaks are reproduced by the
nucleosynthesis in the ejecta expelled by the long-lived DD2 and BLh models.
This demonstrates that the complete solar $r$-process abundances can be recovered 
if the remnant is long-lived and shows the presence of a \swind{}. This is a consequence of the robust 
properties of the latter.
The possibility of short-lived binaries to reproduce the 
solar $1$st and $2$nd r-process peaks, at $A\sim 75$ and $A\sim 125$
respectively, 
strongly depends on the mass-ratio.
Higher-$q$ binaries,
whose dynamical ejecta is mostly of tidal tail origin with very low
electron fraction, show severe underproduction of light $r$-process
material. On the contrary, $q\sim1$ binaries reproduce both
peaks reasonably well. This is the result of inclusion of neutrino
reabsorption as it increases the $Y_e$ of the shocked component of the
ejecta \citep{Radice:2018pdn}.

We find that actinides ($A\sim 230$) are produced in all our
models, but their abundances depend sensitively on the mass ratio.
Very asymmetric binaries produce larger amounts of low-$Y_e$ ejecta which
result in an increased production of actinides, broadly compatible with the solar
pattern. Interestingly, only the highest
mass ratio binaries are able to produce at the same time aboundances close to solar
for the $3$rd r-process peak and for actinides around $^{232}$Th.
This suggests that asymmetric mergers (or, alternatively, BHNS mergers), might play an important
role for the production of the heaviest elements through $r$-process nucleosynthesis.

For long-lived binaries the dynamical ejecta amounts only to a
small fraction of the total mass of material leaving the system, while
the \swind{} is the more massive ejecta in our simulations.
In 
the bottom-right panel of Fig.~\ref{fig:nucle:totalyields}
we show how the inclusion of the
\swind{} changes the abundances of two representative models.
Due to its overall high electron fraction, 
the \swind{} (see Fig. \ref{fig:ejecta:bern:hist}) 
primarily produces first-peak r-process elements $A<95$. 
Since the abundances are normalized to the
$3$rd peak, the relevant differences are those in the $1$st and $2$nd peaks.
We observe that due to the slightly higher average electron fraction of the BLh
outflows (Fig. \ref{fig:ejecta:bern:hist}), it produces more light elements,
$A\sim75$, then DD2 binary. 
Both binaries, however, display abundance pattern noticeably close to solar.

In addition to the dynamical ejecta and \swind{}, the $r$-process
nucleosynthesis occurs in the neutrino-driven wind and the secular
wind from the disk. In the neutrino-driven winds, neutrino irradiation of the expanding
ejecta considerably increases the electron fraction. If the velocity
of the ejecta is sufficiently low, the material reaches the weak
equilibrium with neutrinos in optically thin conditions, and $Y_e\leq 0.45$
\citep{Qian:1996xt}. This will further boost weak $r$-process
nucleosynthesis of light elements $A<130$
\citep{Dessart:2008zd,Perego:2014fma,Just:2014fka,Martin:2015hxa,Foucart:2016rxm}. 
The viscous- and recombination-driven wind is expected to constitute the bulk
of the disk outflow, but this takes place on  longer timescales than those
considered here. Simulations of such systems
\citep{Fernandez:2013tya,Just:2014fka,Wu:2016pnw,Siegel:2017nub,Fujibayashi:2017puw,Fernandez:2018kax}
suggest that this component of the outflow will have a broad range of $Y_e$ and
will synthesize both light and heavy $r$-process nuclei. However, heavy
$r$-process production might be suppressed in the case of long-lived massive NS
remnants \citep{Metzger:2014ila,Lippuner:2017bfm}.

\section{Conclusion}

In this work we discussed the long-term postmerger dynamics of $37$ 
binaries with chirp mass $\mathcal{M}_c=1.188\,\Msun$ compatible to
the source of GW170817, gravitational mass spanning the range
$M\in[2.73, 2.88]\,\Msun$ and mass ratio values $q\in[1,1.8]$. 
Our models were computed with five microphysical EOS compatible with nuclear and
astrophysical constraints. Each binary was simulated at multiple
resolutions for a total of $76$ simulations. Several simulations
were pushed to ${\sim}100$~ms postmerger. Together with our previous data
\citep{Bernuzzi:2015opx,Radice:2016dwd,Radice:2016rys,Radice:2017lry,Radice:2018xqa,Radice:2018pdn,Perego:2019adq,Endrizzi:2020lwl,Bernuzzi:2020txg}
these simulations form the largest sample of merger simulations with microphysics
available to date. 

The outcome of the merger was found to be very sensitive to the assumed
EOS and to the mass ratio \citep{Radice:2020ddv, Bernuzzi:2020tgt,
Bernuzzi:2020txg}. Soft EOSs and/or large mass ratios result in short
lived remnants or prompt collapse to BH. Stiffer EOSs and mass ratio
closer to one result in longer lived, possibly stable remnants. In
agreement with our previous findings, our new simulations also show that
the life time of the remnants and the accretion disk masses are strongly
correlated for comparable mass binaries \citep{Radice:2017lry,
Radice:2018pdn}. Large mass ratio binaries ($q \gtrsim 1.4$) have larger
accretion disks than comparable mass binaries and produce massive
accretion disks and tidal ejecta even when prompt BH formation occurs
\citep[see also][]{Bernuzzi:2020txg}.

The material in the disks can reach high temperatures, $O(10\, {\rm
MeV})$, especially for comparable mass ratio mergers, in which the disk
material is predominantly originating at the collisional interface
between the NSs. Due to the high-temperatures, the disk material is
initially reprocessed to intermediate values of the electron fraction
$Y_e \simeq 0.25$. However, the disks tend to evolve to a lower $Y_e$ of
about $0.1$, as expected from the theory of neutrino dominated accretion
flows \citep{Beloborodov:2008nx,Siegel:2017jug}. 

Over long timescales, the evolution of these remnants is the result of a
complicated interplay between matter accretion, driven by viscous
stresses and neutrino cooling, and matter ejection, driven by neutrino
reabsorption and hydrodynamical torques \citep[spiral waves;][]{Radice:2018xqa}. 
Our results indicate that mass ejection due to
winds can be sufficiently efficient to prevent the collapse of remnants
that have initial masses above the limit supported by uniform rotation,
the so-called hypermassive NSs. The determination of the ultimate fate of
binaries with masses that are intermediate between prompt collapse and the
maximum mass of nonrotating NSs will necessarily require long-term 3D 
neutrino-radiation GRMHD simulations.

We studied the dynamical ejection of matter during the mergers as a
function of the EOS and mass ratio. The main differences with respect to
our previous systematic study \citep{Radice:2018pdn} are that 1) the new
simulations are targeted to GW170817, so they span a smaller range of
total masses; 2) the new simulations were all performed with the M0
scheme for approximate neutrino transport; 3) our new simulations cover
a much broader range of mass ratios. We find that the inclusion of
neutrino reabsorption systematically increases the ejecta mass, as
anticipated in \citep{Sekiguchi:2015dma, Radice:2018pdn}. The ejecta
composition in our simulations is compatible with that of
\citet{Sekiguchi:2016bjd} and \citet{Vincent:2019kor} who use very
different approximation schemes for neutrinos. This suggests that modern
NR simulations are able to capture at least the leading order neutrino
effects reliably. We find that as the mass ratio is increased, the
dynamical ejecta mass increases, while velocity, and $Y_e$ decrease,
although the trend on the ejecta mass is not statistically significant,
given the large inferred numerical uncertainties. This suggests kN
observations could in principle be used to constrain the binary NS mass
ratio. Fits to ejecta and disk masses as a function of the mass ratio
and the tidal parameter $\tilde\Lambda$ are discussed in a companion
paper (Nedora et al.~in prep).

If the remnant does not collapse to a BH, the dominant outflow component
is found to be the \swind{} \citep{Nedora:2019jhl}. This is an outflow
driven by spiral density waves that are launched in the disk by the
remnant NS as it undergoes the bar-mode and one-armed instabilities
\citep{Shibata:1999wm, Paschalidis:2015mla, Radice:2016gym}.  The
\swind{} generates outflows with a rate ${\sim}0.1{-}0.5\, \Msun\, {\rm
s}^{-1}$ which persist for as long as the remnant does not collapse and
until the end of our simulations (up to ${\sim}100$~ms). The ejecta have
a narrow distribution in velocities with $\langle v_\infty \rangle
\simeq 0.2\, c$ and a broad distribution in $Y_e$.

At high latitudes, we observed the emergence of a \nwind{} from the
remnants. This high-$Y_e$ outflow component has characteristics that are
initially similar to those of the $\nu$-winds reported by,
\eg~\citet{Dessart:2008zd, Perego:2014fma, Fujibayashi:2020dvr}.
However, in our simulations the \nwind{} is quickly chocked due to the
presence of high density material that is lifted by thermal pressure
from the disk and pollutes the polar regions. On the other hand, we
remark that previous studies found the emergence of the \nwind{} only at
later times, suggesting that the right conditions for the formation of a
steady \nwind{} might not have been reached in our simulations yet. At
the same time, we cannot exclude that the lack of \nwind{} arises due
to a deficiency in our approximate neutrino treatment. The emergence of
the \nwind{} should be revisited once better neutrino transport schemes
are available.

We performed nucleosynthesis calculations to analyze the $r$-process
yields in the dynamical ejecta and the \swind{}. We find that, because
of the strong dependency of $Y_e$ on $q$, the yields are sensitive to
the binary mass ratio. In particular, very asymmetric binaries produce
larger quantities of actinides. Symmetric binaries, instead, tend to
produce lighter elements. When the \swind{} is included in the
nucleosynthesis calculations, we find that the full solar $r$-process
pattern down to $A \simeq 100$ can be reproduced. However, high
mass-ratio NSNS mergers (or BHNS mergers) appear to be required to
explain the production of actinides.

None of our simulations produce outflows with properties compatible with
those inferred from the direct fitting of simple color light curve
models to AT2017gfo \citep{Villar:2017wcc}. However, anisotropic
multi-components kN models informed with our NR data can reproduce some
of the key features of AT2017gfo \citep{Perego:2017wtu, Nedora:2019jhl}.
In particular, the optical emission at 1 day can be explained with a
combination of dynamical ejecta and \swind{} from long-lived binaries.
However, the rapid collapse of the merger remnant cannot be excluded.
For example, \citet{Fujibayashi:2020qda} found that the kind of
high-$Y_e$ material needed to explain the optical data from AT2017gfo
might also be produced in winds from BH-torus systems. The infrared
emission from AT2017gfo can only be explained assuming that ${\sim}20\%$
of the remnant disk are unbind by viscous processes and nuclear
recombination on a timescale of a few seconds
\citep[\eg][]{Metzger:2008av}.

Future work should address the limitations of this study.
Self-consistent 3D simulations of NS merger systems forming BHs or
massive NSs and spanning even longer timescales up to a few seconds are
needed to confirm whether or not AT2017gfo can be explained from first
principles. Over these timescales, the use of real neutrino transport
schemes, such as grey or spectral M1 \citep{Foucart:2016rxm,
Roberts:2016lzn}, is imperative, since leakage-based schemes, such as
our M0 scheme or the M1-leakage scheme of \citet{Sekiguchi:2015dma,
Fujibayashi:2017puw}, cannot correctly treat the diffusion of neutrinos
from the interior of the remnant. Finally, the impact of MHD effects in
the postmerger still needs to be clarified: they are likely crucial for
the launching of jets in NS mergers \citep{Ruiz:2016rai}, but their
impact on mass ejection and nucleosynthesis is not as clear
\citep{Siegel:2017jug, Fernandez:2018kax}.

\acknowledgments
  S.B. and B.D. acknowledge support by the EU H2020 under ERC Starting
  Grant, no.~BinGraSp-714626.  
  Numerical relativity simulations were performed on the supercomputer
  SuperMUC at the LRZ Munich (Gauss project pn56zo),
  on supercomputer Marconi at CINECA (ISCRA-B project number
  HP10BMHFQQ);
  on the supercomputers Bridges, Comet, and Stampede 
  (NSF XSEDE allocation TG-PHY160025); on NSF/NCSA Blue Waters (NSF
  AWD-1811236); on ARA cluster at Jena FSU.
  This research used resources of the National Energy Research
  Scientific Computing Center, a DOE Office of Science User Facility
  supported by the Office of Science of the U.S.~Department of Energy
  under Contract No.~DE-AC02-05CH11231.
  Data postprocessing was performed on the Virgo ``Tullio'' server 
  at Torino supported by INFN.
  The authors gratefully acknowledge the Gauss Centre for Supercomputing
  e.V. (\url{www.gauss-centre.eu}) for funding this project by providing
  computing time on the GCS Supercomputer SuperMUC at Leibniz
  Supercomputing Centre (\url{www.lrz.de}). 

\input{paper20200812.bbl}
\end{document}

%% file: tabSim.tex
\begin{table*}[t]
\begin{center}
\label{tab:sim}
    \caption{
      Summary table of all the simulations and dynamical ejecta properties. The columns contain
      the following information, starting from the left. Equation of
      state, mass-ratio, available resolutions,
      inclusion of subgrid turbulence, time of the
      simulation end, time of the BH formation for LR, SR, HR
      resolutions separately, time of last output, time the disk mass
      is extracted, disk mass, mass of the
      dynamical ejecta, mass-averaged electron fracton, terminal
      velocity and RMS angle (from the binary plane) for dynamical ejecta. For all
      data except $t_{BH}$, $t_{\rm end}$ and $t_{\rm disk}$, the value that is given is a mean value across resolutions, with an error estimated as one
      standard diviaion from the mean. In case where only one
      resolution is present, the error is assumed to be $20\%$ of the
      value.}
\scalebox{0.9}{
\begin{tabular}{c c c c c c c c c c c c}
    \hline\hline
    EOS & $q$ & Resolution & GRLES & $t_{\text{end}}$ & $t_{\text{BH}}$ & $t_{\text{disk}}$ & $M_{\text{disk}} ^{\text{last}}$ & $\md$ & $\langle \yd \rangle$ & $\langle \vd  \rangle$ & $\langle \theta_{\text{ej}} ^{\text{d}} \rangle$ \\
    &   &   &   & [ms] & [ms] & [ms] &   & $[10^{-2} M_{\odot}]$ &   & $[c]$ &   \\ 
    \hline
    \hline
    BLh & 1.00 & \texttt{LR SR HR} & \cmark & $43.3$ $91.8$ $23.1$ & $>43.3$ $>91.8$ $>23.1$ & 23.1 & $0.166^{+0.052} _{-0.052} $ & $0.14^{+0.02} _{-0.02} $ & $0.27^{+0.01} _{-0.01} $ & $0.17^{+0.01} _{-0.01} $ & $39.65^{+0.35} _{-0.35} $ \\
    BLh & 1.00 & \texttt{LR SR} & \xmark & $36.9$ $15.5$ $ $ & $>36.9$ $>15.5$ $ $ & 36.6 & $0.182^{+0.091} _{-0.091} $ & $0.21^{+0.04} _{-0.04} $ & $0.26^{+0.01} _{-0.01} $ & $0.18^{+0.01} _{-0.01} $ & $36.29^{+0.24} _{-0.24} $ \\
    \hline
    BLh & 1.18 & \texttt{LR} & \cmark & $69.4$ $ $ $ $ & $>69.4$ $ $ $ $ & 69.0 & $0.202^{+0.101} _{-0.101} $ & $0.30^{+0.06} _{-0.06} $ & $0.18^{+0.04} _{-0.04} $ & $0.19^{+0.04} _{-0.04} $ & $33.65^{+6.73} _{-6.73} $ \\
    BLh & 1.18 & \texttt{LR} & \xmark & $16.4$ $ $ $ $ & $>16.4$ $ $ $ $ & 15.9 & $0.229^{+0.115} _{-0.115} $ & $0.25^{+0.05} _{-0.05} $ & $0.16^{+0.03} _{-0.03} $ & $0.20^{+0.04} _{-0.04} $ & $30.86^{+6.17} _{-6.17} $ \\
    \hline
    BLh & 1.34 & \texttt{LR SR} & \cmark & $63.4$ $9.8$ $ $ & $>63.4$ $>9.8$ $ $ & 9.8 & $0.192^{+0.004} _{-0.004} $ & $0.25^{+0.05} _{-0.05} $ & $0.14^{+0.04} _{-0.04} $ & $0.17^{+0.00} _{-0.00} $ & $28.79^{+5.00} _{-5.00} $ \\
    BLh & 1.34 & \texttt{LR} & \xmark & $18.0$ $ $ $ $ & $>18.0$ $ $ $ $ & 18.0 & $0.211^{+0.106} _{-0.106} $ & $0.19^{+0.04} _{-0.04} $ & $0.17^{+0.03} _{-0.03} $ & $0.17^{+0.03} _{-0.03} $ & $33.39^{+6.68} _{-6.68} $ \\
    \hline
    BLh & 1.43 & \texttt{LR SR} & \cmark & $35.1$ $59.6$ $ $ & $>35.1$ $>59.6$ $ $ & 33.8 & $0.265^{+0.001} _{-0.001} $ & $0.27^{+0.08} _{-0.08} $ & $0.19^{+0.03} _{-0.03} $ & $0.16^{+0.00} _{-0.00} $ & $34.49^{+3.59} _{-3.59} $ \\
    \hline
    BLh & 1.54 & \texttt{LR} & \cmark & $45.8$ $ $ $ $ & $>45.8$ $ $ $ $ & 53.8 & $0.324^{+0.162} _{-0.162} $ & $0.20^{+0.04} _{-0.04} $ & $0.17^{+0.03} _{-0.03} $ & $0.13^{+0.03} _{-0.03} $ & $31.21^{+6.24} _{-6.24} $ \\
    BLh & 1.54 & \texttt{LR} & \xmark & $17.4$ $ $ $ $ & $>17.4$ $ $ $ $ & 30.1 & $0.287^{+0.144} _{-0.144} $ & $0.22^{+0.04} _{-0.04} $ & $0.21^{+0.04} _{-0.04} $ & $0.16^{+0.03} _{-0.03} $ & $35.05^{+7.01} _{-7.01} $ \\
    \hline
    BLh & 1.66 & \texttt{LR SR} & \cmark & $64.6$ $20.1$ $ $ & $>64.6$ $1.8$ $ $ & 19.2 & $0.289^{+0.005} _{-0.005} $ & $0.42^{+0.05} _{-0.05} $ & $0.11^{+0.01} _{-0.01} $ & $0.12^{+0.01} _{-0.01} $ & $24.08^{+0.29} _{-0.29} $ \\
    \hline
    BLh & 1.82 & \texttt{LR SR HR} & \cmark & $12.0$ $17.5$ $9.6$ & $1.4$ $1.4$ $1.5$ & 5.9 & $0.170^{+0.001} _{-0.001} $ & $0.81^{+0.04} _{-0.04} $ & $0.03^{+0.01} _{-0.01} $ & $0.11^{+0.00} _{-0.00} $ & $6.53^{+0.65} _{-0.65} $ \\
    BLh & 1.82 & \texttt{LR SR HR} & \xmark & $53.8$ $26.3$ $45.2$ & $1.7$ $1.3$ $1.0$ & 43.2 & $0.098^{+0.049} _{-0.049} $ & $1.07^{+0.07} _{-0.07} $ & $0.03^{+0.01} _{-0.01} $ & $0.12^{+0.00} _{-0.00} $ & $6.27^{+0.53} _{-0.53} $ \\
    \hline
    \hline
    DD2 & 1.00 & \texttt{LR SR HR} & \xmark & $92.0$ $110.2$ $8.1$ & $>92.0$ $>110.2$ $>8.1$ & 9.4 & $0.154^{+0.052} _{-0.052} $ & $0.09^{+0.03} _{-0.03} $ & $0.24^{+0.00} _{-0.00} $ & $0.19^{+0.02} _{-0.02} $ & $37.33^{+1.33} _{-1.33} $ \\
    DD2 & 1.00 & \texttt{LR SR HR} & \cmark & $123.0$ $113.0$ $74.4$ & $>123.0$ $>113.0$ $>74.4$ & 8.2 & $0.111^{+0.040} _{-0.040} $ & $0.12^{+0.03} _{-0.03} $ & $0.27^{+0.01} _{-0.01} $ & $0.16^{+0.00} _{-0.00} $ & $40.03^{+0.71} _{-0.71} $ \\
    \hline
    DD2 & 1.20 & \texttt{LR SR HR} & \xmark & $37.3$ $91.0$ $55.2$ & $>37.3$ $>91.0$ $>55.2$ & 36.6 & $0.261^{+0.028} _{-0.028} $ & $0.21^{+0.08} _{-0.08} $ & $0.18^{+0.03} _{-0.03} $ & $0.17^{+0.01} _{-0.01} $ & $29.07^{+3.75} _{-3.75} $ \\
    DD2 & 1.22 & \texttt{LR SR HR} & \cmark & $42.7$ $107.3$ $19.8$ & $>42.7$ $>107.3$ $>19.8$ & 8.7 & $0.209^{+0.033} _{-0.033} $ & $0.25^{+0.02} _{-0.02} $ & $0.19^{+0.01} _{-0.01} $ & $0.17^{+0.01} _{-0.01} $ & $30.74^{+0.89} _{-0.89} $ \\
    \hline
    DD2 & 1.43 & \texttt{LR SR} & \cmark & $37.7$ $62.0$ $ $ & $>37.7$ $>62.0$ $ $ & 36.7 & $0.304^{+0.051} _{-0.051} $ & $0.70^{+0.64} _{-0.64} $ & $0.14^{+0.05} _{-0.05} $ & $0.14^{+0.01} _{-0.01} $ & $25.51^{+9.58} _{-9.58} $ \\
    \hline
    \hline
    LS220 & 1.00 & \texttt{LR SR} & \cmark & $27.0$ $27.1$ $ $ & $13.7$ $13.7$ $ $ & 16.1 & $0.073^{+0.032} _{-0.032} $ & $0.16^{+0.02} _{-0.02} $ & $0.25^{+0.02} _{-0.02} $ & $0.16^{+0.01} _{-0.01} $ & $35.70^{+0.78} _{-0.78} $ \\
    LS220 & 1.00 & \texttt{LR SR HR} & \xmark & $35.9$ $37.2$ $27.1$ & $33.4$ $16.1$ $15.4$ & 34.6 & $0.072^{+0.006} _{-0.006} $ & $0.16^{+0.06} _{-0.06} $ & $0.22^{+0.00} _{-0.00} $ & $0.16^{+0.01} _{-0.01} $ & $34.99^{+1.68} _{-1.68} $ \\
    \hline
    LS220 & 1.05 & \texttt{SR HR} & \xmark & $ $ $23.3$ $24.1$ & $ $ $17.3$ $13.9$ & 22.3 & $0.107^{+0.054} _{-0.054} $ & $0.16^{+0.02} _{-0.02} $ & $0.21^{+0.01} _{-0.01} $ & $0.16^{+0.01} _{-0.01} $ & $33.28^{+2.37} _{-2.37} $ \\
    LS220 & 1.11 & \texttt{SR HR} & \xmark & $ $ $25.1$ $24.4$ & $ $ $17.0$ $>24.4$ & 24.2 & $0.140^{+0.071} _{-0.071} $ & $0.22^{+0.03} _{-0.03} $ & $0.19^{+0.02} _{-0.02} $ & $0.18^{+0.02} _{-0.02} $ & $30.25^{+4.43} _{-4.43} $ \\
    \hline
    LS220 & 1.16 & \texttt{SR HR} & \cmark & $ $ $95.8$ $11.3$ & $ $ $68.9$ $>11.3$ & 95.5 & $0.306^{+0.153} _{-0.153} $ & $0.34^{+0.00} _{-0.00} $ & $0.22^{+0.00} _{-0.00} $ & $0.16^{+0.00} _{-0.00} $ & $34.08^{+1.00} _{-1.00} $ \\
    LS220 & 1.16 & \texttt{LR SR HR} & \xmark & $29.5$ $36.1$ $28.8$ & $>29.5$ $>36.1$ $24.1$ & - & - & $0.33^{+0.05} _{-0.05} $ & $0.17^{+0.01} _{-0.01} $ & $0.17^{+0.01} _{-0.01} $ & $30.01^{+0.64} _{-0.64} $ \\
    \hline
    LS220 & 1.43 & \texttt{LR SR} & \cmark & $19.8$ $28.5$ $ $ & $15.7$ $12.3$ $ $ & 19.6 & $0.178^{+0.072} _{-0.072} $ & $0.73^{+0.03} _{-0.03} $ & $0.16^{+0.02} _{-0.02} $ & $0.17^{+0.01} _{-0.01} $ & $26.77^{+3.50} _{-3.50} $ \\
    \hline
    LS220 & 1.66 & \texttt{LR SR} & \cmark & $6.8$ $8.0$ $ $ & $1.4$ $2.1$ $ $ & 2.0 & $0.068^{+0.008} _{-0.008} $ & $1.11^{+0.38} _{-0.38} $ & $0.07^{+0.01} _{-0.01} $ & $0.14^{+0.01} _{-0.01} $ & $13.18^{+1.33} _{-1.33} $ \\
    \hline
    \hline
    SFHo & 1.00 & \texttt{SR HR} & \cmark & $ $ $25.3$ $11.6$ & $ $ $6.0$ $4.0$ & 50.0 & $0.023^{+0.012} _{-0.012} $ & $0.40^{+0.07} _{-0.07} $ & $0.21^{+0.00} _{-0.00} $ & $0.19^{+0.01} _{-0.01} $ & $32.48^{+1.79} _{-1.79} $ \\
    SFHo & 1.00 & \texttt{LR SR HR} & \xmark & $3.2$ $7.7$ $9.0$ & $>3.2$ $4.1$ $3.8$ & 7.2 & $0.019^{+0.007} _{-0.007} $ & $0.28^{+0.07} _{-0.07} $ & $0.23^{+0.01} _{-0.01} $ & $0.21^{+0.01} _{-0.01} $ & $31.66^{+1.80} _{-1.80} $ \\
    \hline
    SFHo & 1.13 & \texttt{SR HR} & \cmark & $ $ $14.2$ $14.3$ & $ $ $6.3$ $>14.3$ & - & - & $0.44^{+0.12} _{-0.12} $ & $0.18^{+0.01} _{-0.01} $ & $0.23^{+0.01} _{-0.01} $ & $33.20^{+0.78} _{-0.78} $ \\
    SFHo & 1.13 & \texttt{LR SR HR} & \xmark & $16.5$ $19.3$ $15.2$ & $5.5$ $11.6$ $3.9$ & 15.1 & $0.046^{+0.041} _{-0.041} $ & $0.42^{+0.03} _{-0.03} $ & $0.17^{+0.03} _{-0.03} $ & $0.22^{+0.01} _{-0.01} $ & $29.63^{+4.39} _{-4.39} $ \\
    \hline
    SFHo & 1.43 & \texttt{LR} & \cmark & $19.6$ $ $ $ $ & $4.8$ $ $ $ $ & 18.9 & $0.201^{+0.101} _{-0.101} $ & $0.38^{+0.08} _{-0.08} $ & $0.14^{+0.03} _{-0.03} $ & $0.20^{+0.04} _{-0.04} $ & $29.20^{+5.84} _{-5.84} $ \\
    SFHo & 1.43 & \texttt{SR} & \cmark & $ $ $46.5$ $ $ & $ $ $>46.5$ $ $ & 50.8 & $0.241^{+0.121} _{-0.121} $ & $0.24^{+0.05} _{-0.05} $ & $0.19^{+0.04} _{-0.04} $ & $0.14^{+0.03} _{-0.03} $ & $32.86^{+6.57} _{-6.57} $ \\
    \hline
    SFHo & 1.66 & \texttt{LR SR} & \cmark & $11.2$ $16.8$ $ $ & $1.3$ $1.3$ $ $ & 11.6 & $0.177^{+0.153} _{-0.153} $ & $0.15^{+0.00} _{-0.00} $ & $0.07^{+0.00} _{-0.00} $ & $0.12^{+0.01} _{-0.01} $ & $10.39^{+1.14} _{-1.14} $ \\
    \hline
    \hline
    SLy4 & 1.00 & \texttt{LR SR} & \cmark & $10.5$ $13.1$ $ $ & $2.8$ $2.8$ $ $ & - & - & $0.09^{+0.02} _{-0.02} $ & $0.23^{+0.02} _{-0.02} $ & $0.27^{+0.02} _{-0.02} $ & $30.81^{+2.81} _{-2.81} $ \\
    SLy4 & 1.00 & \texttt{LR SR} & \xmark & $12.7$ $22.0$ $ $ & $2.7$ $13.8$ $ $ & 12.5 & $0.071^{+0.175} _{-0.175} $ & $0.36^{+0.13} _{-0.13} $ & $0.23^{+0.04} _{-0.04} $ & $0.26^{+0.06} _{-0.06} $ & $35.67^{+0.03} _{-0.03} $ \\
    \hline
    SLy4 & 1.13 & \texttt{LR SR} & \xmark & $8.4$ $20.3$ $ $ & $>8.4$ $13.0$ $ $ & 8.0 & $0.164^{+0.023} _{-0.023} $ & $0.59^{+0.07} _{-0.07} $ & $0.16^{+0.00} _{-0.00} $ & $0.24^{+0.01} _{-0.01} $ & $29.67^{+1.97} _{-1.97} $ \\
    \hline
    SLy4 & 1.43 & \texttt{SR} & \cmark & $ $ $40.3$ $ $ & $ $ $>40.3$ $ $ & 45.2 & $0.200^{+0.100} _{-0.100} $ & $0.20^{+0.04} _{-0.04} $ & $0.21^{+0.04} _{-0.04} $ & $0.15^{+0.03} _{-0.03} $ & $34.03^{+6.81} _{-6.81} $ \\
    \hline
    SLy4 & 1.66 & \texttt{SR} & \cmark & $ $ $7.2$ $ $ & $ $ $1.2$ $ $ & 3.9 & $0.138^{+0.069} _{-0.069} $ & $0.28^{+0.06} _{-0.06} $ & $0.05^{+0.01} _{-0.01} $ & $0.12^{+0.02} _{-0.02} $ & $8.43^{+1.69} _{-1.69} $ \\
    \hline\hline
\end{tabular}
}
\end{center}
\end{table*}

%% file: tabFitsPoly22.tex

\begin{table}[t]
  \centering
  \caption{\label{tab:fitpoly22coefs}Coefficients for the polynomial
    regression with Eq.~\eqref{eq:fit:poly22} of the data with chirp
    mass $\M_c=1.188\Msun$ in this paper. The last rows reports 
    coefficient of determination $R^2$ of the fit.}
  \scalebox{0.85}{
    \begin{tabular}{c|cccc}
      \hline\hline
      & $\amd$ $[10^{3}\Msun]$ & $\avd$ $[c]$ & $\ayd$ & $M_{\rm disk}$ $[\Msun]$\\
      \hline
      $b_0$ & $54.247$               & $0.677$               & $-6.607\times10^{-2}$ & $-1.752$\\
      $b_1$ & $-57.32$               & $-0.182$              & $0.318$               & $2.272$\\
      $b_2$ & $-6.887\times10^{-2}$  & $-1.083\times10^{-3}$ & $6.084\times10^{-4}$  & $1.139\times10^{-3}$\\
      $b_3$ & $13.604$               & $-4.912\times10^{-2}$ & $-0.155$              & $-0.730$\\
      $b_4$ & $4.831\times10^{-2}$                & $3.893\times10^{-4}$  & $-2.055\times10^{-4}$ & $-2.921\times10^{-4}$\\
      $b_5$ & $1.067\times10^{-5}$   & $4.239\times10^{-7}$  & $-2.896\times10^{-7}$ & $-5.532\times10^{-7}$\\
      \hline
  $R^2$ & $0.716$ & $0.779$ & $0.769$ & $0.498$ \\
  \hline\hline
\end{tabular}
}
\end{table}

%% file: tabWind.tex
\begin{table*}[t]
  \centering
  \caption{%
    Summary table of the \swind{} properties of long-lived remnants. The columns contain
    the following information, starting from the left. Equation of
    state, mass-ratio, available resolutions,
    inclusion of subgrid turbulence, time of the
    simulation end, mass of the
    \swind{}, mass-loss rate via \swind, mass-averaged electron fracton, terminal
    velocity and, finally, RMS angle for \swind{}. For these four
    quantities we give the mean value among the resolutions and
    one-sigma deviations. For binaries for which only one
    resolution is present, the error is assumed to be $20\%$ of the value.}
  \begin{tabular}{c c c c c c c c c c}
    \hline\hline
    EOS & $q$ & Resolution & GRLES & $t_{\text{end}}$ & $\amw$ & $\amw/\Delta t$ & $\ayw$ & $\avw$ & $\langle \theta_{\text{ej}}^{\text{w}} \rangle$ \\
    &   &   &   & [ms] & $[10^{-2} M_{\odot}]$ & $[M_{\odot}/s]$ &   & $[c]$ &   \\ 
    \hline
    \hline
    BLh & 1.18 & \texttt{LR} & \cmark & $69.4$ $ $ $ $ & $1.28^{+0.64} _{-0.64} $ & $1.23^{+0.25} _{-0.25} $ & $0.33^{+0.01} _{-0.01} $ & $0.11^{+0.02} _{-0.02} $ & $14.98^{+2.00} _{-2.00} $ \\
    \hline
    BLh & 1.43 & \texttt{LR SR} & \cmark & $35.1$ $59.6$ $ $ & $0.75^{+0.18} _{-0.18} $ & $1.06^{+0.67} _{-0.67} $ & $0.27^{+0.01} _{-0.01} $ & $0.09^{+0.01} _{-0.01} $ & $19.43^{+2.22} _{-2.22} $ \\
    \hline
    BLh & 1.54 & \texttt{LR} & \cmark & $45.8$ $ $ $ $ & $0.63^{+0.32} _{-0.32} $ & $0.44^{+0.09} _{-0.09} $ & $0.32^{+0.01} _{-0.01} $ & $0.10^{+0.02} _{-0.02} $ & $21.46^{+2.00} _{-2.00} $ \\
    \hline
    BLh & 1.66 & \texttt{LR SR} & \cmark & $64.6$ $20.1$ $ $ & $0.12^{+0.09} _{-0.09} $ & $0.37^{+0.34} _{-0.34} $ & $0.33^{+0.05} _{-0.05} $ & $0.13^{+0.01} _{-0.01} $ & $52.08^{+20.89} _{-20.89} $ \\
    \hline
    \hline
    DD2 & 1.00 & \texttt{LR SR HR} & \cmark & $123.0$ $113.0$ $74.4$ & $1.25^{+0.14} _{-0.14} $ & $1.30^{+0.19} _{-0.19} $ & $0.30^{+0.01} _{-0.01} $ & $0.17^{+0.00} _{-0.00} $ & $14.88^{+0.87} _{-0.87} $ \\
    \hline
    DD2 & 1.20 & \texttt{LR SR HR} & \xmark & $37.3$ $91.0$ $55.2$ & $0.48^{+0.09} _{-0.09} $ & $0.74^{+0.24} _{-0.24} $ & $0.26^{+0.01} _{-0.01} $ & $0.15^{+0.00} _{-0.00} $ & $24.54^{+2.23} _{-2.23} $ \\
    \hline
    DD2 & 1.43 & \texttt{LR SR} & \cmark & $37.7$ $62.0$ $ $ & $0.60^{+0.02} _{-0.02} $ & $0.51^{+0.06} _{-0.06} $ & $0.23^{+0.12} _{-0.12} $ & $0.16^{+0.00} _{-0.00} $ & $21.74^{+0.03} _{-0.03} $ \\
    \hline
    \hline
    SFHo & 1.43 & \texttt{SR} & \cmark & $ $ $46.5$ $ $ & $0.58^{+0.30} _{-0.30} $ & $0.43^{+0.09} _{-0.09} $ & $0.31^{+0.01} _{-0.01} $ & $0.17^{+0.02} _{-0.02} $ & $22.67^{+2.00} _{-2.00} $ \\
    \hline
    \hline
    SLy4 & 1.43 & \texttt{SR} & \cmark & $ $ $40.3$ $ $ & $0.53^{+0.27} _{-0.27} $ & $0.38^{+0.08} _{-0.08} $ & $0.29^{+0.01} _{-0.01} $ & $0.18^{+0.02} _{-0.02} $ & $23.52^{+2.00} _{-2.00} $ \\
    \hline\hline
\end{tabular}
\label{tab:spiralwavewind}
\end{table*}